\pdfoutput=1
\documentclass[cernpreprint, atlasdraft=false, texlive=2023, UKenglish, texmf, orcidlogo]{atlasdoc}
 
\usepackage{atlaspackage}
\usepackage{atlasbiblatex}
 
\usepackage{atlasphysics}
\usepackage{tabularx}
\usepackage{multirow}
\usepackage{dcolumn}
\usepackage{amsfonts}
\usepackage{array}   
\usepackage{setspace}
 
\graphicspath{{logos/}{figures/}{figs/}}
 
\usepackage{ANA-HIGG-2019-11-PAPER-defs}

\AtlasTitle{Searches for lepton-flavour-violating decays of the Higgs boson into $e\tau$ and $\mu\tau$
in $\sqrt{s}=13$\,TeV $pp$ collisions with the ATLAS detector }
 
\AtlasAbstract{
This paper presents direct searches for lepton flavour violation in
Higgs boson decays, $H\rightarrow e\tau$ and $H\rightarrow\mu\tau$,
performed using data collected with the ATLAS detector at the LHC. The
searches are based on a data sample of proton--proton collisions at a
centre-of-mass energy $\sqrt{s} = 13$\,TeV, corresponding to an
integrated luminosity of 138~fb$^{-1}$.
Leptonic ($\tau \to \ell \nu_\ell \nu_\tau$) and hadronic ($\tau \to \text{hadrons } \nu_\tau$) decays of the $\tau$-lepton are considered.
Two background estimation techniques are
employed: the MC-template method, based on data-corrected simulation
samples, and the Symmetry method, based on exploiting the symmetry
between electrons and muons in the Standard Model backgrounds.
No significant excess of events is observed and the results are interpreted as
upper limits on lepton-flavour-violating branching ratios of the Higgs
boson.  The observed (expected) upper limits set on the branching ratios at
95\% confidence level, $\mathcal{B}(H\to e\tau)<0.20\%$ ($0.12\%$)
and $\mathcal{B}(H\to \mu\tau)<0.18\%$ ($0.09\%$), are obtained with
the MC-template method from a simultaneous measurement of potential
$H \rightarrow e\tau$ and $H \rightarrow\mu\tau$ signals.  The
best-fit branching ratio difference,
$\mathcal{B}(H\to \mu\tau)- \mathcal{B}(H\to e\tau)$,
measured with the Symmetry method in the channel where the $\tau$-lepton
decays to leptons, is $(0.25 \pm 0.10)\%$, compatible with a value of zero within $2.5\sigma$.
}
 
\author{The ATLAS Collaboration}
 
\AtlasRefCode{HIGG-2019-11}

\AtlasJournal{JHEP}
\AtlasJournalRef{JHEP \textbf{07} (2023) 166}
\AtlasDOI{10.1007/JHEP07(2023)166}
\PreprintIdNumber{CERN-EP-2022-279}

\hypersetup{pdftitle={ATLAS document},pdfauthor={The ATLAS Collaboration}}
 
\begin{document}
 
\maketitle
 
\tableofcontents

\clearpage
 
\section{Introduction}

One of the main goals of the Large Hadron Collider (LHC) physics
programme at CERN is to discover physics beyond the Standard Model
(SM).  The discovery of a scalar Higgs boson at the
LHC~\cite{ATLASHiggsJuly2012, CMSHiggsJuly2012} has provided important
insight into the mechanism of electroweak symmetry
breaking~\cite{ENGLERT,HIGGS,HIGGS2,HIGGS3,Guralnik:1964eu,Kibble:1967sv} and made it possible to search
for physics phenomena beyond the SM (BSM physics phenomena) in the Higgs sector.
A possible sign of new physics would be the observation of lepton
flavour violation (LFV) in decays of the Higgs boson into a pair of
leptons with different flavours.
 
The observation of neutrino oscillations indicates that LFV is
realised in nature and that lepton flavour is not an exact symmetry,
making it possible for BSM physics to participate in flavour-changing dynamics.
LFV decays of the Higgs boson occur naturally in models with more than one Higgs
doublet~\cite{Bjorken:1977vt,DiazCruz:1999xe,Arana-Catania:2013xma,Arhrib:2012ax,Aloni:2015wvn},
composite Higgs models~\cite{Agashe:2009di,Azatov:2009na}, models with
flavour symmetries~\cite{Ishimori:2010au} or warped extra
dimensions~\cite{Perez:2008ee,Azatov:2009na,Albrecht:2009xr} and
other models~\cite{Goudelis:2011un,McKeen:2012av}.  The flavour anomalies
measured by BaBar, Belle and LHCb~\cite{BaBar1,PhysRevLett.115.111803,LHCb:2017smo,LHCb:2015tgy,Belle1}
could be linked to LFV decays of the Higgs boson or other massive
particles~\cite{Ko:2017lzd,Carmona:2015ena,Botella:2015hoa}.
 
The most stringent bounds on the LFV decays of the Higgs boson, \hte{} and \htm{},
are derived from direct searches.  These include a previous ATLAS
search~\cite{HIGG-2017-08} which placed 95\% confidence
level (CL) upper limits on the branching ratios (\BR) of $H\to e\tau$ and
$H\to\mu\tau$ at $0.47\%$ and $0.28\%$, respectively, using data
collected at $\sqrt{s}=13\,\TeV$, corresponding to an integrated
luminosity of $36.1\,\ifb$.  Likewise, the CMS Collaboration set
95\% CL upper limits restricting the branching ratios to
$\BR(H \to e \tau) < 0.22\%$ and
$\BR(H \to \mu \tau) < 0.15\%$ using data collected at
$\sqrt{s}=13\,\TeV$, with an integrated luminosity of
$137\,\ifb$~\cite{CMS-HIG-20-009}. The ATLAS Collaboration performed a direct search
for $H \to e \mu$ decay and obtained a 95\% CL upper limit on the
branching ratio value of $\BR(H \to e \mu) <  6.1\times 10^{-5}$~\cite{HIGG-2018-58}, using data collected at $\sqrt{s}=13\,\TeV$, with an integrated luminosity of $139\,\ifb$.
The most stringent indirect constraint on the $H \to e \mu$ decay is derived
from the results of searches for $\mu \to e \gamma$ decays~\cite{MEG:2013oxv},
and a bound of $\BR(H \to e \mu) < O(10^{-8})$ is
obtained~\cite{Harnik:2012pb,Blankenburg:2012ex}.
 
This document presents searches for two LFV decays of the Higgs boson,
$H\rightarrow e\tau$ and $H\rightarrow\mu\tau$, with the ATLAS experiment at the LHC.
The two decay modes \leplep{} and \lephad{} illustrated in Figure~\ref{fig:decay} are considered
for each search, where $\ell/\ell^\prime$ is used to denote electrons and muons,\footnote{Throughout this document the inclusion of
charge-conjugate decay modes is implied.} also referred to as \enquote{light leptons}.
The former exploits the leptonic $\tau$-decay mode $\tau\to\ell^\prime\nu\bar{\nu}$,
while the latter exploits the hadronic $\tau$-decay mode $\tau\to\mathrm{hadrons}+\nu$.
 
\begin{figure}[!htb]
\centering
\subfloat[]{\includegraphics[width=0.45\textwidth]{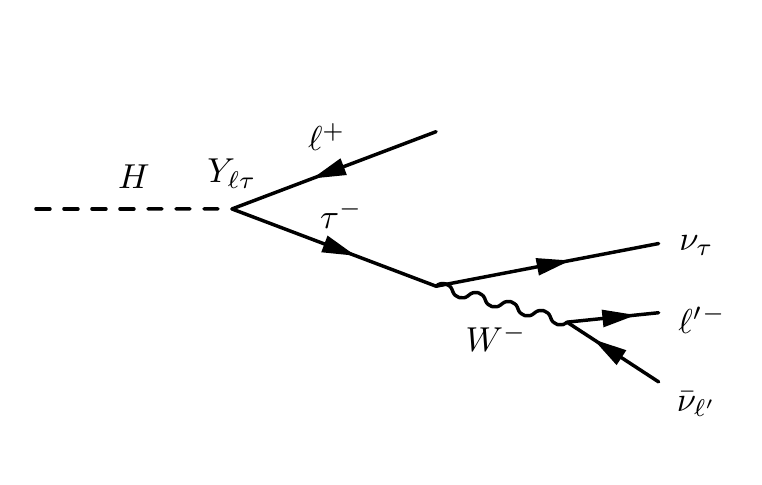}}\hspace{10mm}
\subfloat[]{\includegraphics[width=0.45\textwidth]{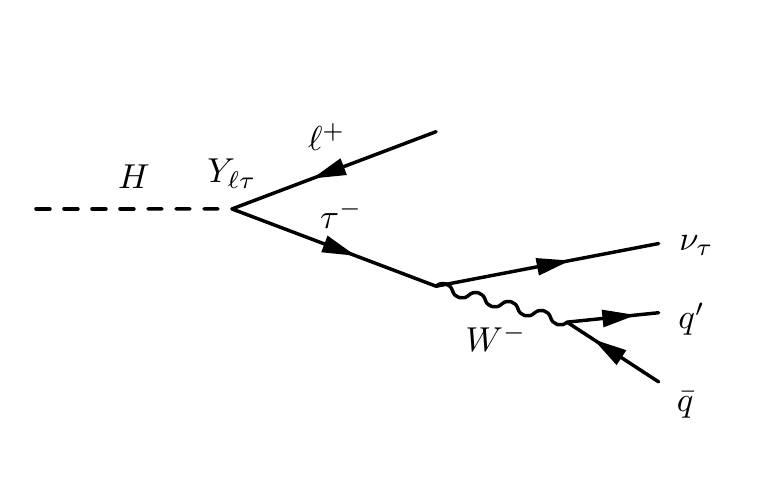} } \\
\caption{LFV decay schemes of the Higgs boson for the (a) \leplep{} and (b) \lephad{} final states.  The off-diagonal Yukawa
coupling term is indicated by the $Y_{\ell \tau}$
symbol.}\label{fig:decay}
\end{figure}
 
Due to the large background of same-flavour lepton pairs produced by Drell--Yan processes,
the \leplep{} final state only considers pairs of different-flavour leptons.
 
Two independent methods are exploited to estimate the background in the \leplep{}
final state: \enquote{\MCbased} and \enquote{\Symbased}.  The first method
uses templates from Monte Carlo (MC) simulation, where the
normalisation of the two main backgrounds is obtained from data, and a
data-driven estimate of the \enquote{\fakebkg}.  The
second method relies on the assumption that the
prompt-lepton\footnote{Leptons from the decay of the Higgs boson,
heavy vector bosons, and $\tau$-leptons are considered
prompt leptons.} backgrounds in the SM are symmetric under the exchange of electrons
and muons to derive a data-driven background estimate for the main
backgrounds~\cite{Bressler:2014jta}.  A separate data-driven estimate
is employed for the \fakebkg{}, and simulation is used for the remaining minor
background contributions.
For the \lephad{} final state, only the \MCbased{} method is used.
The \MCbased{} method targets the measurement of the actual values of
\BR(\hte) and \BR(\htm) individually, while the \Symbased{} method is
sensitive to the difference of the branching ratios.
A multivariate analysis (MVA) technique is developed for each final state in
both methods to achieve maximum separation between signal and
background.
 
Three statistical analyses are performed: one independent
search for each of the \hte{} and \htm{} processes,
and one simultaneous determination of the \hte{} and \htm{} signals.
In the independent search for the \hte{} process,
the \htm{} signal is assumed to be zero,
and vice versa in the case of the \htm{} search.
For each event category used in the statistical analysis,
the  \MCbased{} method in the  \lephad{} final state is combined
with either the \Symbased{} method or the  \MCbased{} method in
the  \leplep{} final state. The method having the higher
expected sensitivity is chosen.
In the simultaneous determination of the \hte{} and \htm{} signals,
the assumption about the absence of one of the two signals is removed.
Consequently, the  \MCbased{} method is used for both the \leplep{}
and \lephad{} final states.


\section{ATLAS detector}
 
\newcommand{\AtlasCoordFootnote}{
ATLAS uses a right-handed coordinate system with its origin at the nominal interaction point (IP)
in the centre of the detector and the \(z\)-axis along the beam pipe.
The \(x\)-axis points from the IP to the centre of the LHC ring,
and the \(y\)-axis points upwards.
Cylindrical coordinates \((r,\phi)\) are used in the transverse plane,
\(\phi\) being the azimuthal angle around the \(z\)-axis.
The pseudorapidity is defined in terms of the polar angle \(\theta\) as \(\eta = -\ln \tan(\theta/2)\).
Angular distance is measured in units of \(\Delta R \equiv \sqrt{(\Delta\eta)^{2} + (\Delta\phi)^{2}}\).}
 
The ATLAS detector~\cite{PERF-2007-01} at the LHC covers nearly the entire solid angle around the collision point.\footnote{\AtlasCoordFootnote}
It consists of an inner tracking detector surrounded by a thin superconducting solenoid, electromagnetic and hadron calorimeters,
and a muon spectrometer incorporating three large superconducting air-core toroidal magnets.
 
The inner-detector system is immersed in a \SI{2}{\tesla} axial magnetic field
and provides charged-particle tracking in the range \(|\eta| < 2.5\).
The high-granularity silicon pixel detector covers the vertex region and typically provides four measurements per track,
the first hit normally being in the insertable B-layer installed before Run~2~\cite{ATLAS-TDR-19,PIX-2018-001}.
It is followed by the silicon microstrip tracker, which usually provides eight measurements per track.
These silicon detectors are complemented by the transition radiation tracker (TRT),
which enables radially extended track reconstruction up to \(|\eta| = 2.0\).
The TRT also provides electron identification information based on a likelihood method.
 
The calorimeter system covers the pseudorapidity range \(|\eta| <
4.9\). Within the region \(|\eta|< 3.2\), electromagnetic calorimetry
is provided by barrel and endcap high-granularity lead/liquid-argon
(LAr) calorimeters, with an additional thin LAr presampler covering
\(|\eta| < 1.8\) to correct for energy loss in material upstream of
the calorimeters.  Hadron calorimetry is provided by the
steel/scintillator-tile calorimeter, segmented into three barrel
structures within \(|\eta| < 1.7\), and two copper/LAr hadron endcap
calorimeters.  The solid angle coverage is completed with forward
copper/LAr and tungsten/LAr calorimeter modules optimised for
electromagnetic and hadronic energy measurements respectively.
 
The muon spectrometer comprises separate trigger and
high-precision tracking chambers measuring the deflection of muons in
a magnetic field generated by the superconducting air-core toroidal
magnets.  The field integral of the toroids ranges between \num{2.0}
and \SI{6.0}{\tesla\metre} across most of the detector.
The region \(|\eta| < 2.7\) is covered with three layers of precision chambers
composed of monitored drift tubes, complemented by cathode-strip
chambers in the forward region, where the background is highest.
The muon trigger system covers the range \(|\eta| < 2.4\) with
resistive-plate chambers in the barrel, and thin-gap chambers in the
endcap regions.
 
Interesting events are selected by the first-level trigger system
implemented in custom hardware, followed by selections made by
algorithms implemented in software in the high-level
trigger~\cite{TRIG-2016-01}.  The first-level trigger accepts events
from the \SI{40}{\MHz} bunch crossings at a rate below \SI{100}{\kHz},
which is reduced to about \SI{1}{\kHz} by the high-level trigger and these
events are recorded to disk.
 
An extensive software suite~\cite{ATL-SOFT-PUB-2021-001} is used in data simulation,
in the reconstruction and analysis of real and simulated data, in detector
operations, and in the trigger and data acquisition systems of the experiment.


\section{Collision data and simulation samples}
 
\newcommand{\SamplesFootnote}{Additional \Vjets (QCD/EW) samples used by the \Symbased{} method.}
 
The dataset used for the searches consists of the LHC proton--proton collision data recorded
by the ATLAS experiment at $\sqrt{s}=13$~\TeV{} during
the period from 2015 to 2018. Events are selected for analysis only if they
are of good quality and if all the relevant detector components are
known to have been in good operating condition~\cite{DAPR-2018-01}.
The total integrated luminosity of the analysed data is \ilumi.
The events considered were accepted by
single-lepton or dilepton
triggers~\cite{TRIG-2018-05,TRIG-2018-01,TRIG-2019-03,TRIG-2019-02}.
The \pt thresholds of the single-lepton triggers were $\pt^{e} > 27\,(25)$~\GeV\
and $\pt^{\mu} > 27.3\,(21)$~\GeV\ for the 2016--2018 (2015) data-taking period.
The \pt thresholds of the dilepton triggers were $\pt^{e} >
18$~\GeV\ and $\pt^{\mu} > 14.7$~\GeV.
 
Simulated events are used to model $H \to \ell \tau$ signal processes, as well as most of the backgrounds from SM processes.
A summary of all the generators used for the simulation of the signal
and background processes is shown in \cref{tab:generators}, and more details can be found in Ref.~\cite{HIGG-2019-09}.
The measured Higgs boson mass of \SI{125.09}{\GeV}~\cite{HIGG-2014-14} is
assumed in the calculation of the expected cross-sections and
branching fractions. The Higgs-boson production cross-sections are fixed to the SM
predictions~\cite{LHCXS} throughout this measurement.
 
The main Higgs boson production modes at the LHC are, in descending order of predicted cross-section, gluon--gluon
fusion (\ggF), followed by vector-boson fusion (VBF), and associated
$VH$ and \ttH{} production.  For the LFV signal, \hte{} and \htm{},
the \ggF, VBF and $VH$ production mechanisms are considered. The \ttH{} production process is not considered for the LFV signal due to its negligible contribution.
The background contribution originating from the SM \Htt{} and \HWW{} decays is small and
the SM predictions are assumed for the branching ratios.  These two processes were modelled using the same
simulation strategy as for the LFV signal, but the \ttH{}
production mode was also included. Other background processes involve
electroweak production of $W/Z$ bosons via VBF, Drell--Yan production
of $W/Z$ in association with jets, and diboson, single
top-quark and top-quark pair (\ttbar{}) production.
 
All samples of simulated events were processed through the ATLAS
detector simulation~\cite{SOFT-2010-01} based on
$\GEANT$~\cite{Geant4}.  The effects of multiple interactions in the
same and neighbouring bunch crossings (pile-up) were modelled by overlaying
minimum-bias events, simulated using the soft QCD processes of
\PYTHIA[8.186]~\cite{Pythia8} with the A3~\cite{ATL-PHYS-PUB-2016-017}
set of tuned parameters and the \NNPDF[2.3lo]~\cite{Ball:2012cx} parton
distribution functions (PDF).
 
\begin{table}[h]
\caption{Overview of the MC generators used for the main signal and background samples. The last column, labelled \enquote{Order}, specifies the order of the cross-section calculation used for the normalisation of the simulated samples. More details can be found in Ref.~\cite{HIGG-2019-09}. The \enquote{$^*$} symbol denotes additional \Vjets (QCD/EW) samples used by the \Symbased{} method.}
\label{tab:generators}
\resizebox{\textwidth}{!}{
\begin{tabular}{l c c c c l l}
\toprule
Process               & \multicolumn{2}{c}{Generator}                                   & \multicolumn{2}{c}{PDF set}                                      & Tune  & Order        \\
& ME                                & PS                          & ME                                        & PS                   &       &              \\
\midrule
Higgs boson     &                                   &                             &                                           &                      &                      \\
\midrule
\ggF                   & \POWHEGBOX[v2]~\cite{Frixione:2007nw,Nason:2004rx,Frixione:2007vw,Alioli:2010xd,Hartanto:2015uka}                           & \PYTHIA[8]~\cite{Sjostrand:2014zea}                     & \PDFforLHC[15nnlo]~\cite{Butterworth:2015oua}                & \CTEQ[6L1]~\cite{Pumplin:2002vw}              & AZNLO~\cite{STDM-2012-23} & N$^{3}$LO QCD + NLO EW    \\
VBF                   & \POWHEGBOX[v2]                           & \PYTHIA[8]                    & \PDFforLHC[15nlo]                                & \CTEQ[6L1]              & AZNLO & NNLO QCD + NLO EW        \\
$VH$                  & \POWHEGBOX[v2]                           & \PYTHIA[8]                    & \PDFforLHC[15nlo]                               & \CTEQ[6L1]              & AZNLO & NNLO QCD + NLO EW          \\
\ttH                  & \POWHEGBOX[v2]                        & \PYTHIA[8]                    & \NNPDF[3.0nnlo]~\cite{Ball:2014uwa} & \NNPDF[2.3lo]~\cite{Ball:2012cx}              & A14~\cite{ATL-PHYS-PUB-2014-021}                  & NLO QCD + NLO EW                  \\
\midrule
Background            &                                   &                             &                                           &                      &                      \\
\midrule
\Vjets (QCD/EW)      & \multicolumn{2}{c}{\SHERPA[2.2.1]~\cite{Bothmann:2019yzt} }& \multicolumn{2}{c}{\NNPDF[3.0nnlo]} & \SHERPA~\cite{Schumann:2007mg}                                   & NNLO QCD + LO EW                                       \\
\Vjets (QCD/EW)$^*$ & \POWHEGBOX[v2]             & \PYTHIA[8] & \CT[10nlo]~\cite{Lai:2010vv}  & \CTEQ[6L1]  & AZNLO &    NNLO \\
Diboson               & \multicolumn{2}{c}{\SHERPA[2.2.1]} & \multicolumn{2}{c}{\NNPDF[3.0nnlo]} & \SHERPA                                   & NLO                                        \\
\ttbar                & \POWHEGBOX[v2]                           & \PYTHIA[8]                    & \NNPDF[3.0nnlo] & \NNPDF[2.3lo]               & A14                  & NNLO + NNLL          \\
Single top            & \POWHEGBOX[v2]                           & \PYTHIA[8]                    & \NNPDF[3.0nnlo] & \NNPDF[2.3lo]               & A14                  & NLO                  \\
\bottomrule
\end{tabular}
}
\end{table}


\section{Object reconstruction and event selection}
The topology of the $\htm$ and $\hte$ events requires the
reconstruction of electrons, muons, visible products of hadronically
decaying $\tau$ leptons (\tauhadvis{}), jets and missing transverse
momentum.
 
\subsection{Object reconstruction}
The tracks measured in the inner detector are used to reconstruct
the interaction vertices. The vertex with the highest
sum of squared transverse momenta of associated tracks is taken
as the primary vertex~\cite{PERF-2015-01}.
 
Electrons and photons are reconstructed from energy deposits in the
electromagnetic calorimeters~\cite{EGAM-2018-01}. Electron candidates
are matched to inner-detector tracks. They are required
to satisfy the `Loose' likelihood-based identification criterion defined in
Ref.~\cite{EGAM-2018-01}, which has an efficiency of about
$93\%$, and to have $\pT > 15\,\GeV$ and $|\eta| < 1.37$ or $ 1.52 <
|\eta| < 2.47$, thus excluding the transition region between the barrel and
endcap calorimeters.
Selected electrons
are also required to satisfy the `Gradient' isolation, which has
an efficiency of $90\%$ at $\pt=\SI{25}{\GeV}$ and $99\%$ at
$\pt=\SI{60}{\GeV}$, and `Medium' identification criteria, which have
an average efficiency of $88\%$ for typical electroweak
processes~\cite{EGAM-2018-01}.
 
Muons are reconstructed from track candidates in the muon spectrometer
matched with tracks in the inner detector.  They are required to
satisfy the `Loose' identification criterion~\cite{MUON-2018-03}, and $\pt > 10\,\GeV$
and $|\eta|< 2.5	$. These criteria have an efficiency of about
$98\%$ with good uniformity in $\pT$.
Selected muons are also required to satisfy `Medium'
identification and `Tight' isolation
criteria~\cite{MUON-2018-03}. The latter are based on calorimetric and
track information for the \MCbased{} method and on track information
only for the \Symbased{} method, where a slightly larger
track-\pt{} contribution around the muon is allowed in order to increase the number of
events available for that method's data-driven background estimation.
 
Jets are reconstructed with the \antikt algorithm~\cite{Cacciari:2008gp,Fastjet} using a radius parameter $R=0.4$.
The jet-clustering input objects are based on particle flow~\cite{PERF-2015-09} in the inner detector and the calorimeter.
Cleaning criteria are used to identify jets arising from non-collision
backgrounds or noise in the calorimeters~\cite{ATLAS-CONF-2015-029},
and events containing such jets are removed. Only jets with
$\pT>\SI{20}{\GeV}$ and $|\eta|<4.5$ are considered.
To identify and reject jets that are not associated with the primary vertex of the hard interaction
(pile-up jets), a \enquote{jet-vertex tagger} (JVT)~\cite{PERF-2014-03} algorithm is applied to
jets with $\pT<\SI{60}{\GeV}$ and $|\eta|<2.5$.
To suppress pile-up jets in the forward region, a \enquote{forward JVT}~\cite{PERF-2016-06} algorithm
which exploits jet shapes and topological jet correlations in pile-up interactions is applied
to all jets with $\pT<\SI{60}{\GeV}$ and $|\eta|>2.5$.
Jets with $|\eta|<2.5$ containing $b$-hadrons are identified using the
DL1r $b$-tagging algorithm~\cite{FTAG-2018-01,ATL-PHYS-PUB-2017-013,FTAG-2019-07}.
The fixed $85\%$ efficiency (measured in simulated $t\bar{t}$ events)
working point is used. The rejection factors for $b$-tagging a jet
initiated by a $c$-quark or light parton are 2.6 and 29 respectively
at the $85\%$ efficiency working point.
 
Leptonic $\tau$-decays are reconstructed as electrons or muons.
The \tauhad{} decays are composed of a neutrino and a set of visible
decay products, most frequently one or three charged pions and up to
two neutral pions.  The reconstruction of the \tauhadvis{} is seeded
by jets reconstructed by the \antikt
algorithm~\cite{Cacciari:2008gp}, using calibrated topological
clusters~\cite{PERF-2014-07} as inputs, with a radius parameter of
$R=0.4$~\cite{PERF-2013-06,ATLAS-CONF-2017-029}.  The jets form
\tauhadvis{} candidates and are additionally required to have $\pT >
\SI{10}{\GeV}$ and $|\eta|<2.5$.  Reconstructed tracks are matched to
\tauhadvis{} candidates.
To separate the \tauhadvis{} candidates originating from hadronic
$\tau$ decays from quark/gluon-initiated jets, a recurrent neural
network (RNN) \tauhadvis{} identification
algorithm~\cite{ATL-PHYS-PUB-2019-033} is used.  The \tauhadvis{}
objects are required to satisfy the `Very Loose' \tauhadvis{}
identification criterion, which has an efficiency of 95\% for
simulated \tauhad{} decays.
A separate multivariate discriminant (eBDT)~\cite{ATL-PHYS-PUB-2015-045}
is employed to reject backgrounds arising from electrons that are
misreconstructed as single-track \tauhadvis{}.

The reconstructed electrons, muons, \tauhadvis{} and jets used in this
analysis are not built from a set of mutually exclusive tracks or
calorimeter clusters; it is therefore possible that two different
objects share most of their constituents. To resolve this ambiguity,
an overlap removal procedure is applied.  The \pt{} threshold
of muons considered in the overlap removal with  \tauhadvis{} is
lowered to 2~\GeV. More details can be found
in Ref.~\cite{HIGG-2019-09}.

The missing transverse momentum vector, $\vecmet$, is reconstructed as
the negative vector sum of the transverse momenta of light leptons, photons,
\tauhadvis, jets, and the \enquote{soft-term}.  The soft-term is
calculated as the vectorial sum of the \pT of tracks matched to the
primary vertex but not to a reconstructed light lepton,
\tauhadvis{} or jet~\autocite{PERF-2016-07}.  The magnitude of
$\vecmet$ is referred to as the missing transverse momentum, \met.

\subsection{Event selection and categorisation}
\label{sec:evt_sel}
The two analysis channels are defined according to the $\tau$ decay
mode. Events in the {\leplep} channel contain exactly two light leptons of opposite-sign electric charges and different flavours, while events in the {\lephad} channel contain exactly one light lepton
and a {\tauhadvis} with opposite-sign electric charges.
To preserve event separation between the channels, events containing
\tauhadvis{} are vetoed in the \leplep{} channel.
 
For each channel, a \Baseline{} selection is applied, based on the properties of the light leptons, {\tauhadvis}, missing transverse momentum, and on event properties as the absence of $b$-tagged jets.
Events satisfying the \Baseline{} selection criteria are further classified into two statistically independent categories, \VBF{} and \nonVBF{}, based on the kinematic properties of the jets produced in association with the Higgs boson candidate.
The event selection for all regions is summarised in
Table~\ref{tab:event_selection}.
Background control regions (CRs) used in the analyses are described in Section~\ref{sec:background} and summarised in Table~\ref{tab:control_regions}.

\begin{table}
\caption{\Baseline{} event selection in the {\leplep} and {\lephad}
channels and further splitting into the \VBF{} and \nonVBF{}
categories. The opposite-sign charge of the
final-state particles is labelled as OS.} \centering \resizebox{\textwidth}{!}{
\begin{tabular}{lcc}
\toprule
Selection & \leplep & \lephad \\
\midrule
\multirow{6}{*}{\Baseline{}}
& exactly $1e$ and $1\mu$, OS                         & exactly $1\ell$ and $1\tauhadvis$, OS \\
& \multicolumn{1}{c}{\tauhad-veto}                    & \multicolumn{1}{c}{\tauhad Tight ID}\\
& \multicolumn{1}{c}{$-$}                                & \multicolumn{1}{c}{Medium eBDT ($e\tauhad$)}\\
& \multicolumn{1}{c}{$b$-veto } & \multicolumn{1}{c}{$b$-veto } \\
& $\pt^{\ell_1} > 45\,(35)\,\GeV$ \MCbased{} (\Symbased{} method)                           & $\pt^{\ell} > 27.3\,\GeV$\\
& $\pt^{\ell_2} > 15\,\GeV$                           & $\pt^{\tauhadvis} > 25\,\GeV$, $|\eta^{\tauhadvis}| < 2.4$ \\
& $30\,\GeV < m_{\ell_1\ell_2} < 150\,\GeV$               & $\sum\limits_{i=\ell,\tauhadvis}\!\!\!\!\!\!\!\!\cos\Delta\phi(i,\MET) > -0.35$ \\
&  $0.2< \pt^{\textrm{track}}(\ell_{2}=e)/\pt^{\textrm{cluster}}(\ell_{2}=e) < 1.25$  (\MCbased{})    &  $|\Delta\eta(\ell,\tauhadvis)|<2$ \\
& track $d_{0}$ significance requirement (see text)        & \\
& $|z_0 \sin{\theta}| < 0.5\,\mathrm{mm}$  & \\
 
\midrule
\multirow{3}{*}{\VBF} & \multicolumn{2}{c}{\Baseline{}} \\
& \multicolumn{2}{c}{$\ge 2$ jets, $\pt^{\mathrm{j_1}} > 40\,\GeV$, $\pt^{\mathrm{j_2}}>30\,\GeV$} \\
& \multicolumn{2}{c}{$|\detajj| > 3$, $\mjj > 400\,\GeV$ }\\
\midrule
\multirow{3}{*}{$\nonVBF$} & \multicolumn{2}{c}{\Baseline{} plus fail \VBF{} categorisation} \\
&  \multirow{2}{*}{$-$} & veto events if \\
&  & $90 < \mvis(e,\tauhadvis) < 100$ \GeV\\
\bottomrule
\end{tabular}
}
\label{tab:event_selection}
\end{table}
 
In the {\leplep} channel, the definition of the leading ({\lepone}) and subleading
({\leptwo}) light leptons is based on the \pt-ordering of
the light leptons in the laboratory frame.
This ordering is used in the context of the rejection and estimation of backgrounds, where the \pt in the laboratory frame is relevant.
Additionally, an approach based on the \emph{approximate Higgs boson rest frame}
is used for the classification of the events.
The four-momentum of the Higgs boson rest frame
is built  from the two leptons and \MET{} four-momenta.
The pseudorapidity component of the missing-momentum vector is assumed to be
the same as that of the system formed by the two charged light leptons,
and the resulting invariant mass of the Higgs boson system is
constrained to be $\SI{125}{\GeV}$.
The light lepton having the higher \pt{} in the approximate Higgs boson rest frame is called  {\lephiggs}, and is assumed to be the lepton originating from the
Higgs boson decay.
The other charged light lepton is called {\leptau}, and is assumed to be the light lepton originating from the $\tau$ decay.
Events are divided into the $\mu\tau_e$
and $e\tau_\mu$ final states depending on whether {\lephiggs} is the muon or the electron.
Using this approach, the lepton misassignment
for the \Baseline{} event selection is reduced to about 5\% (7\%) for the
\ggF{} (VBF) production mode of the LFV signal, compared with a 11\% (19\%)
misassignment obtained when the laboratory frame is used.
 
In the {\leplep} channel of the \MCbased{} method, the leading light lepton ($\ell_1$) is required to
have $\pt^{\ell_{1}} > 45~\GeV$, while events with lower $\pt^{\ell_{1}}$ values
are used to obtain a region enriched with \Ztt{} events, as described in Section~\ref{sec:background_leplep}.
The events must satisfy a requirement on the
invariant mass of the two final-state leptons, $\SI{30}{\GeV} <
m_{\ell_1\ell_2} < \SI{150}{\GeV}$, to reduce the single-top-quark
and \ttbar{} (hereafter collectively labelled \enquote{top-quark}) background.
Additionally, to reduce the top-quark background contribution, events with one or more identified $b$-tagged jets are rejected.
In order to suppress the contribution from backgrounds with non-prompt light
leptons such as heavy-flavoured hadron decays and \Ztautau{} processes,
and to ensure compatibility with the primary vertex, additional
requirements are imposed on the transverse impact parameter ($d_{0}$)
significance and the longitudinal impact parameter ($z_0$), weighted by the normalized transverse momentum of the track ($\sin{\theta}$).
The $d_{0}$ significance
\cite{ATL-PHYS-PUB-2017-013} of the \lepone{} track is required to
be smaller than 5, the combination $|z_0 \sin{\theta}|$ is required to be smaller than $0.5\,\mathrm{mm}$.
If \leptwo{} is an electron, the $d_{0}$
significance of its track is required to be smaller than 10.
When $\ell_{2}=e$, the requirement $0.2< \pt^{\textrm{track}}/\pt^{\textrm{cluster}} < 1.25$
on the ratio of the $\pt$ measured using only the inner detector, $\pt^{\mathrm{track}}$, to
the $\pt$ measured in the calorimeter, $\pt^{\mathrm{cluster}}$,
aims to reduce the number of \Zmumu{} background events, in which
one of the muons deposits a significant fraction of its energy in the
calorimeter.
 
In the {\leplep} channel of the \Symbased{} method, the requirement on
$\pt^{\ell_{1}}$ is reduced relative to the \MCbased{} method to $\pt^{\ell_{1}} > 35~\GeV$
to increase the number of events used in the Neural Network (NN) training, introduced
in Section~\ref{sec:mva}.  To preserve a symmetric selection, electron
$|\eta|$ requirements from the object reconstruction are also applied
to muons.  Therefore, selected muons are required to have $|\eta| <
1.37$ or $ 1.52 <|\eta| < 2.47$. The requirement $\SI{30}{\GeV} <
m_{\ell_1\ell_2} < \SI{150}{\GeV}$ is used to reduce the top-quark
background.  The $d_{0}$ significance of the tracks of both light
leptons is required to be smaller than 10. No requirements on
$\pt^{\textrm{track}}(\ell_{2})/\pt^{\textrm{cluster}}(\ell_{2})$ are applied when
$\ell_{2}=e$ because it would break the symmetry between electrons and muons.
 
In the {\lephad} channel, angular selection criteria
$(\cos\Delta\phi(\ell,\MET) + \cos\Delta\phi(\tauhadvis,\MET))> -0.35$
and ${|\Delta\eta(\ell,\tauhadvis)| < 2}$ are imposed to reject
$\Wjets$ and multi-jet production processes.
To reduce the top-quark background contribution,
events with one or more identified $b$-tagged jets are rejected.
Selected $\tauhadvis$ are required to satisfy the `Tight' \tauhadvis{}
identification criterion.
In the $\ethad$ channel, the reconstructed
\tauhadvis{} is required to pass the~eBDT~\enquote{Medium} working point in
order to suppress $\Zeejets$ processes where one of the electrons is
reconstructed as a $\tauhadvis$ object.
 
The \VBF{} and \nonVBF{} categories in each of the \leplep{} and \lephad{}
final states give rise to a total of four signal regions (SRs)
in each search.
The \VBF{} category is designed to enhance the sensitivity to the VBF
Higgs boson production mode.  Dedicated requirements are applied to
the jet kinematics and topology of the two jets to separate VBF
Higgs boson production from the other production modes.
The leading and subleading \pT-ordered jets are denoted by
$\mathrm{j}_1$ and $\mathrm{j}_2$, respectively.  The {\mjj} and {\detajj}
variables represent the invariant mass and $\eta$-separation of the two leading
jets, respectively.  The \nonVBF{} category contains events failing
the \VBF{} selection, and in the $\ethad$ channel, an additional
requirement on $\mvis(e,\tauhadvis)$ reduces the
$\Zeejets$ background component.


\section{Background estimation}
\label{sec:background}
 
The $Z\to\tau\tau$ process is one of the dominant backgrounds in all
the categories and channels.  Other relevant backgrounds originate
from \Wjets\ and multi-jet events with at least one jet misidentified
as an electron, muon or \tauhad{} (referred to as \enquote{\fake} hereafter).
The \fake{} background in the two methods (\MCbased{} and
\Symbased) and the two analysis channels (\leplep{} and \lephad{}) is evaluated using
data-driven techniques.
Additionally, top-quark decays form a non-negligible contribution,
particularly in the \leplep{} channel. Other background
components such as diboson production ($WW$, $WZ$ and $ZZ$),
$H\rightarrow\tau\tau$ and $H\rightarrow WW$ are also considered.
 
In the following, the background estimates are discussed in more detail for the \MCbased{} \leplep{} channel in Section~\ref{sec:background_leplep}, for the \MCbased{} \lephad{} channel in Section~\ref{sec:background_lephad}, and for the \Symbased-based \leplep{} channel in Section~\ref{sec:lep_taulep_symm}.

\subsection{\MCbased{}{ \leplep} channel}
\label{sec:background_leplep}
In the \MCbased{} \leplep{} channel, the main background contributions
arise from top-quark, \Ztt{} and diboson processes, and events with
misidentified leptons. Sources of smaller backgrounds are \Zll{} events
and SM Higgs boson decays. In addition to simulation, data control and validation regions
are used to estimate the background contributions, where possible.
Backgrounds from Higgs boson processes such as \Htautau and \HWW are expected
to be small and are normalised to their SM predictions.

The top-quark processes contribute 34\%--54\% of the total background, depending on the category.
For each category, the top-quark simulation is validated with data in a top-quark CR, statistically independent
of the SRs.
These CRs are defined by applying all the baseline criteria for \nonVBF{} or \VBF{} categorisation
except for the $b$-tagged jet veto, and 95\% of their events are top-quark events.
Good modelling by the top-quark simulation is
observed, as shown in Figure~\ref{fig:leplep_CRs} (a, b).
To account for possible theoretical uncertainties in the production cross-section for the simulated samples, the top-quark CRs
and two normalisation factors, one per category, are included in the
statistical analysis (Section~\ref{sec:stat}). Each normalisation factor is determined
during the signal extraction.
 
The \Ztt{} events account for  23\% (11\%)  of the total background in the  \nonVBF{} (\VBF{}) SR.
For each category, the \Ztt{} simulation is validated with data in a \Ztt{} CR, statistically independent
of the SRs.
These CRs are defined by requiring $\SI{35}{\GeV} < \pt^{\ell_1} < \SI{45}{\GeV}$,
and the \Ztt{} process contributes ${\sim}65\%$ (${\sim}32\%$)
of all events in the \nonVBF{} (\VBF{}) \Ztt{} CR.
Good modelling in the \Ztt{} CRs is illustrated for the \nonVBF{} case in Figure~\ref{fig:leplep_CRs} (c, d).
In the statistical analysis, the \Ztt{} CRs are used jointly with two independent normalisation factors to constrain the \Ztt{} normalisation
in the SRs.
\begin{figure}[!htb]
\centering
\subfloat[]{\includegraphics[width=0.43\textwidth]{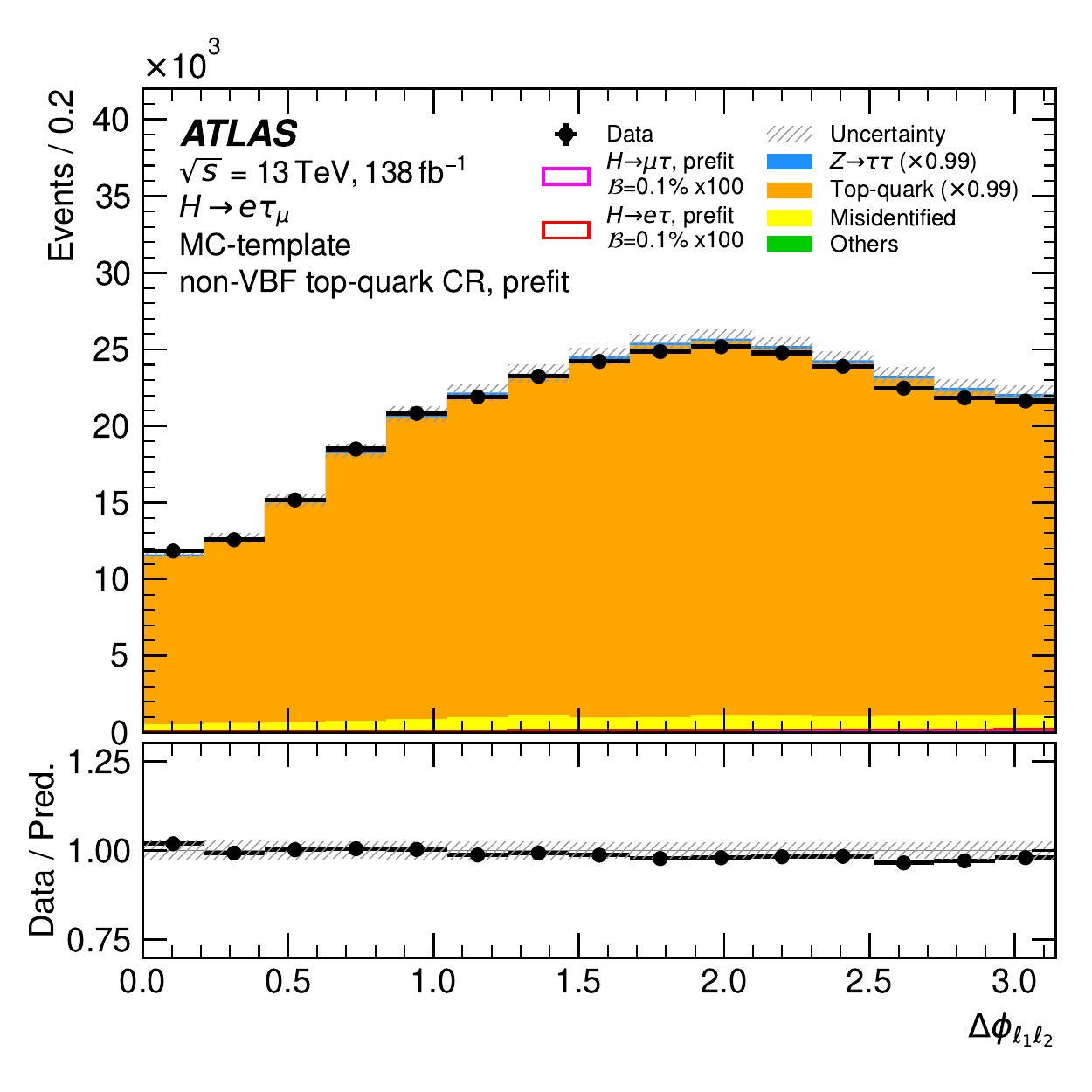}} \hspace{10mm}
\subfloat[]{\includegraphics[width=0.43\textwidth]{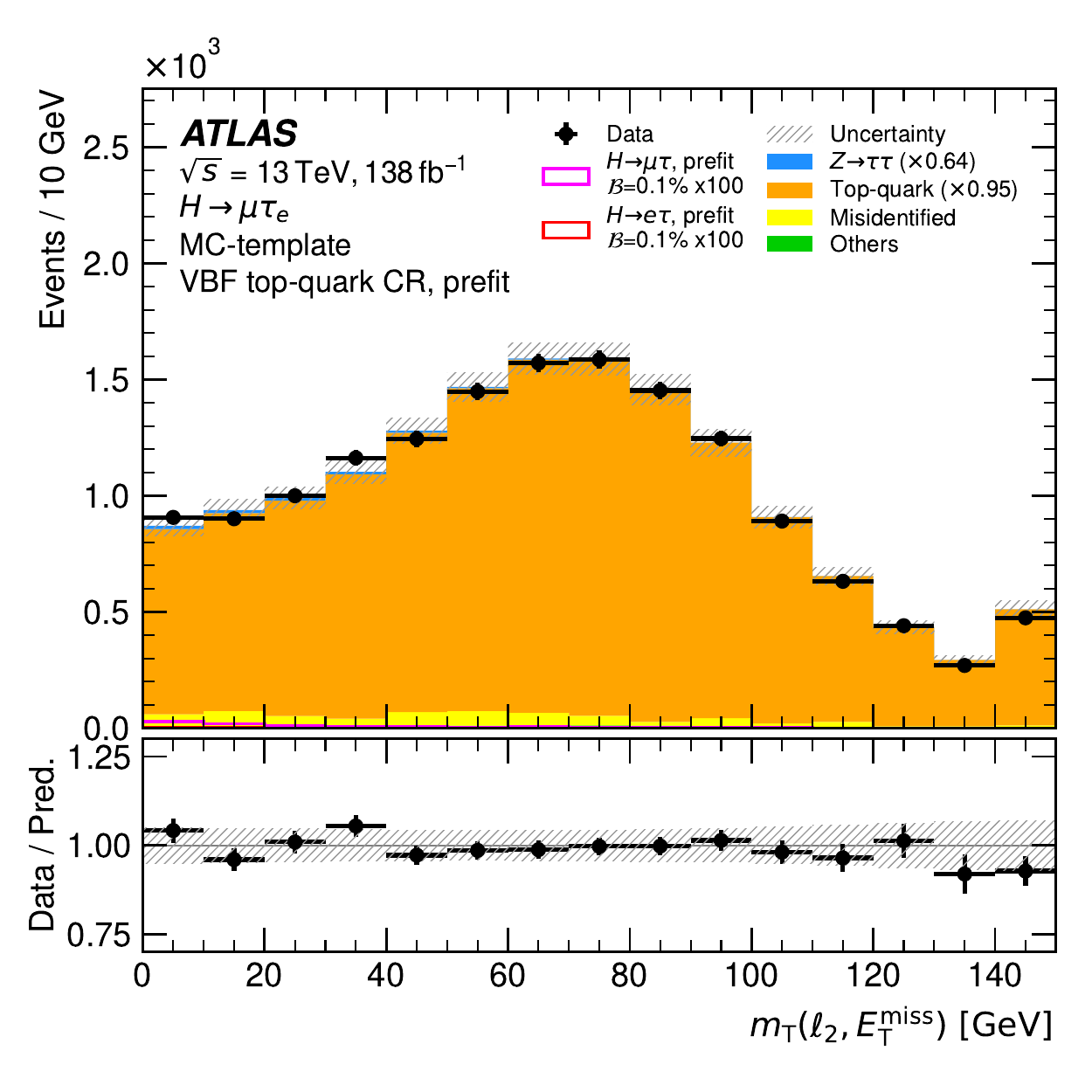}} \\
\subfloat[]{\includegraphics[width=0.43\textwidth]{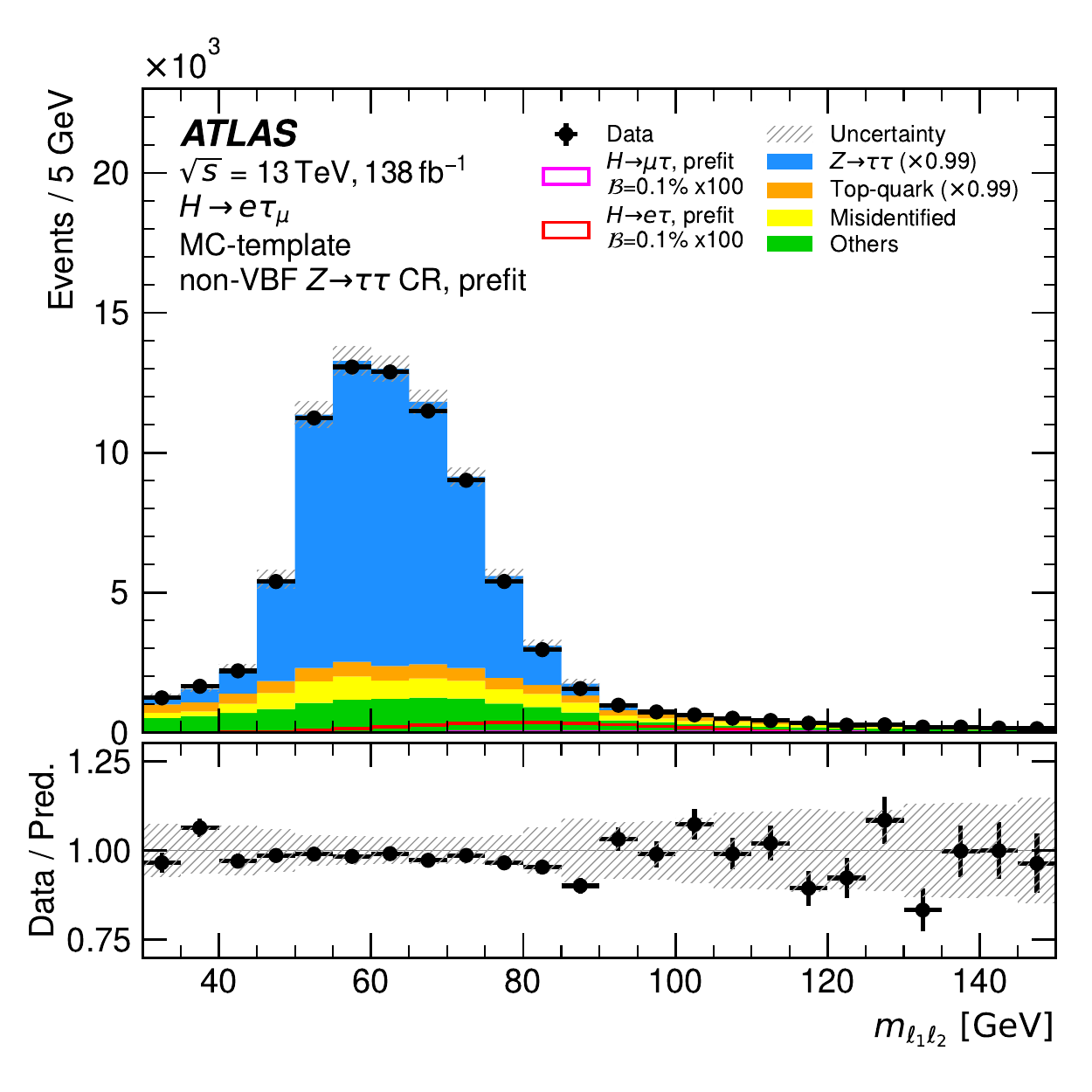}} \hspace{10mm}
\subfloat[]{\includegraphics[width=0.43\textwidth]{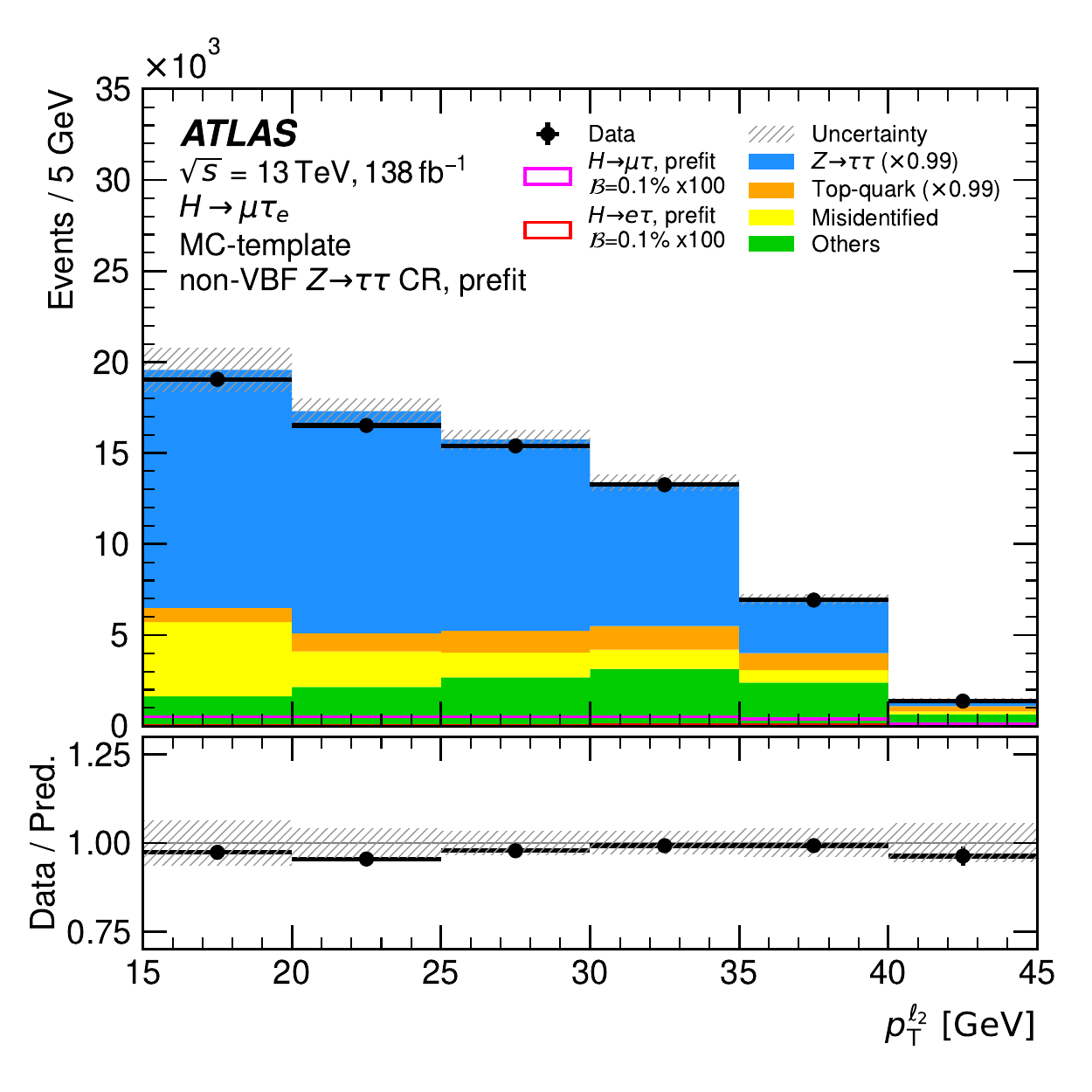}} \\
\caption{Distributions of representative quantities for the top-quark and
\Ztt{} CRs before the statistical analysis (prefit) in the \etm{} (left)
and \mte{} (right) final states: (a) $\Delta\phi_{\ell_1,\ell_2}$ and
(b) $m_{\mathrm{T}}(\ell_2,\met)$ for the top-quark CRs, (c)
$m_{\ell_1\ell_2}$ and (d) $\pT^{\ell_2}$ for the \Ztt{} CRs.  Entries
with values that would exceed the horizontal axis range are included in
the last bin of each distribution.
The hashed band indicates the prefit statistical, experimental and theoretical uncertainties.
The \Ztt{} and top-quark contributions are scaled by a normalisation factor (indicated in the legend)
obtained from the likelihood fit performed independently
in \nonVBF{} and \VBF{} categories of the \MCbased{} \leplep{} channel.
Overlaid prefit signal shapes  assume $\BR(H \to \ell \tau) =
0.1\%$ and are enhanced by a factor 100 for visibility.}
\label{fig:leplep_CRs}
\end{figure}
 
Diboson events form 19\%--32\% of the total background, depending on the category.
The shape and normalisation of diboson process distributions are estimated from the
simulation and validated with data in a dedicated validation region,
where the diboson processes contribute ${\sim}67\%$ of the total background.  This region is
defined by applying the \Baseline{} selection criteria and requiring
$\pt^{\ell_2} > \SI{30}{\GeV}$ and the mass of the two leptons to satisfy
$\SI{100}{\GeV} < m_{\ell_1\ell_2} < \SI{150}{\GeV}$.
The transverse mass, $m_{\mathrm{T}}$, calculated from the transverse momentum of the subleading light lepton and the \met{},
is required to be greater than \SI{30}{\GeV} and jets with $\pT >30$~{\GeV} are rejected.
Good modelling in the diboson validation region is observed, as illustrated in~\Fig{\ref{fig:leplep_VR}}.
\begin{figure}[!htb]
\centering
\subfloat[]{\includegraphics[width=0.43\textwidth]{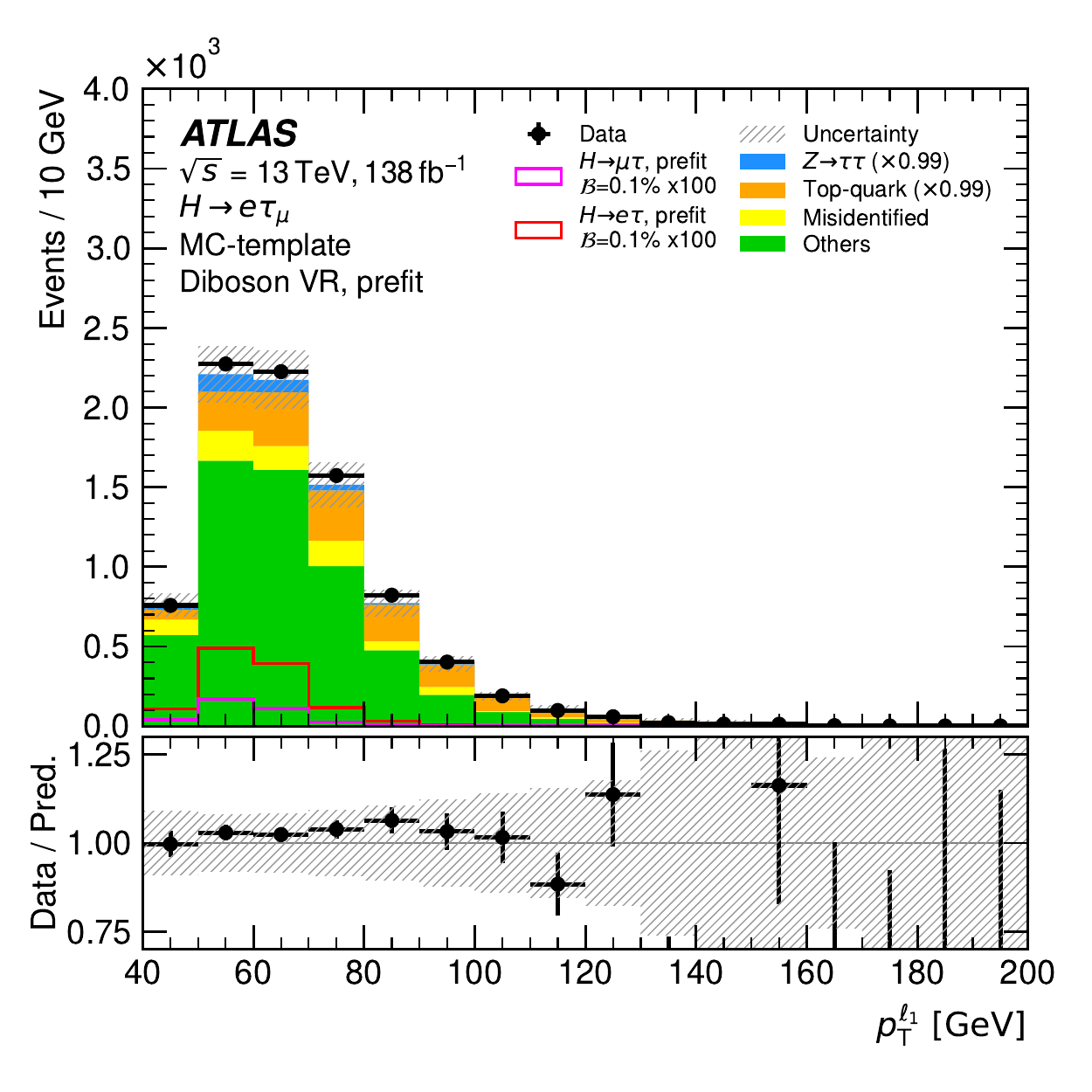}} \hspace{10mm}
\subfloat[]{\includegraphics[width=0.43\textwidth]{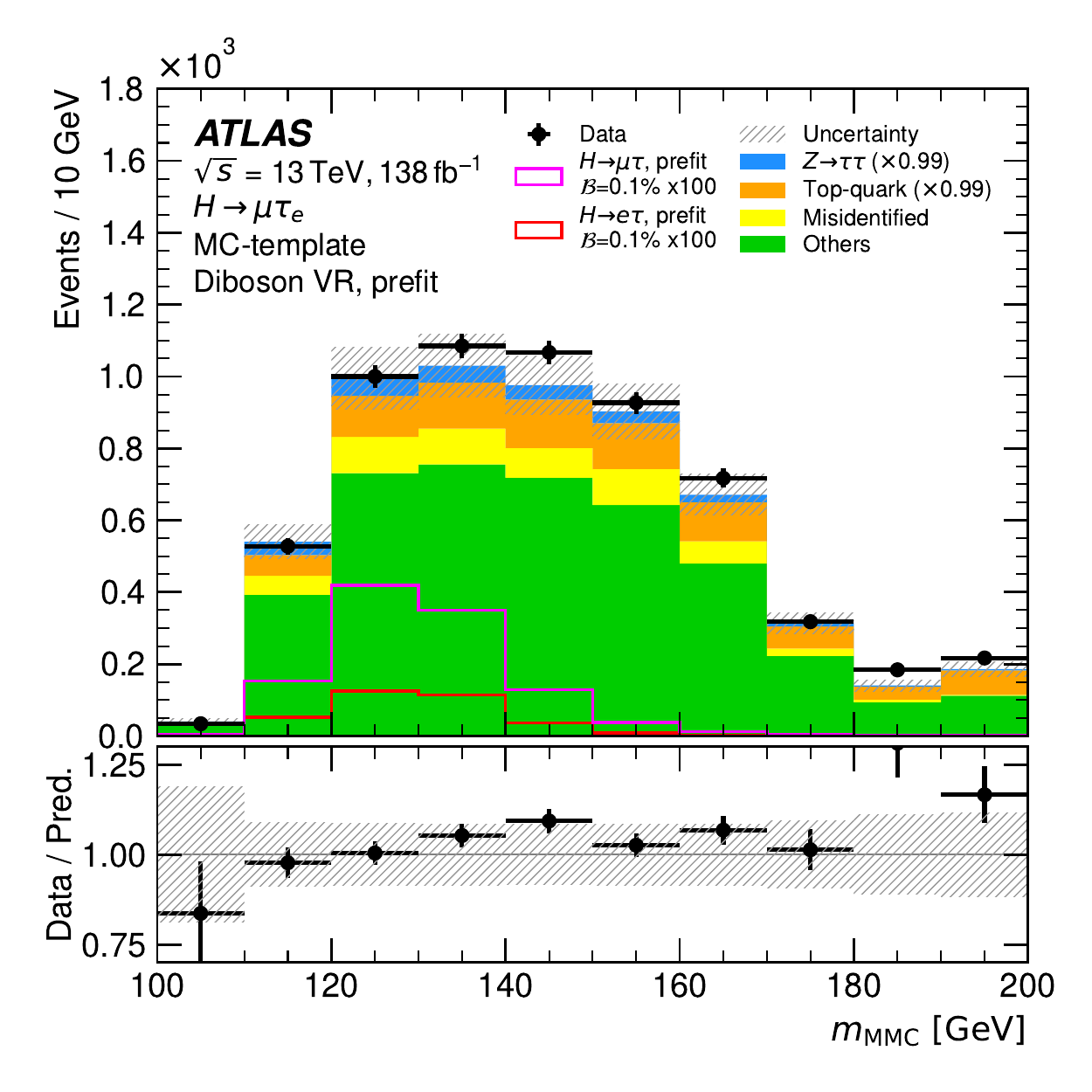}} \\
\caption{Distributions of representative quantities for the diboson validation region
before the statistical analysis (prefit): (a) distribution of $\pt^{\ell_1}$ in the \etm{} final state,
(b) distribution of  the \mmmc{}~\cite{MMCpaper} mass in the \mte{} final state. Entries
with values that would exceed the horizontal axis range are included in
the last bin of each distribution.
The hashed band indicates the prefit statistical, experimental and theoretical uncertainties.
The \Ztt{} and top-quark contributions are scaled by a normalisation factor (indicated in the legend)
obtained from the likelihood fit performed independently
in \nonVBF{} and \VBF{} categories of the \MCbased{} \leplep{} channel.
Overlaid prefit signal shapes  assume $\BR(H \to \ell \tau) =
0.1\%$ and are enhanced by a factor 100 for visibility.}
\label{fig:leplep_VR}
\end{figure}
 
The \Zmumu{} process contributes sizeably only in the $\mu \tau_e$
channel, where it represents up to 2\% of the total background.
The modelling of the \Zmumu{} background by the simulation is validated in a dedicated \Zmumu{} CR,
which is not included in the statistical analysis.
This CR is obtained by applying the \Baseline{} selection, but with
$\SI{35}{\GeV}<\pt^{\ell_1} < \SI{45}{\GeV}$ and with the dilepton pair mass near
the $Z$ boson mass peak, $\SI{75}{\GeV} < m_{\ell_1\ell_2} < \SI{100}{\GeV}$.
Additionally, the $\phi$ separation between the subleading light lepton and
the $\vecmet$ is required to satisfy $|\Delta \phi(\ell_2, \met)|<1.5$.
To ensure that the \Zmumu{} CR is statistically independent of the SRs and the \Ztt{} CRs,
the events are required to have
$1.25<\pt^{\mathrm{track}}(\ell_2)/\pt^{\mathrm{cluster}}(\ell_2)<3$.
While no mismodelling of the MC template shape is found in the \Zmumu{} CR,
a global normalisation offset of $25\%$ is observed. Therefore, the \Zmumu{}
event yields in the SRs are decreased by $25\%$ and
a normalisation uncertainty of that size is assigned to the \Zmumu{}
contribution in the statistical analysis.
 
\begin{figure}[!htb]
\centering
\subfloat[]{\includegraphics[width=0.43\textwidth]{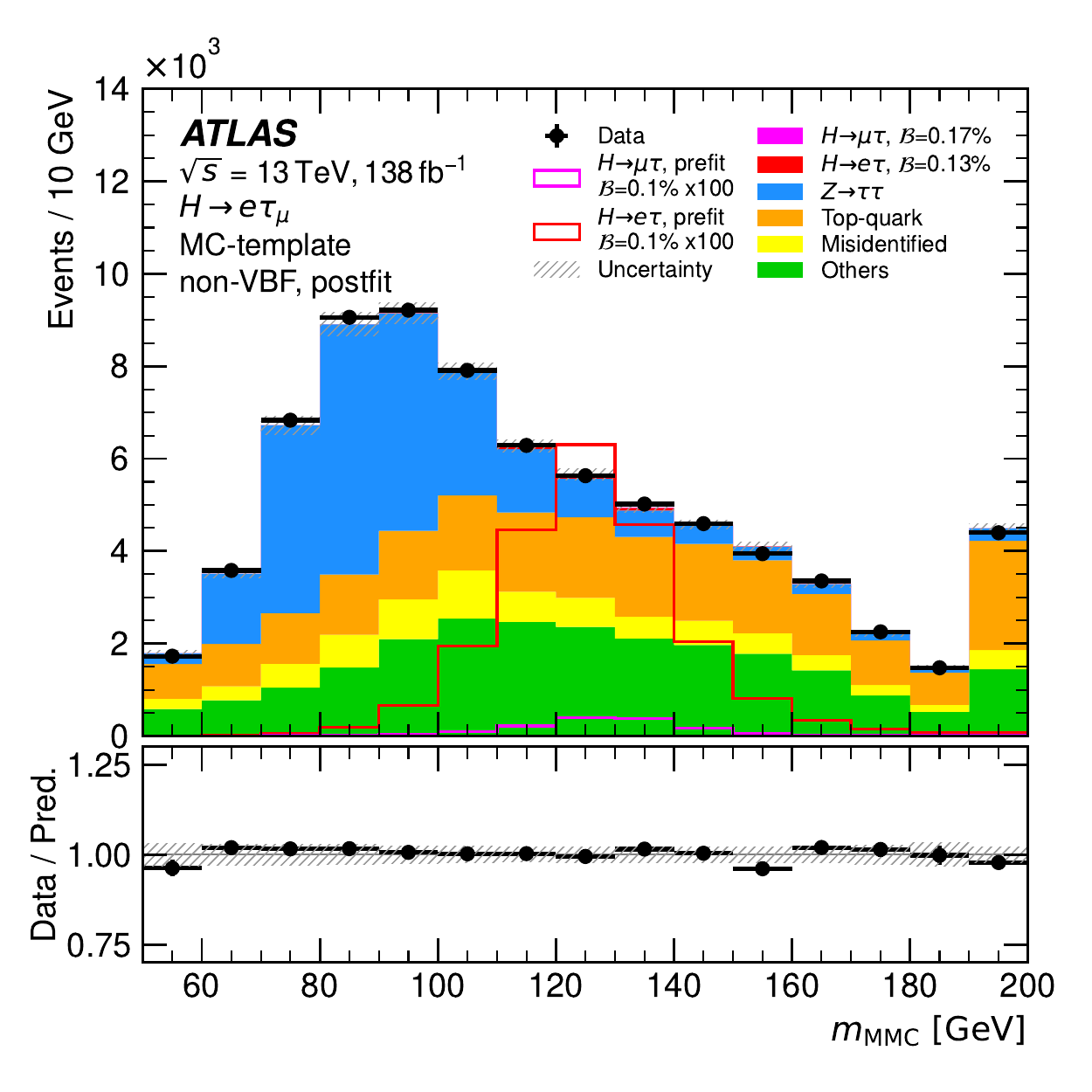}} \hspace{10mm}
\subfloat[]{\includegraphics[width=0.43\textwidth]{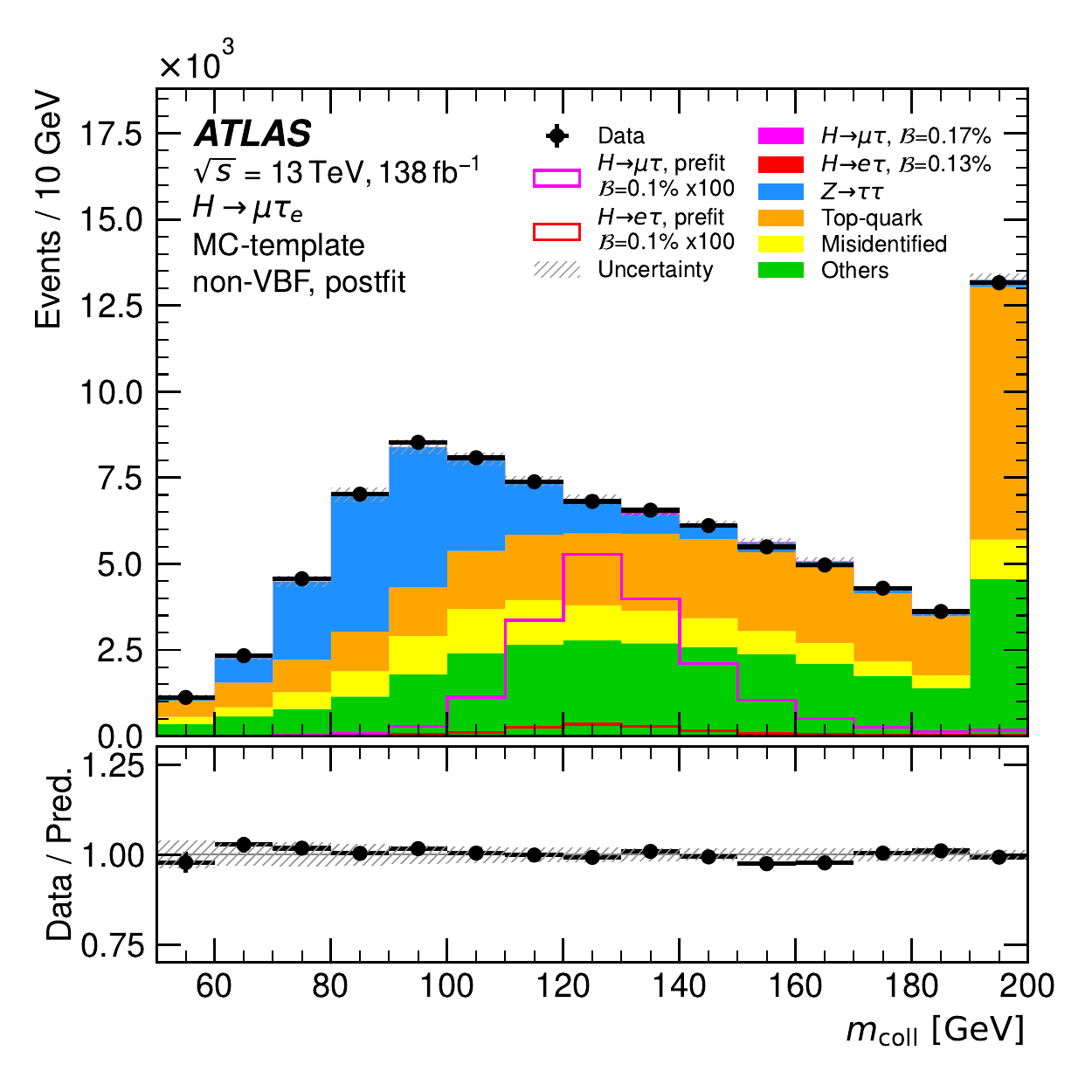}} \\
\subfloat[]{\includegraphics[width=0.43\textwidth]{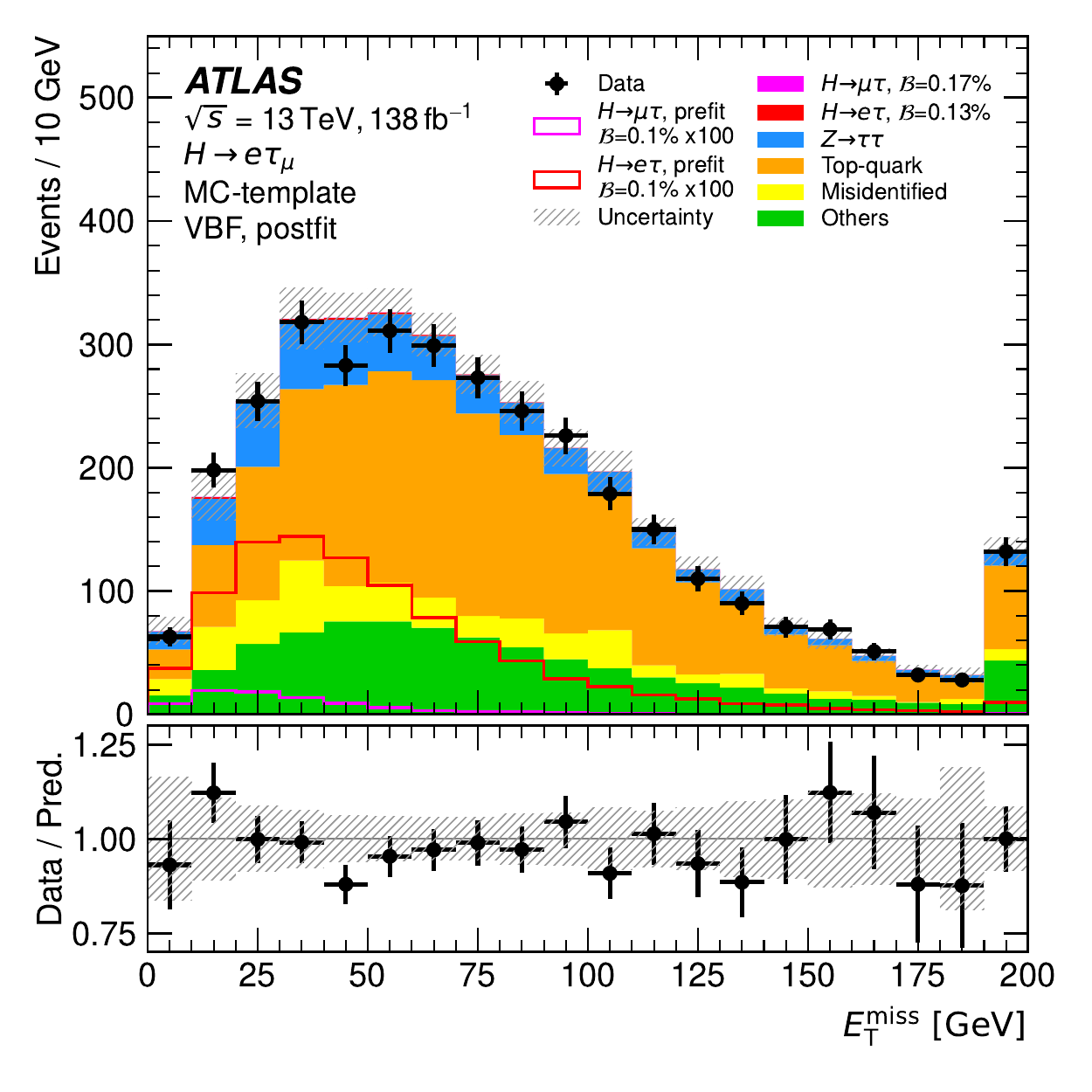}} \hspace{10mm}
\subfloat[]{\includegraphics[width=0.43\textwidth]{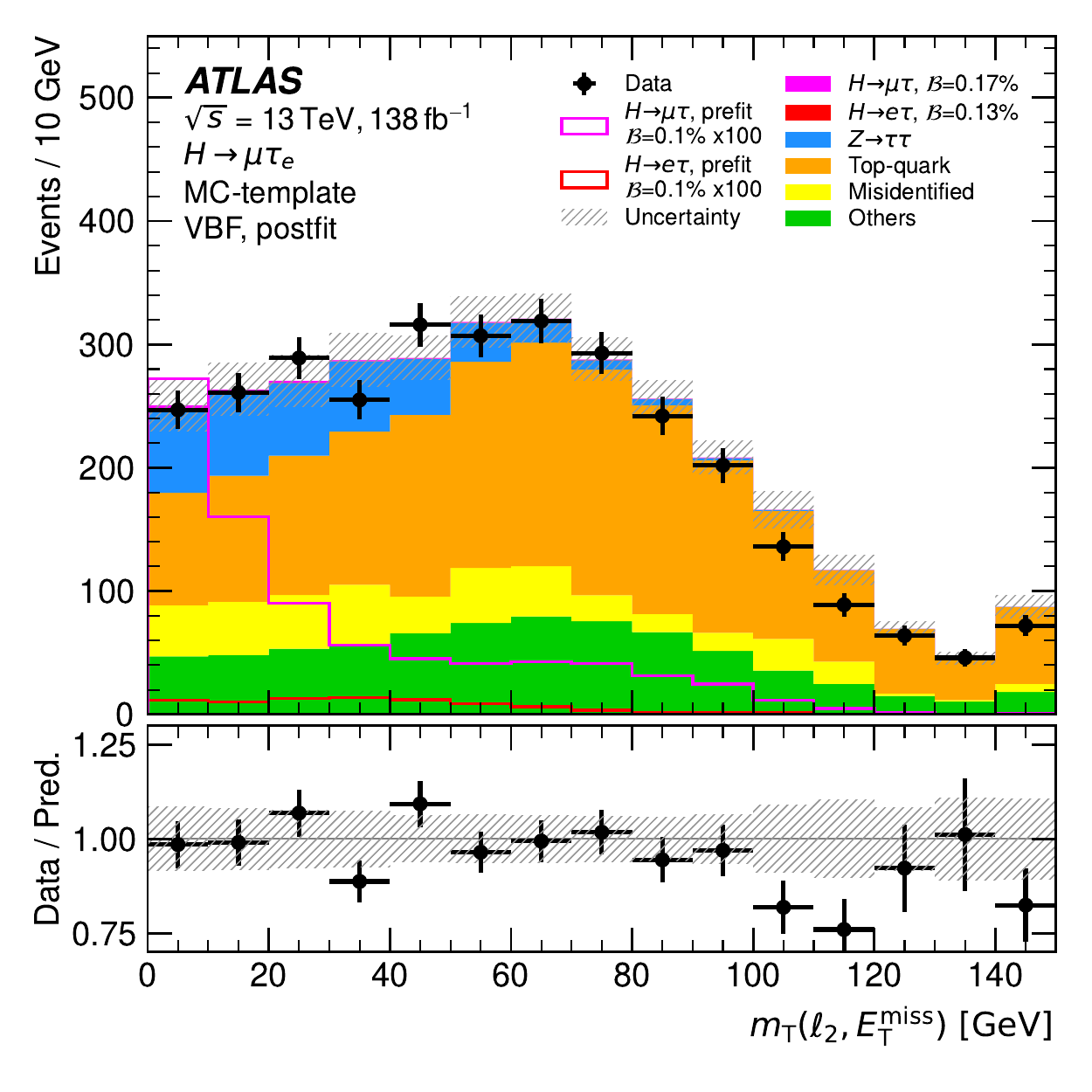}} \\
\caption{ Distributions of representative kinematic quantities for
\etm{} (left) and \mte{} (right) final states,
after a simultaneous fit of the \hte{} and \htm{} signals,
obtained by fitting the data of the \MCbased{} \leplep{} channel:
(a) the mass \mmmc{} and (b)
\mcol{}~\cite{Ellis:1987xu} in the \nonVBF{} category, (c)
the missing transverse momentum \met{} and (d) the transverse mass
$\mT(\ell_2,\met)$ in the \VBF{} category.  Entries with values
that would exceed the horizontal axis range are included in the last bin
of each distribution.  The hashed band represents the prefit
statistical uncertainty, and the experimental and theoretical
uncertainties obtained from the likelihood fit.  Overlaid prefit
signal shapes assume $\BR(H \to \ell \tau) = 0.1\%$ and are
enhanced by a factor 100 for visibility. The postfit signal contributions are considered as part of the predictions.}
\label{fig:leplep}
\end{figure}
 
A data-driven method is used to estimate the \fakebkg{} contribution
from events having at least one light lepton originating from heavy-flavour decays,
photon conversion, a jet or a \tauhad{}
misidentified as a light lepton. These events originate mostly from
\Wjets, multi-jet production and top-quark processes.
The method used is based on the assumption that the ratio
of opposite-sign to same-sign
light-lepton events is approximately constant when inverting the
isolation quality requirement for the subleading light
lepton~\cite{HIGG-2017-08}. The ratio, named a transfer factor, is
measured in regions enriched in \fakebkg{}. These regions use the SRs'
selection criteria, but the lepton isolation requirement is inverted for
the subleading light lepton, and the electron is allowed to fail the
\enquote{Medium} identification requirement while still
passing the \enquote{Loose} requirement. To obtain the \fakebkg{} prediction in the SRs,
the transfer factor is applied to same-sign regions. Events in the same-sign regions are required to satisfy the
same selection criteria as in the SRs, but have two light leptons with same-sign electric charges.
Events containing two prompt light leptons (mostly from diboson events)
form a small component of this background and are subtracted using
simulation. The transfer factors are calculated independently in the
$e\tau_\mu$ and $\mu\tau_e$ final states. In each of the two cases,
transfer factors are obtained separately for events passing the single-lepton
or dilepton trigger requirement, and separately for events
passing or failing the $b$-veto requirement.
 
The statistical uncertainty of the  \fakebkg{} is separated into four uncorrelated components, based on whether the event is triggered by a single-lepton or dilepton trigger and whether the event passed or failed the $b$-veto requirement.
The systematic components account for several effects. The uncertainty related to the subtraction of the
prompt-lepton processes in the same-sign region is determined to be between 6\% and 8\%.
The systematic uncertainty due to possible differences between the jet flavour composition in the \fakebkg{} enriched CRs and the SRs is taken into account.
To estimate this uncertainty, the transfer factors are calculated in \fakebkg{} enriched CRs using the simulated samples of the main sources of fake leptons (\Wjets\ and $V\gamma$).
The transfer factors are applied to the simulated events passing the \Baseline{} selection, except that two same-sign leptons instead of opposite-sign leptons are required. The obtained prediction for the BDT distribution (see Section~\ref{sec:mva}) is compared with the simulated events passing the \Baseline{} selection. The difference between the two BDT distribution shapes is taken as the magnitude of the systematic uncertainty, which is found to vary between 10\% and 30\%.
To cover potential residual differences between the simulation and data,
a non-closure uncertainty is obtained by defining additional opposite-sign \fakebkg{} enriched CRs.
These regions are statistically independent of the SRs and use the same selection criteria as the \Baseline{} selection but invert the requirement on $d_0$ significance.
The transfer factors are recalculated using these additional regions and the background predictions are re-evaluated using the new transfer factors.
The difference between data and the background prediction in the additional opposite-sign \fakebkg{} enriched CRs is assigned as a systematic uncertainty. The magnitude of the uncertainty ranges from 2\% to 12\%.
The dependence of the transfer factor on the $d_0$ significance requirement was investigated and found to be negligible.
 
Figure~\ref{fig:leplep} shows the level of agreement between the data and simulation in the
\nonVBF{} and \VBF{} categories for a subset of variables with the
highest discrimination power used in the MVA
as described in Section~\ref{sec:mva}.
The distributions show clear separation between signal and background, as well as good modelling of the data within statistical and systematic uncertainties.
The collinear mass of the two final-state leptons (\mcol)~\cite{Ellis:1987xu}, the Higgs boson mass
obtained with the Missing Mass Calculator (MMC) technique
(\mmmc)~\cite{MMCpaper} and $m_{\ell_1\ell_2}$ show the highest discrimination power.
The MMC algorithm is tuned specifically to reconstruct the mass of the LFV Higgs boson.
The prediction for each sample is determined from the likelihood fit performed to measure \hte{} and \htm{} signals, based only on data in the \leplep{} final state.
Corresponding SR and CR event yields are detailed in Tables~\ref{tab:MC_yields_leplep_etau_2poi}
and~\ref{tab:MC_yields_leplep_mutau_2poi} in Appendix~\ref{app:yieldsMCleplep}.


\subsection{\MCbased{}{ \lephad} channel}
\label{sec:background_lephad}
 
In the \MCbased{} \lephad{} channel, the main background contributions
arise from the \Ztt{} events and are estimated using the
simulated events. The \Ztt{} events form 48\%--67\% of the total background
depending on the category.  The second-largest
background consists of events containing a jet misidentified as
\tauhadvis{} and contributes 22\%--30\% of the total background. This \fakebkg{}
is evaluated using the \ff{}
technique~\cite{HIGG-2013-32,HIGG-2013-21}. Contributions with \tauhad{} and
a jet misidentified as an electron or a muon are estimated from simulation and found to be negligible. Backgrounds from other processes such as
top-quark, \Zll{}, dibosons, \Htautau and \HWW  are
much smaller and are estimated from simulation.
 
The shape of the \Ztt{} background distribution is modelled with
simulations, and its normalisation in the
\VBF{} and \nonVBF{} categories is constrained by data in the BDT score
distributions (see Section~\ref{sec:mva}) in the SRs.
The shape of the top-quark background distribution is modelled with
simulation. The corresponding normalisation factors in the \VBF{} and
\nonVBF{} categories are constrained using the \leplep{} top-quark CRs in the
simultaneous fit of the \lephad{} and \leplep{} channels.
 
Diboson events form 2\%--5\% of the total background, depending on the category.
In the $\mu\tauhad$ channel, the \Zmumu{} process contributes about
5\%--6\% of the total background in the \nonVBF{} category, and
3\%--4\% in the \VBF{} category. Its modelling is
validated in a dedicated validation region, where the \Zmumu{} event fraction is ${\sim}65\%$.
The \Zmumu{} validation region is defined by applying the \Baseline{} selection
criteria and also requiring $|\eta(\tau)|<0.1$ and the
collinear mass~\cite{Ellis:1987xu} to be near the $Z$ peak: $\SI{90}{\GeV} <
\mcol(\mu,\tau) < \SI{110}{\GeV}$.
While no template mismodelling
is observed in this region, a global normalisation offset of $13\%$ is
observed. Thus, an uncertainty of that size is assigned
to the \Zmumu{} contribution in the statistical analysis described in Section~\ref{sec:stat}.
 
\begin{figure}[!htb]
\centering
\subfloat[]{\includegraphics[width=0.43\textwidth]{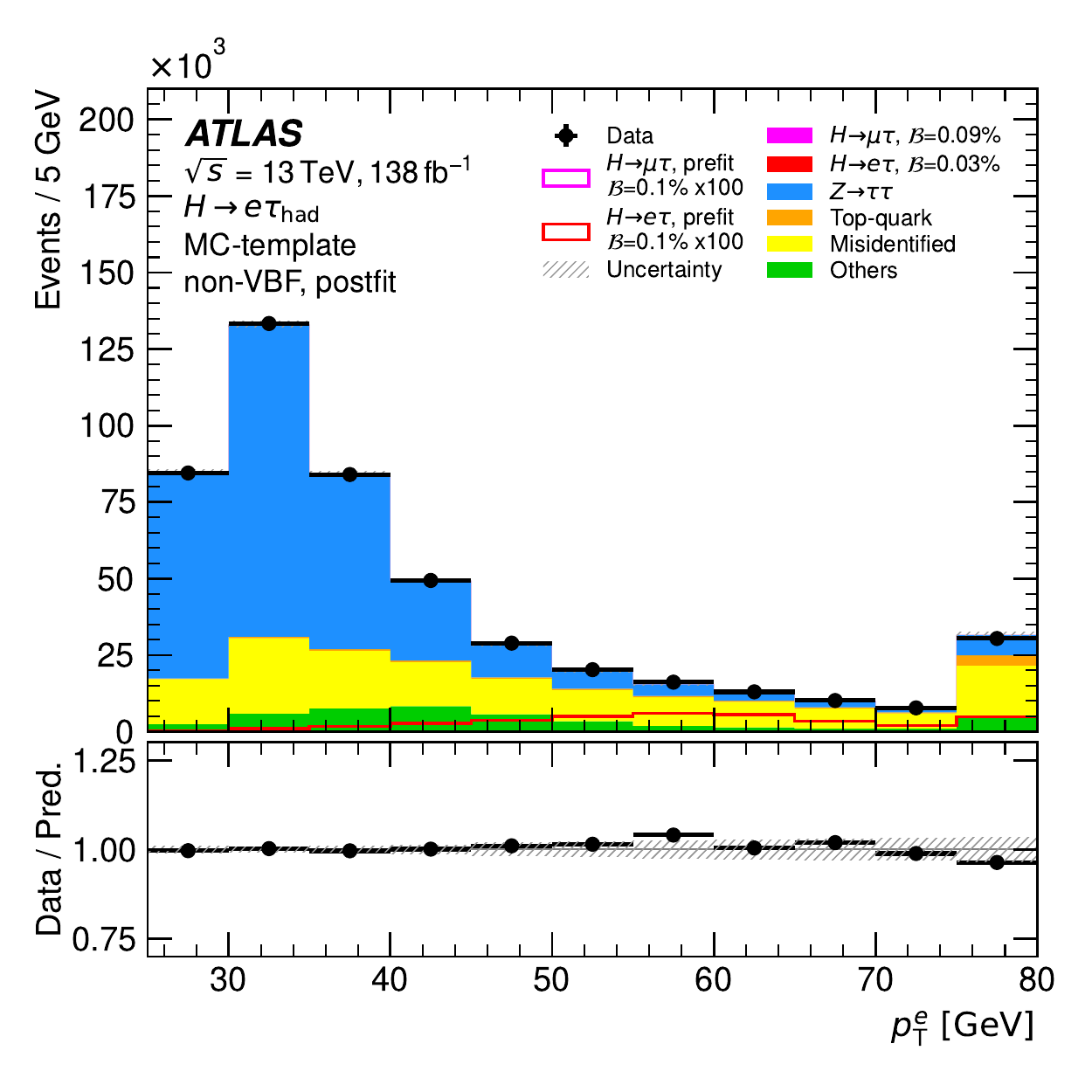}}
\subfloat[]{\includegraphics[width=0.43\textwidth]{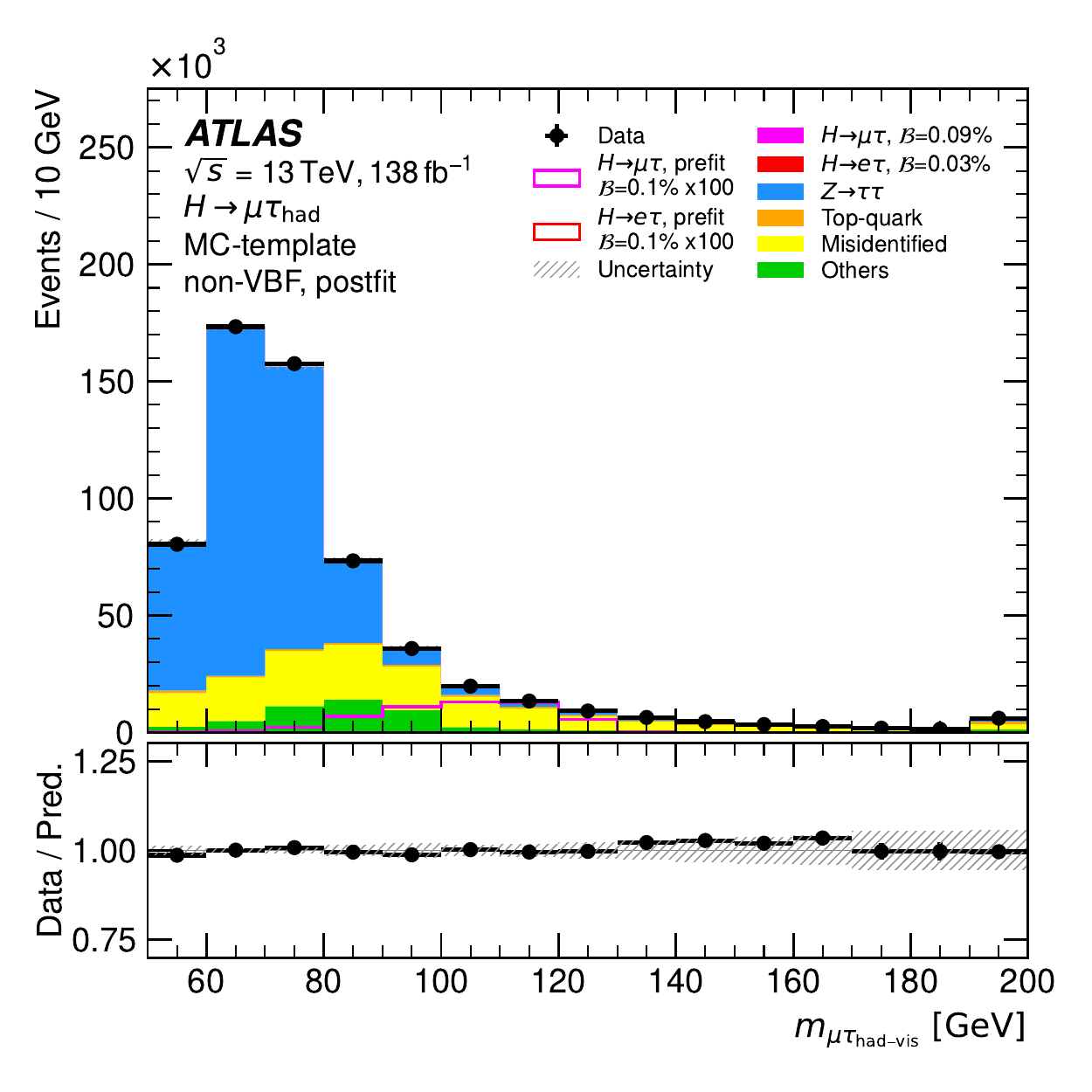}} \\
\subfloat[]{\includegraphics[width=0.43\textwidth]{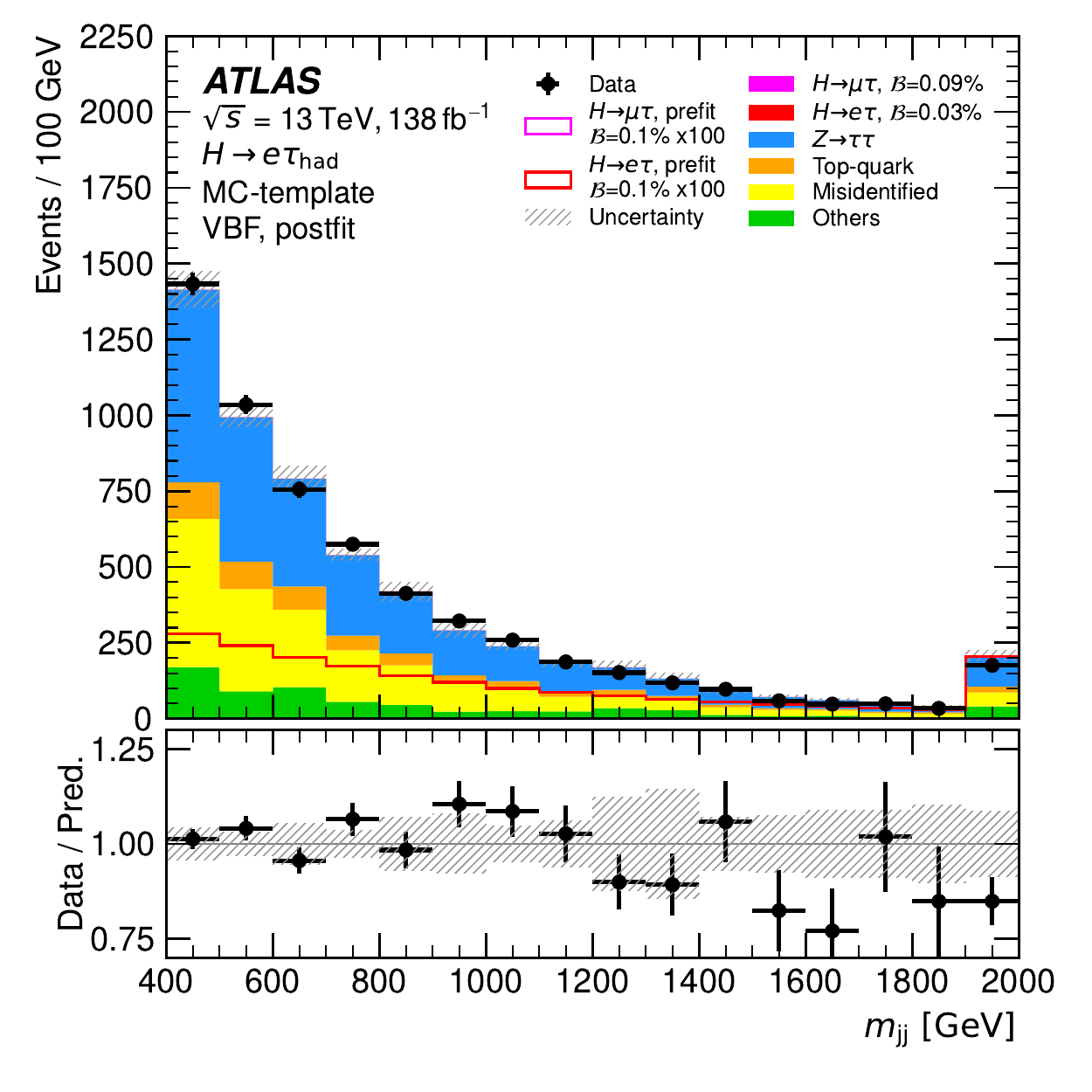}}
\subfloat[]{\includegraphics[width=0.43\textwidth]{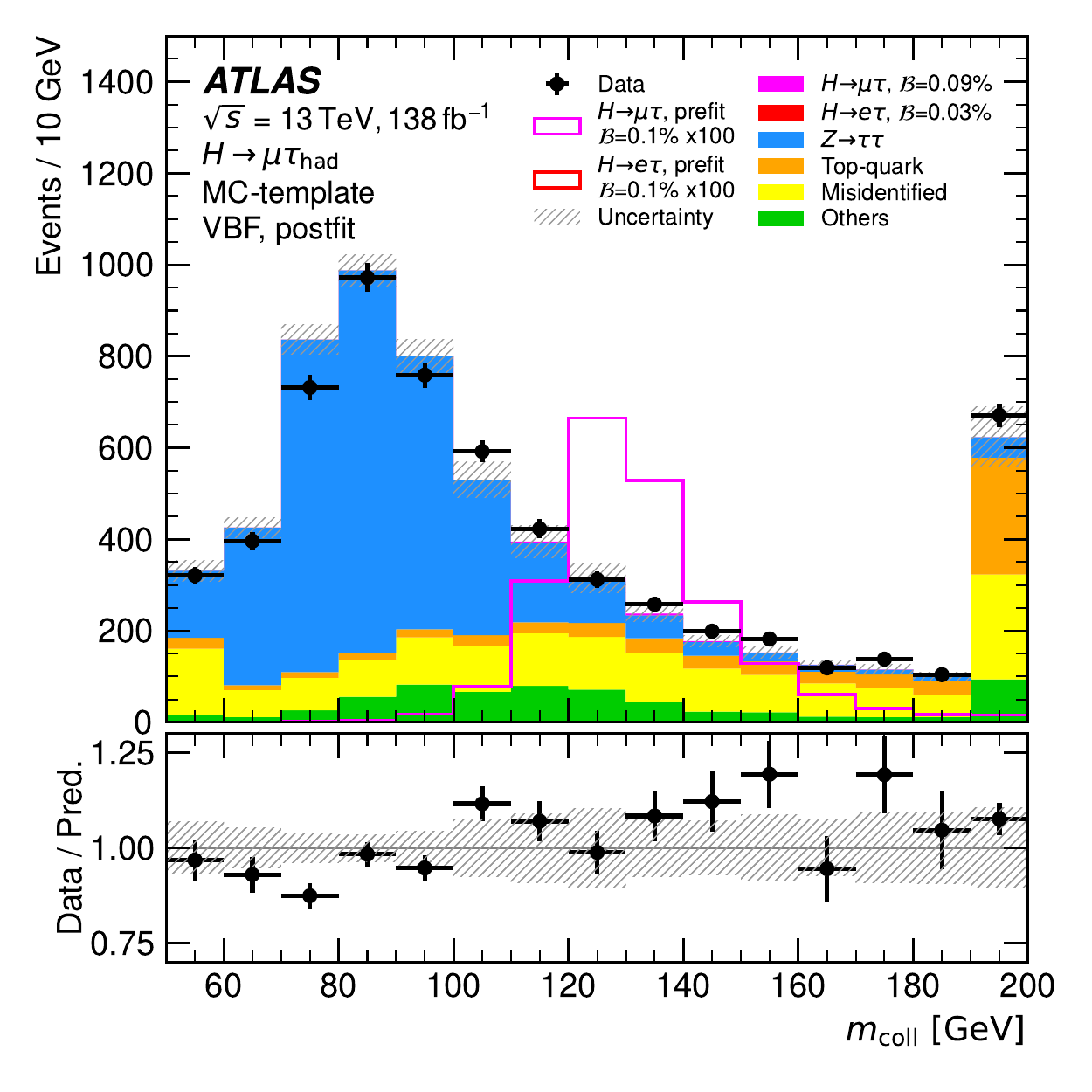}}
\caption{Distributions of representative kinematic quantities for
$e\thad$ (left) and $\mu\thad$ (right) final states,
after a simultaneous fit of the \hte{} and \htm{} signals,
obtained by fitting the data of the \MCbased{} \lephad{} channel:
(a) the electron transverse momentum, $p_{\mathrm{T}}^{e}$, and (b) the
visible mass $\mvis$ in the \nonVBF{} category, (c) the dijet
invariant mass \mjj{} and (d) collinear mass \mcol{}
in the \VBF{} category.  Entries with values that would exceed the
horizontal axis range are included in the last bin of each distribution.
The hashed band represents the prefit statistical uncertainty,
and the experimental and theoretical uncertainties obtained from the likelihood fit.
Overlaid prefit signal shapes assume $\BR(H \to \ell \tau) = 0.1\%$ and are enhanced by a factor 100 for visibility.
The postfit signal contributions are considered as part of the predictions.}
\label{fig:lephad}
\end{figure}

The distribution of the \fakebkg{} component in the SR is obtained by multiplying
the number of events that satisfy the SR selection criteria but fail the `Tight' \tauhadvis{}
identification requirement by a \ff{}.
The \ff{} is defined as the ratio of the number of jets misidentified as
\tauhadvis{} that satisfy the `Tight' \tauhadvis{} identification criteria to the number that do not, but still satisfy the `Very Loose' identification requirement. The \ff{} is parameterised as a function of the \pT{} and track multiplicity of the \tauhadvis{}.
Since the \fakebkg{} originates from \Wjets{} and multi-jet production
processes, two independent \ffs{}, $F_{W}$ and $F_\mathrm{QCD}$, are measured in
dedicated \Wjets{} and multi-jet production CRs, respectively.
These regions are defined to be statistically independent of the SRs by
inverting several selection criteria. In particular, $(\cos\Delta\phi(\ell,\MET) + \cos\Delta\phi(\tauhadvis,\MET))< -0.35$, $\mT(\ell,\MET)>60$~\GeV{} and $\mT(\tau,\MET)>40$~\GeV{} for the $\Wjets$ CR, and $|\Delta \eta(\ell,\tauhadvis)| \geq 2$ and $\mT(\ell,\MET)<60$~\GeV{} for the multi-jet CR.
To obtain the final estimate of the \fakebkg{} contribution in the SRs, the combined \ff{}
$F = R_\mathrm{QCD} F_\mathrm{QCD}+(1-R_\mathrm{QCD})F_{W}$ is used.
The fraction of multi-jet events in the \fakebkg{}, $R_\mathrm{QCD}$,
is obtained  by scaling the number of events in the multi-jet CR by
the ratio of the number of events where the light lepton passes the isolation requirements to the number where it does not~\cite{HIGG-2019-09}.
This ratio is measured in another region enriched in multi-jet events where the $\ell$ and $\tauhad$ have the same electric charge.
 
The statistical uncertainties obtained from the \ff{} and $R_\mathrm{QCD}$ calculations are included
in the statistical analysis. They are considered as independent sources of uncertainty and their
magnitude varies between 2\% and 3\% in the  \VBF{} category, while it is $< 0.4\%$ in the  \nonVBF{} category.
An additional uncertainty accounts for the residual difference between
the \fakebkg{} modelling and the data in the \fakebkg{} enriched same-sign region
defined by applying the SR selection criteria, but requiring the \tauhad{} and $\ell$ to have the same charge.
The magnitude of the difference between the two regions varies from 2\% to 10\%, depending on the category, and it
is assigned to both the positive and negative component of the systematic uncertainty.
 
Figure~\ref{fig:lephad} shows the level of agreement between the data and
background modelling in the \nonVBF{} and \VBF{} categories for a
subset of variables with the highest discrimination power used in the
MVA as described in Section~\ref{sec:mva}.
Here the normalisation of signal and background yields is obtained by performing the simultaneous fit of the \hte{} and \htm{} signals (discussed in Section~\ref{sec:2poi}),
based only on data in the \lephad{} final state.
The distributions show different levels of separation between signal and background, and reasonable modelling of the data within statistical and systematic uncertainties.
The corresponding SR event yields are detailed in Table~\ref{tab:MC_yields_lephad_2poi} in Appendix \ref{app:yieldsMClephad}.


\subsection{\Symbased-based \leplep{} channel}
\label{sec:lep_taulep_symm}
 
The data-driven $e\slash\mu$-symmetry method was first employed in
Ref.~\cite{HIGG-2015-09}. This method is sensitive to the difference
between \BR(\hte) and \BR(\htm), and it is based on the following
two assumptions:
\begin{enumerate}
\item SM processes
are symmetric under the
exchange of prompt electrons with prompt muons to a good
approximation. As a consequence, the kinematic distributions
of prompt electrons and prompt muons are approximately the
same;
\item Flavour-violating decays of the Higgs boson break this
symmetry.
\end{enumerate}
 
According to the $e\slash\mu$-symmetry assumption, the SM background
is split equally between the \mte{} and \etm{} datasets.
However, the
\htm{}~signal is mostly present in the \mte{} dataset, since the \pt{}
of the muon originating from the Higgs boson is typically larger
than the \pt{} of the electron originating from the decay of the
$\tau$-lepton. Similarly, the \hte{} signal is mostly present in the
\etm{} dataset.
Thus, in a search for
\htm{} decays, the SM background can be estimated using the \etm{}
dataset, and vice versa.
 
Detector-related effects cause a distortion of
the original $e\slash\mu$-symmetry and need to be accounted for.
The first effect is due to events containing misidentified and non-prompt
light leptons, collectively referred to as \fakebkg{} in the following.
These particles originate from misidentified jets, misclassified
light leptons, hadronic decays within a jet, and electrons from photon conversion.
They contribute differently to the \mte{} and
\etm{} datasets.  The second effect is due to the muons and
electrons having different trigger, reconstruction, identifications and isolation efficiencies
which depend differently on kinematic properties such as \pt, |$\eta$|, and $\phi$.
 
\subsubsection*{Events containing \fake{} light leptons}
 
The \fake{} light leptons are subdivided into two contributions based
on their estimation method: \mcfakes{} and \fffakes{} for
contributions estimated from the simulation and from the data using
the \ff{} method, respectively.
 
Events contributing to \mcfakes{} include \tauhad{} that are
misidentified as light leptons, muons that are misidentified as
electrons, and photons that are misidentified as electrons.  Events
containing one or more \fake{} light leptons are estimated from simulation,
taking into account the following processes: $\Ztt$, top-quark, diboson, $\Zmumu$, $\Zee$ and $V\gamma$.  Contributions
from $H\rightarrow\tau\tau$ and $H\rightarrow WW$ are negligible. The
most significant contribution to \mcfakes{} is from
$\Zmumu$ where low-\pt muons are misidentified as
electrons. The modelling of this process (total cross-section and
shapes of various kinematic distributions) is validated in a dedicated
validation region dominated by $\Zmumu$ events. This region's definition
is similar to the one described in
Section~\ref{sec:background_leplep}, replacing the requirement
$\Delta\phi(\leptau,\met)<1.5$ with a requirement of $\mcol
< 115$~\GeV{}.  The systematic uncertainty associated with the
normalisation of this background is determined to be 16\%.
Conservatively, this uncertainty is applied
to all the  \mcfakes{} contributions.
 
Events contributing to \fffakes{} include jets misidentified as light
leptons and light leptons originating from hadronic decays within a
jet. These mostly originate from \Wjets{} processes.  A smaller
contribution originates from multi-jet production. This background is
estimated using the \ff{} method
where
the \ffs{} are measured in a dedicated $\Zjets$ CR enriched in
\fake{} light leptons.
This $\Zjets$ CR is based on the selection of exactly three electrons or muons where two are required to be consistent with a $Z$-boson decay, while the third is assumed to originate from a jet which is misidentified either as an electron or muon and is used for the determination of the \ffs{}.
A difference in jet flavour composition between the $\Zjets$ and
\Baseline{} regions could result in different \ffs.
This is taken into account by using a set of correction
factors measured in simulation as the ratio of the \ffs in the
two regions.
 
The \ffs{} and correction factors are binned as a function of lepton \pT, with electron \ffs{} also being parameterised with respect to $\Delta\phi \left(
e, \met \right)$.
The product of the \ffs{} and correction factors is applied as event weights,
depending on the kinematic quantities of the non-prompt light leptons that fail to satisfy the lepton identification criteria.
Contributions, where either the electron or muon is misidentified, are estimated separately and summed; the numbers of events where both light leptons are misidentified are treated specifically to avoid double counting.
 
The uncertainties associated with the \ff method consist of
statistical and systematic components.  The statistical component is
estimated for each \ff bin separately and consists of the statistical
uncertainties from the data and the subtracted prompt-lepton simulated yields within
the $\Zjets$ CR propagated to the \ffs.  The systematic component
corresponds to the uncertainty from the subtraction of events
containing three real light leptons, i.e.\ from $WZ$ and $ZZ$ processes where
the subtracted yields are varied by the theory cross-section uncertainties of
both processes separately.  Uncertainties associated with the
correction factors include a contribution from the limited number of simulated
events in the $\Zjets$ and \Baseline{} regions, and from the choice of a
specific MC generator.  In the statistical model (described in
Section~\ref{sec:stat}), the statistical sources are treated as
uncorrelated between \ff{} or correction factor bins, the systematic sources as
correlated between bins, and the MC subtraction-uncertainty also
as correlated between electron and muon \fakes{} background.
 
\subsubsection*{Efficiency correction}
 
The efficiency to measure a \mte{} event ($\epsilon^{\mu\tau_e}$)
or a \etm{} event ($\epsilon^{e\tau_\mu}$) is given by the product of
the event trigger efficiency and the muon and electron
efficiencies.
The trigger efficiency calculation and the associated systematic
uncertainties are described in~Ref.~\cite{PERF-2017-01}.
Simulated events are scaled using dedicated scale factors
to account for the modelling of trigger efficiency in the simulation.
 
The lepton efficiency is a product of the reconstruction,
identification and isolation efficiencies. For the muons, these are measured
in~Ref.~\cite{PERF-2015-10} along with their uncertainties
and found to be independent of the event selection.
For the electrons, regardless of the parameterisation used, the actual efficiencies
are strongly dependent on the event selection, and need to be measured for the analysis-specific phase space.
Simulated events are
corrected to match the data efficiencies by using dedicated scale factors.
 
In this study, the electron efficiencies are estimated from the simulated events
passing the \Baseline{} event selection
with the following two changes: the requirement $\ptOne > \SI{35}{\GeV}$ is
relaxed to  $\ptOne > \SI{20}{\GeV}$, in order to increase the number of events,
and only events that contain either one electron and one muon candidate, or two electron candidates, are considered.

The number of events available after applying this selection
determines the statistical uncertainty associated with the electron
efficiency measurement.  Systematic uncertainties associated with this
measurement are estimated by comparing the efficiencies obtained when
using either the electron or muon as the tag lepton, and by raising the
\pt{} selection for \lepone{} back to $\SI{35}{\GeV}$. The efficiencies
are scaled to account for possible differences observed between data
and simulation, using the set of scale factors and associated
uncertainties described in~Ref.~\cite{EGAM-2018-01} available for each
efficiency separately.
 
The efficiency corrections are applied per event, depending on
the kinematic properties of the electron and muon.  Each event in the
\etm{} sample is used to mimic an event in the \mte{} sample
with the same kinematic properties by scaling it with the efficiency ratio
$R_\epsilon = \epsilon^{\mu\tau_e}/\epsilon^{e\tau_\mu}$.
The same procedure is applied with the inverted ratio when scaling \mte{}
to \etm{} events.
The procedure is tested in simulated events, as shown in Figure~\ref{fig:symmEffectiveValidation}, by applying the efficiency
correction to \etm{} events and comparing them with \mte{} events.
By construction, the efficiency correction is only valid for
events containing prompt leptons. In the simulation, events
containing \fake{} leptons are vetoed.  In the data, contributions
from both \mcfakes{} and \fffakes{} are subtracted before the
efficiency correction is applied. As shown in Figure~\ref{fig:symmEffectiveValidation}, this procedure restores the symmetry between the two channels in simulation.
 
\begin{figure}[tb]
\centering
\subfloat[]{\includegraphics[width=0.43\textwidth]{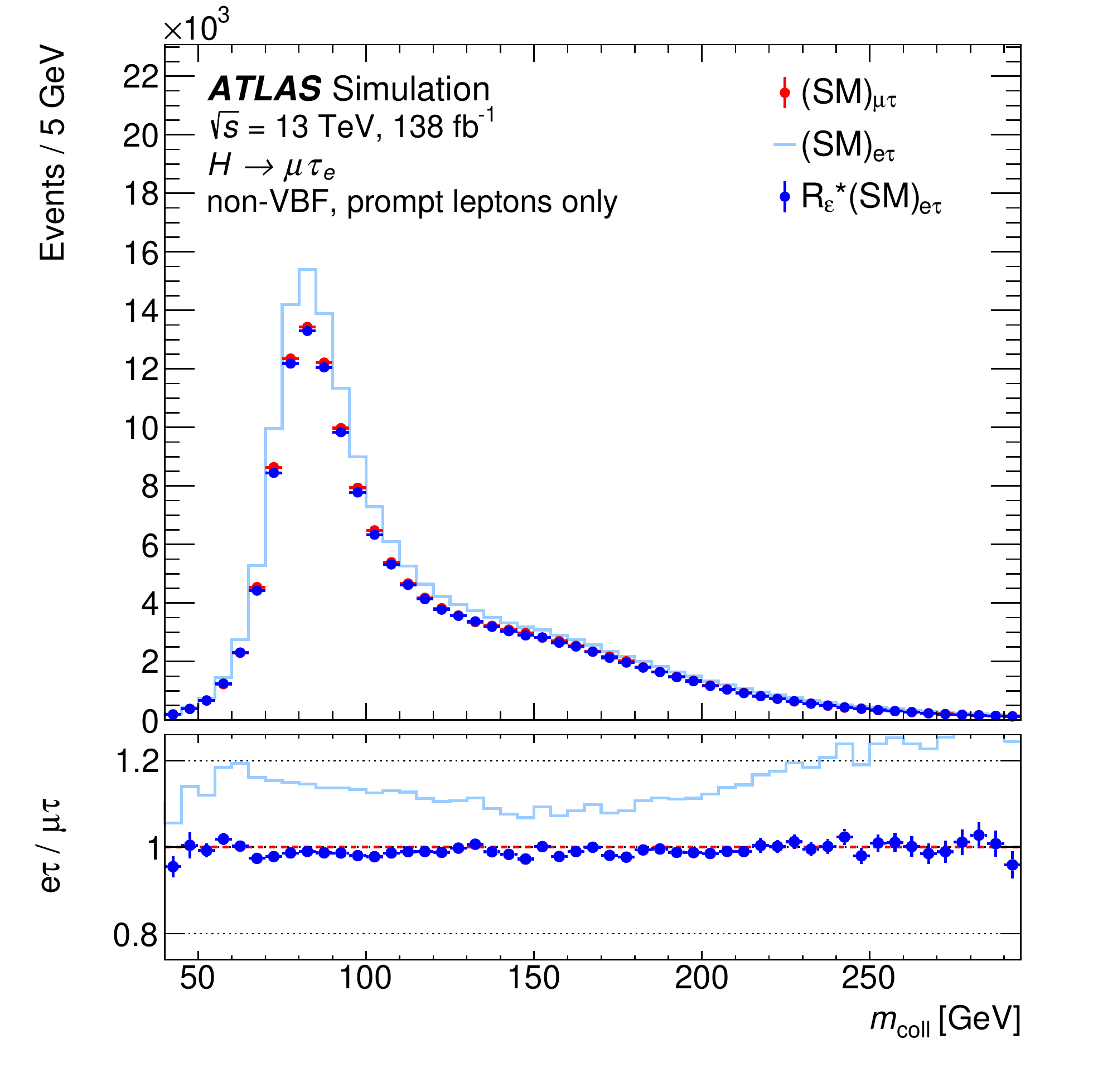}}
\subfloat[]{\includegraphics[width=0.43\textwidth]{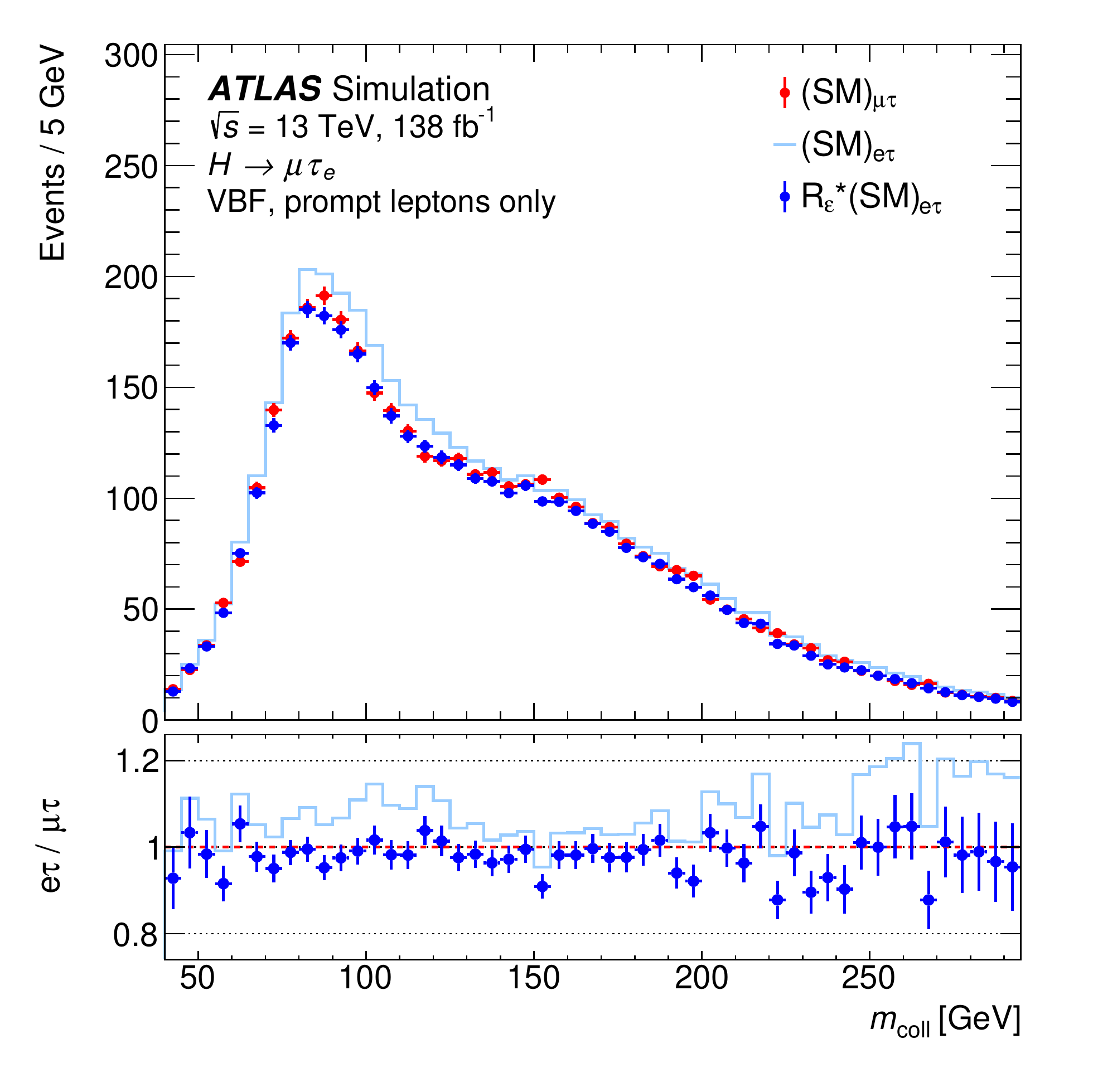}} \\
\caption{The collinear mass~\cite{Ellis:1987xu} distributions,
\mcol{}, in simulated events satisfying the (a) \nonVBF{}
and (b) \VBF{} selection criteria. Events with a leading
muon and a subleading electron from the $\tau$-lepton decay
($\mathrm{(SM)_{\mu\tau}}$) are compared with events with a leading electron and a
subleading muon without ($\mathrm{(SM)_{e\tau}}$) or with the
efficiency correction applied ($R_\epsilon^*\mathrm{(SM)_{e\tau}}$). The bottom panel shows the ratio of the $e\tau$ and $\mu\tau$ predictions.
The ratio for the $e\tau$ prediction with the efficiency correction applied ($R_\epsilon^*\mathrm{(SM)_{e\tau}}$/$\mathrm{(SM)_{\mu\tau}}$) is shown by dark blue points, while the $e\tau$ prediction without the efficiency correction applied ($\mathrm{(SM)_{e\tau}}$/$\mathrm{(SM)_{\mu\tau}}$) is shown by the light blue line. Uncertainties included are statistical only.
}
\label{fig:symmEffectiveValidation}
\end{figure}
 
\FloatBarrier
 
\subsubsection*{Background estimation}
 
\begin{figure}[!tb]
\centering
\subfloat[]{\includegraphics[width=0.43\textwidth]{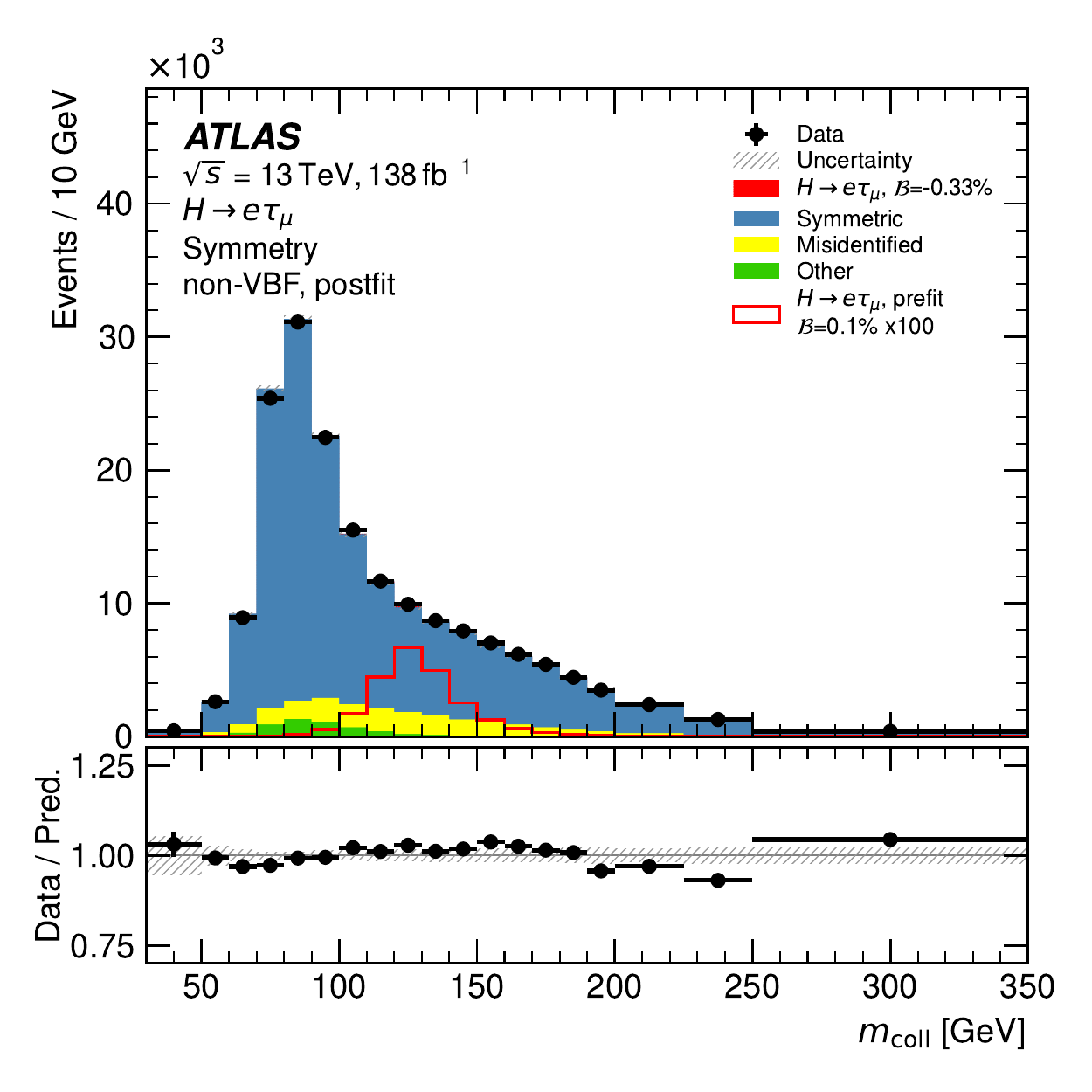}}  \hspace{10mm}
\subfloat[]{\includegraphics[width=0.43\textwidth]{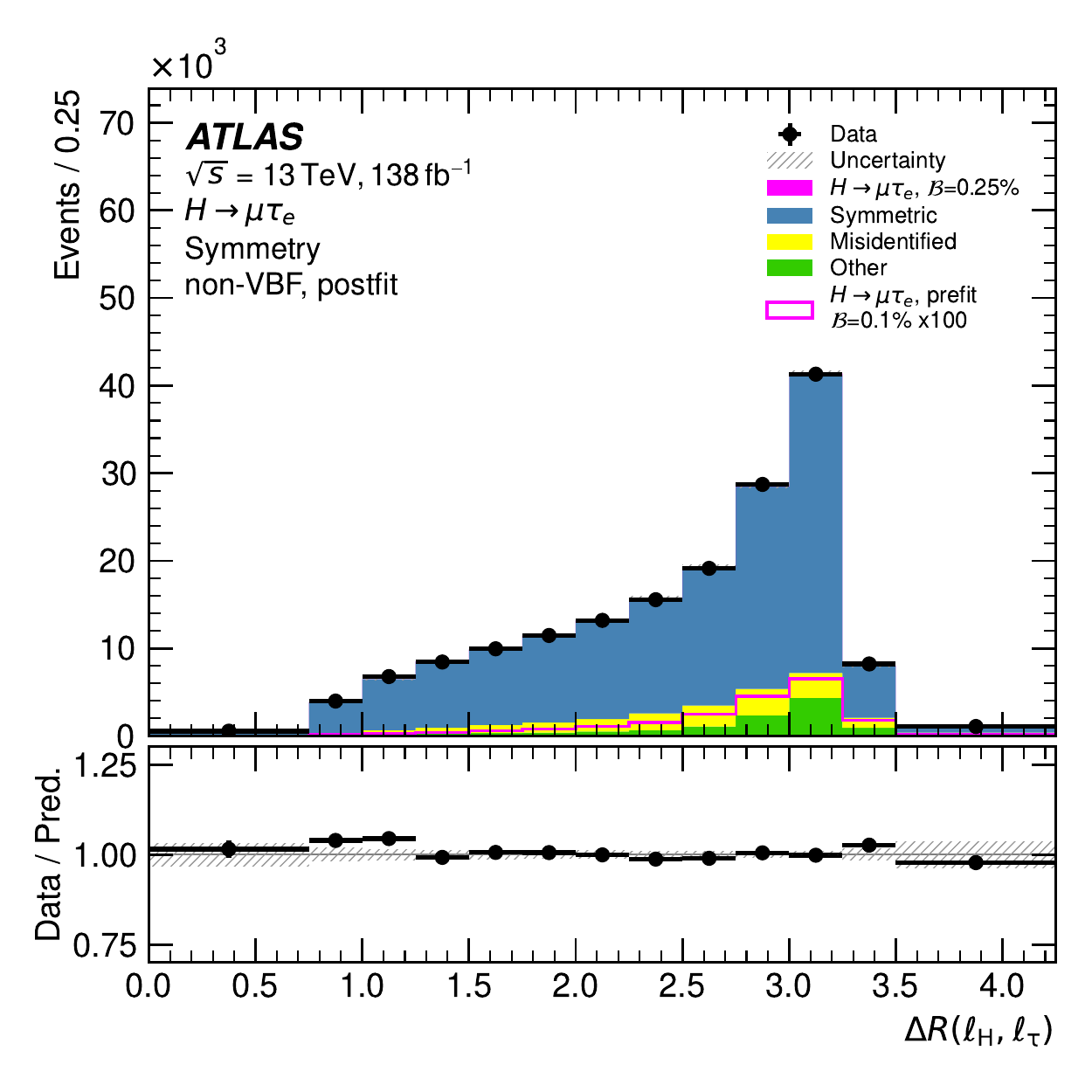}} \\
\subfloat[]{\includegraphics[width=0.43\textwidth]{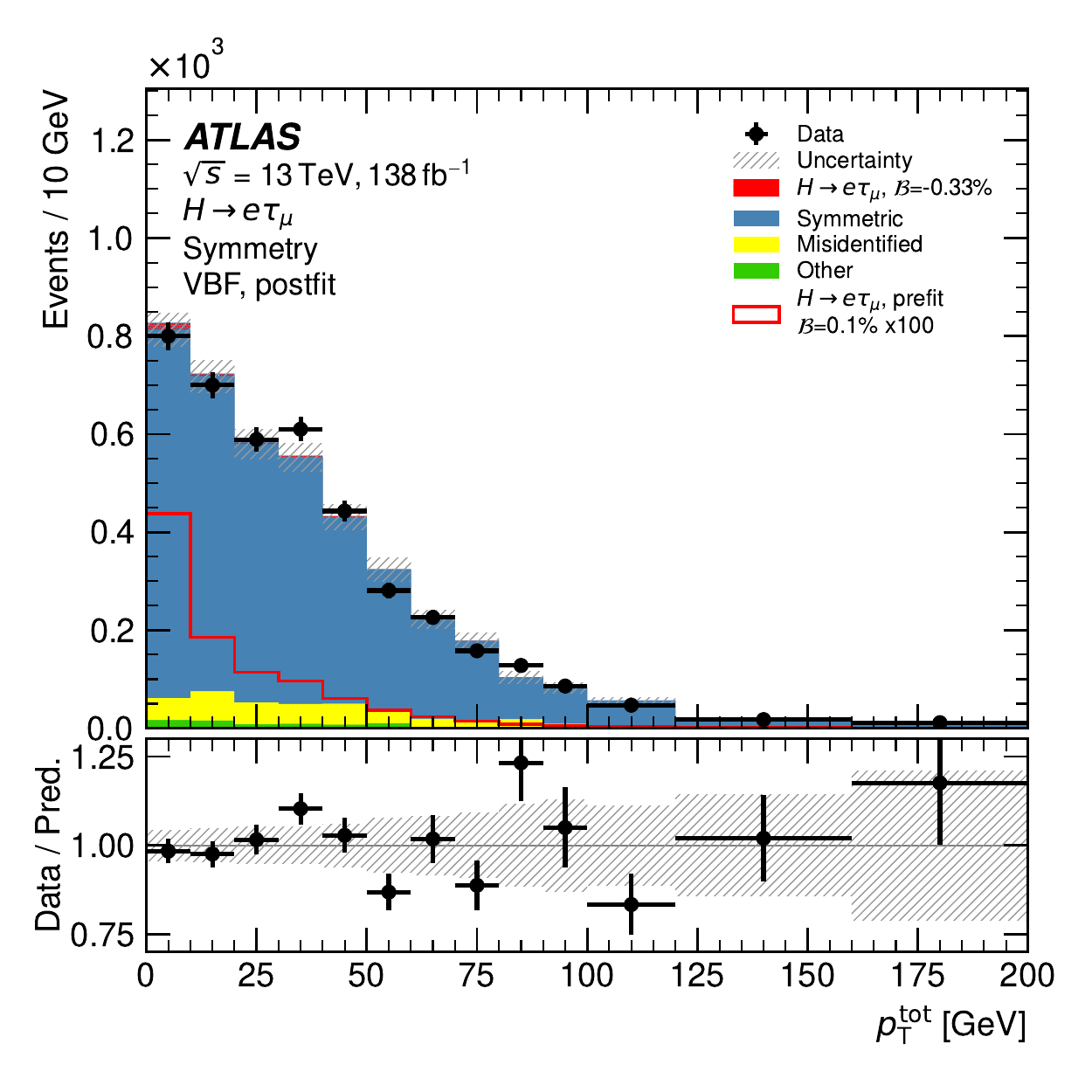}} \hspace{10mm}
\subfloat[]{\includegraphics[width=0.43\textwidth]{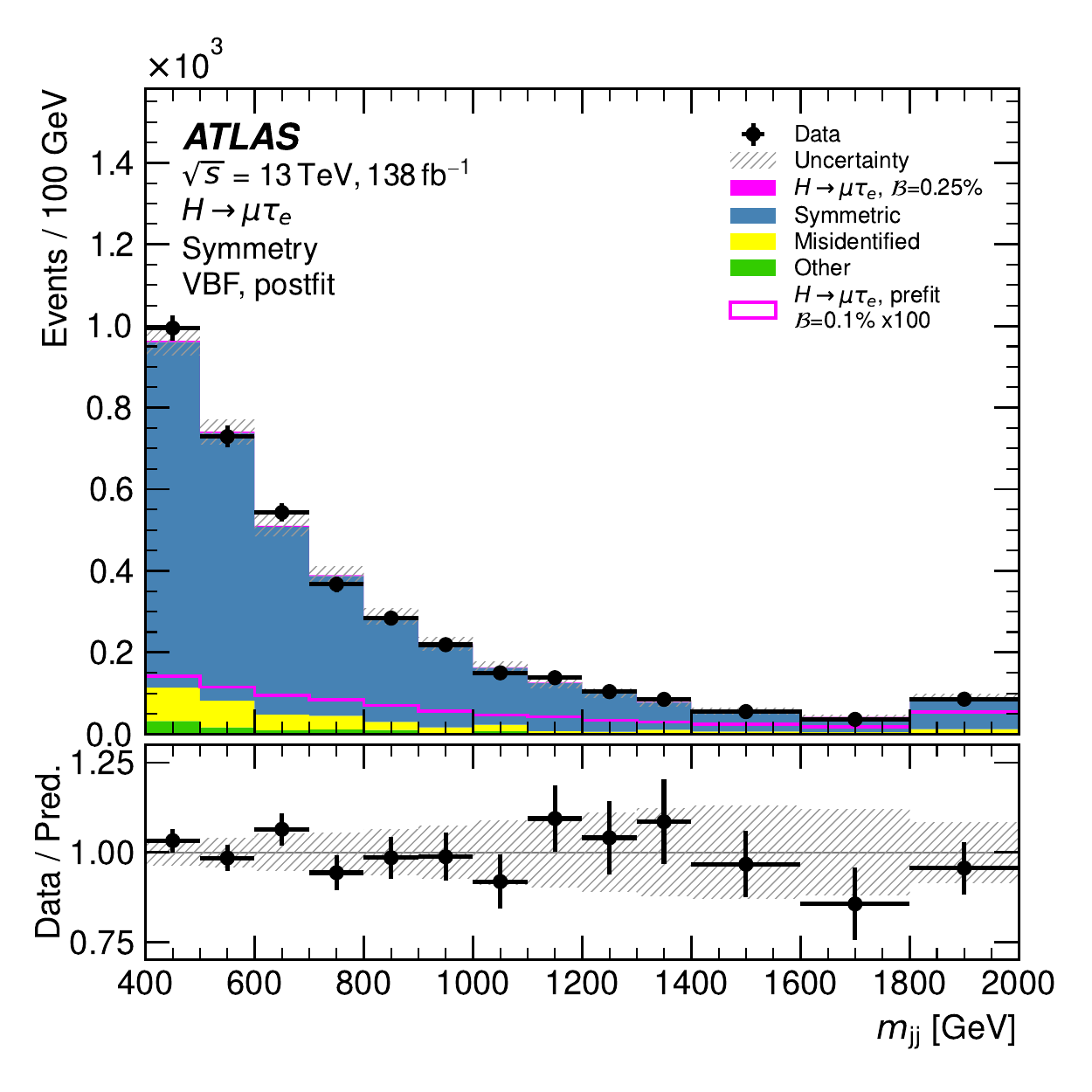}} \\
\caption{ Distributions of representative kinematic quantities
for \etm{} (left) and \mte{} (right) final states,
after an independent fit of the \hte{} and \htm{} signals, respectively,
obtained by fitting the data of the \Symbased{} \leplep{} channel: (a) \mcol{} and (b) the angular separation between \lephiggs{} and \leptau{}, $\Delta R(\lephiggs,\leptau)$ in the \nonVBF{} category, (c) $\pT^{\text{tot}}$, which is
the absolute value of the vectorial sum of $\pT$ of the two
leptons, the two leading jets and the $\met$, and (d) the invariant
mass of the dijet system, $\mjj$, in the \VBF{} category.  Entries with values that would
exceed the horizontal axis range are included in the last bin of
each distribution.
The hashed band represents the prefit statistical uncertainty,
and the experimental and theoretical uncertainties obtained from the likelihood fit.
Overlaid prefit signal shapes assume $\BR(H \to \ell \tau) = 0.1\%$ and are enhanced by a factor 100 for visibility. The postfit signal contributions are considered as part of the predictions.
}
\label{fig:symm-background-prefit}
\end{figure}

The number of background events in each dataset (\etm{} and \mte) is
estimated as a sum of three contributions: a symmetric component
estimated by applying the efficiency correction procedure to the
events in the other dataset, \mcfakes{} and \fffakes{}.
Since the efficiency correction is only valid for true
leptons, the contribution from events containing \fake{} leptons is
subtracted from the other dataset before applying the efficiency
correction.
 
Figure~\ref{fig:symm-background-prefit} compares the data and the
background prediction for a selected set of kinematic variables with
strong signal-vs-background discrimination power for the \etm{} and \mte{} datasets in the \nonVBF{} and \VBF{}
categories. In the figures, the \fffakes{} contribution is labelled as
\enquote{Misidentified}, while \mcfakes{} contribution is labelled as \enquote{Other}.
The normalisation of signal and background yields is obtained by
performing the independent fits of the \hte{} and \htm{} signals (discussed in Sections~\ref{sec:1poi} and \ref{sec:comp-leplep}),
based only on data in the \leplep{} final state.
The distributions show either \hte{} or \htm{} signal, with clear shape differences to the backgrounds, particularly for mass-based variables, and good agreement of the prediction with data within statistical and systematic uncertainties.
The corresponding SR event yields are shown in Table~\ref{tab:symmYields} in Appendix~\ref{app:yieldsSymm}.
 
\FloatBarrier



\section{Multivariate analysis strategy}
\label{sec:mva}
 
In order to separate signal from various background contributions,
two MVA techniques are used:
boosted decision trees (BDTs), exploited in
the \MCbased{} method, and fully connected deep neural networks (NNs),
used in the \Symbased{} method.
Different MVA techniques are used for the \nonVBF{} and \VBF{} categories to exploit the VBF topology.
The MVA technique with the best expected sensitivity was chosen for each background estimation method.
The BDTs are trained using the Toolkit for Multivariate Data Analysis~\cite{TMVA}, while the NNs
are trained using Keras (v2.2)~\cite{chollet2015keras}, with Tensorflow (v1.12)~\cite{abadi2016tensorflow} as the backend.
 
In the \leplep{} channel of the \MCbased{} method, three different BDTs are utilised:
$\mathrm{BDT}_1^{\leplep{}}$ is devoted to separating the signal from events
with \fake{} leptons; $\mathrm{BDT}_2^{\leplep{}}$ separates
the signal from top-quark background processes, diboson production
and $H\rightarrow WW$ decays; and $\mathrm{BDT}_3^{\leplep{}}$ separates the signal from
$Z\rightarrow \tau\tau$, $Z\rightarrow \ell \ell$ and $H\rightarrow
\tau \tau$ backgrounds.
The three BDTs were utilised because they provided better expected signal significance than a single BDT.
To increase the number of events in the BDT training, the \etm{}
and the \mte{} datasets are combined.
The resulting BDT scores are combined into a single score using a
linear weighted sum, with the weights optimised using the expected signal significance
as a figure-of-merit.
 
In the \lephad{} channel of the \MCbased{} method, two BDTs are used in all cases
except for the \nonVBF{} category in the $\ethad$ channel, where three BDTs are used.
Two/three BDTs were chosen instead of a single BDT because they provide better expected signal significance.
In the case where three BDTs are trained, $\mathrm{BDT}_1^{\lephad{}}$ discriminates the signal from
the \Ztt{} events; $\mathrm{BDT}_2^{\lephad{}}$ separates the signal from \fakebkg{}; and $\mathrm{BDT}_3^{\lephad{}}$ separates the signal from the rest of the  backgrounds.
In the case where two BDTs are trained, $\mathrm{BDT}_{1'}^{\lephad{}}$ separates the signal from \Ztt{} events and
$\mathrm{BDT}_{2'}^{\lephad{}}$ separates the signal from the rest of the backgrounds.
The resulting BDT scores are combined into a single score using a
linear weighted sum in the \nonVBF{} categories, and a quadratic
weighted sum in the \VBF{} categories. The latter provided a 4\% improvement in the expected signal significance, used as a figure-of-merit, for the \VBF{} categories; in the \nonVBF{} categories, the linear combination provided the better expected signal significance.
 
In the \leplep{} channel of the \Symbased{} method,
the requirement on the jet invariant mass ($\mjj$) in the \VBF{} category is lowered to 300~\GeV{} to increase the size
of the training set. The depth of the NN is chosen taking into account the number of events
available for the training.
For the \nonVBF{} category, a multiclass classifier, with three output classes, is used.
The three classes correspond to signal,
flavour-symmetric SM background and \fakebkg{}. The \mcfakes{}
contribution is added to the flavour-symmetric SM background class
since the distributions of its main processes, $Z\rightarrow\mu\mu$,
$Z\rightarrow\tau\tau$ and $V\gamma$, are more similar to the
distributions of the flavour-symmetric SM background processes than to
the ones originating from jets misidentified as a lepton.
For the \VBF{} category, three single binary classification networks are
trained and the resulting output distributions are combined via a
linear combination to obtain a single discriminant. The three networks
are trained to separate signal events from:
(i) \mcfakes{}, $Z\rightarrow\tau\tau$ and $H\rightarrow\tau\tau$,
(ii) top-quark, diboson and $H\rightarrow WW$ and
(iii) \fffakes{} events.
 
Details of the MVA optimisation procedure, as well as the input variables used
in the training, are described in Appendix~\ref{app:MVA}.
The distributions of the resulting MVA discriminants are shown in
Figures~\ref{fig:BDT_scores_leplep}, \ref{fig:BDT_scores_lephad}
and \ref{fig:symm-background-postfit} for
the \MCbased{} \leplep{}, \MCbased{} \lephad{} and \Symbased{} \leplep{}
channels, respectively.
The \MCbased{} method distributions are obtained after performing the
simultaneous fit of the \hte{} and \htm{} signals described in
Section~\ref{sec:2poi}.
The \Symbased{} method distributions use the set-up for
independent searches for \hte{} and \htm{} (see Section~\ref{sec:1poi}), setting the other signal to zero
in each case, and are based on \leplep{} final-state data, which pass the \Symbased-method-specific selection.
The conversion of the fit of the branching ratio difference into individual branching ratios results in positive
contributions for \htm{} and negative contributions for \hte{}, as discussed in Section~\ref{sec:lep_taulep_symm}.
This is indicated with a positive or negative sign in Figure~\ref{fig:symm-background-postfit}.
 
\begin{figure}[!hptb]
\centering
\subfloat[]{\includegraphics[width=0.45\textwidth]{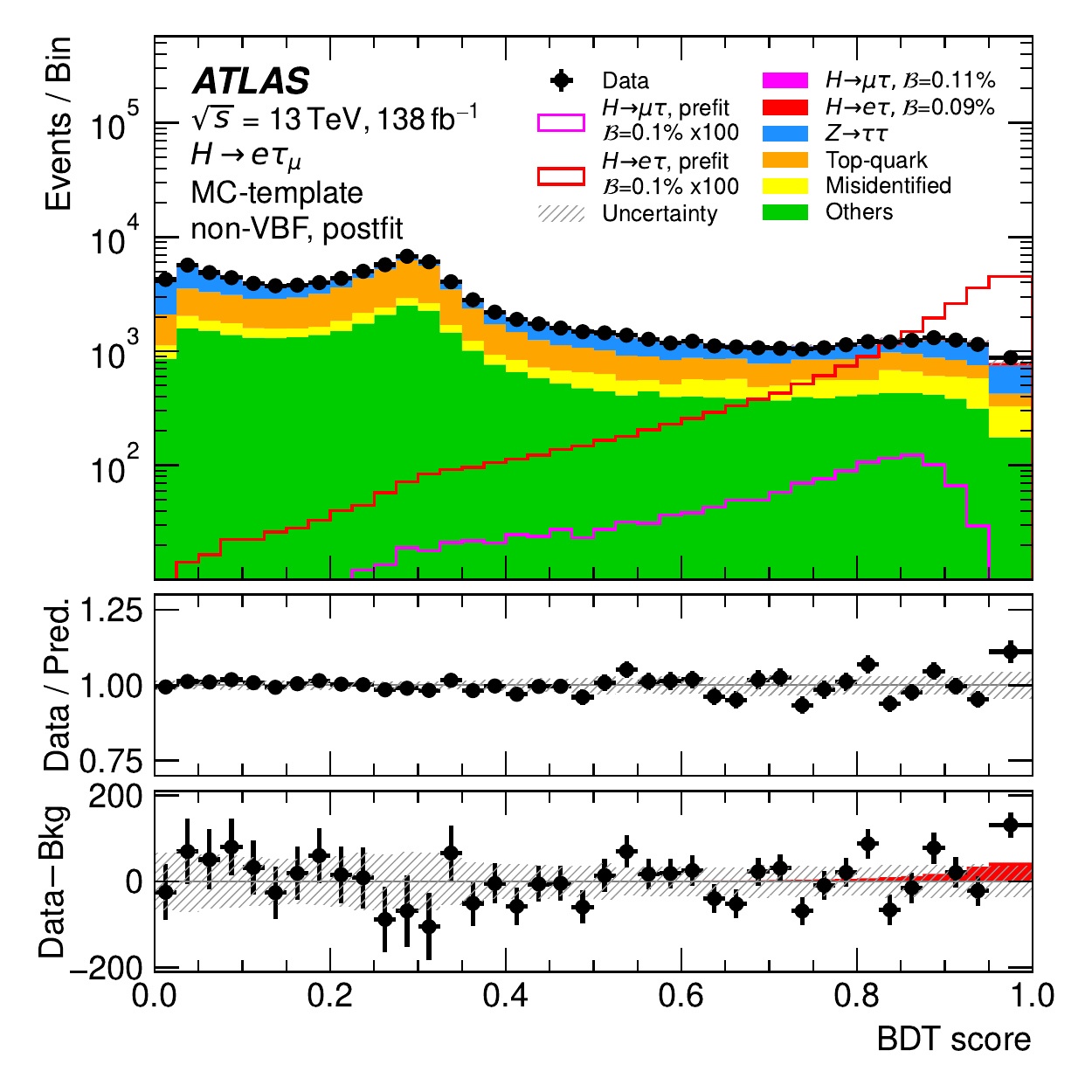}}
\subfloat[]{\includegraphics[width=0.45\textwidth]{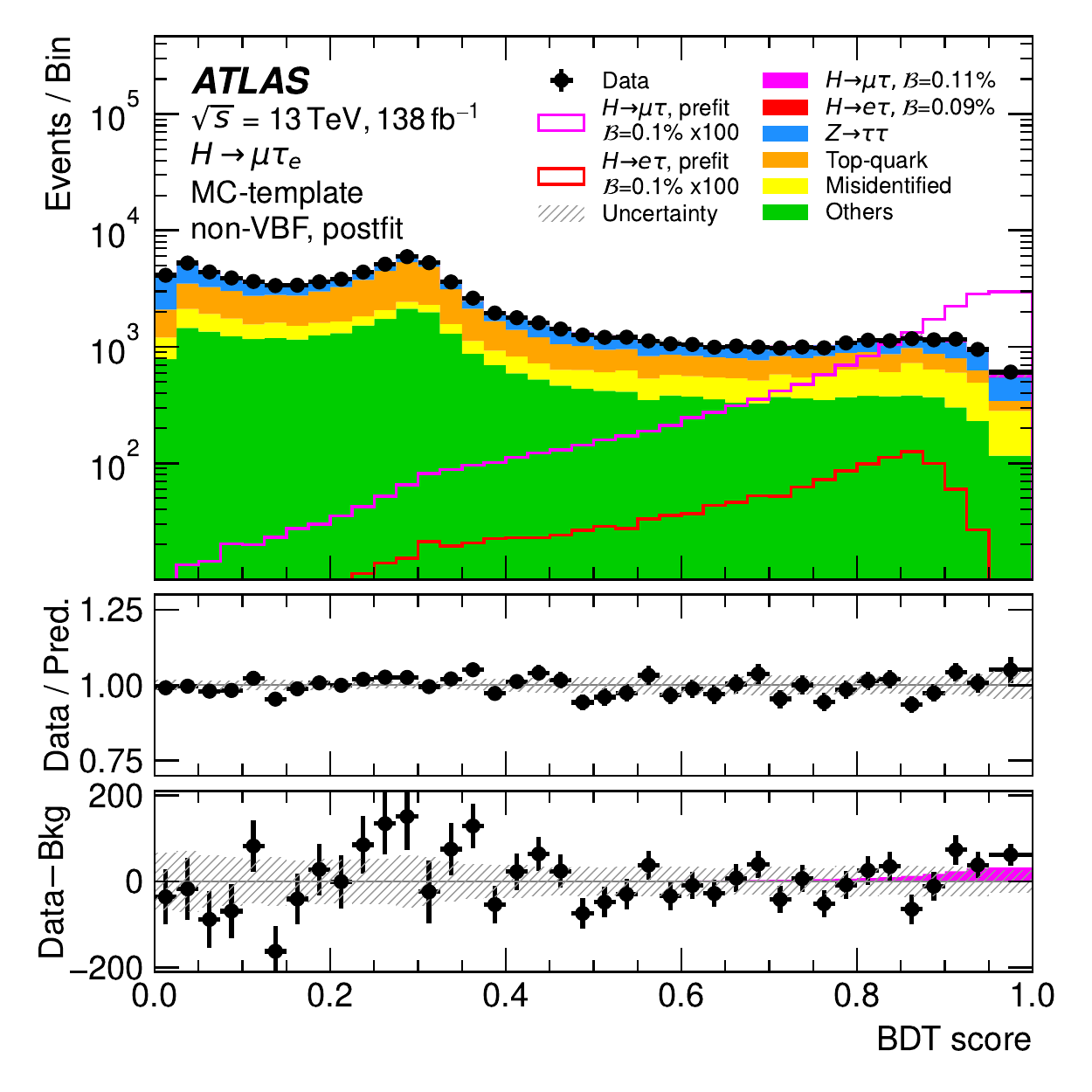}} \\
\subfloat[]{\includegraphics[width=0.45\textwidth]{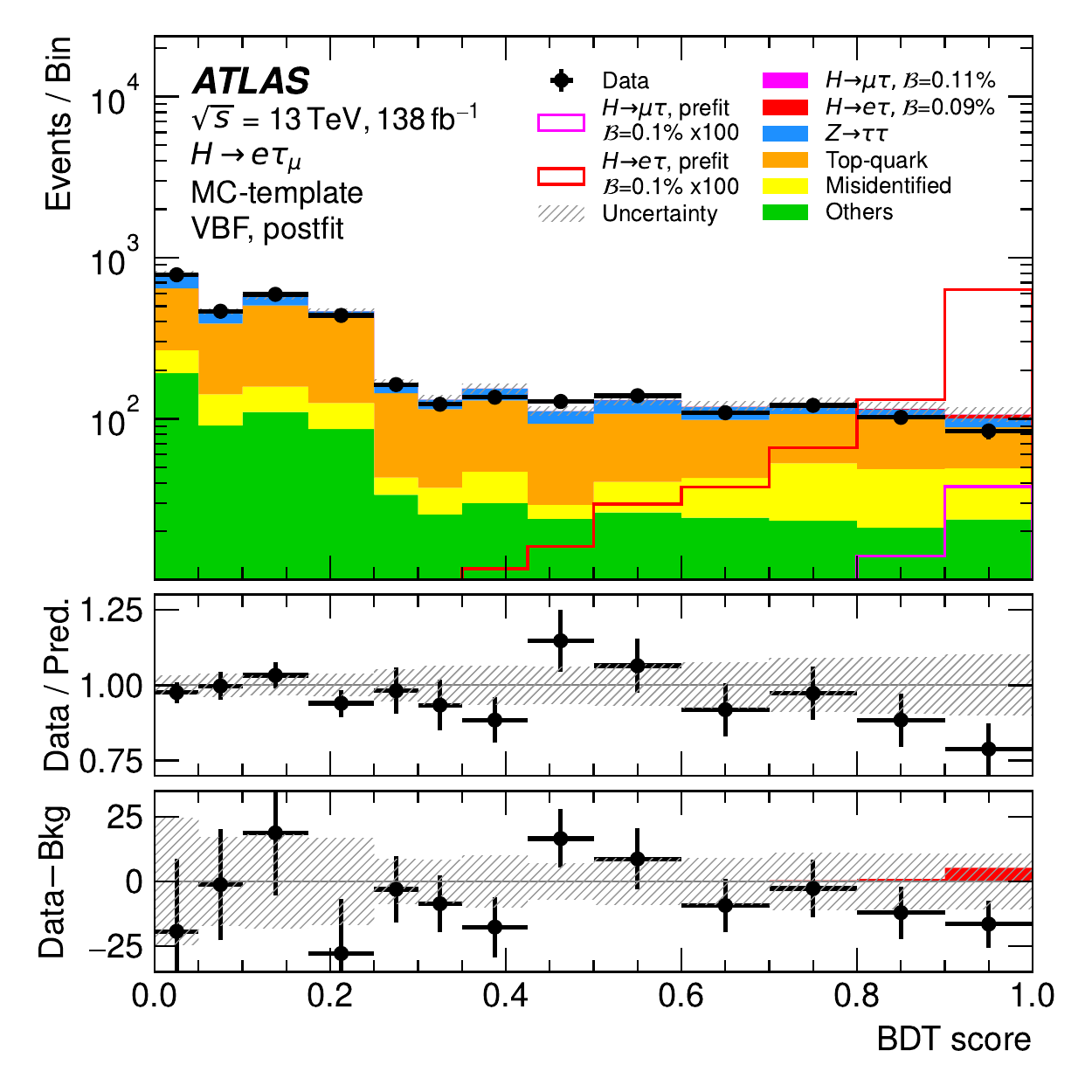}}
\subfloat[]{\includegraphics[width=0.45\textwidth]{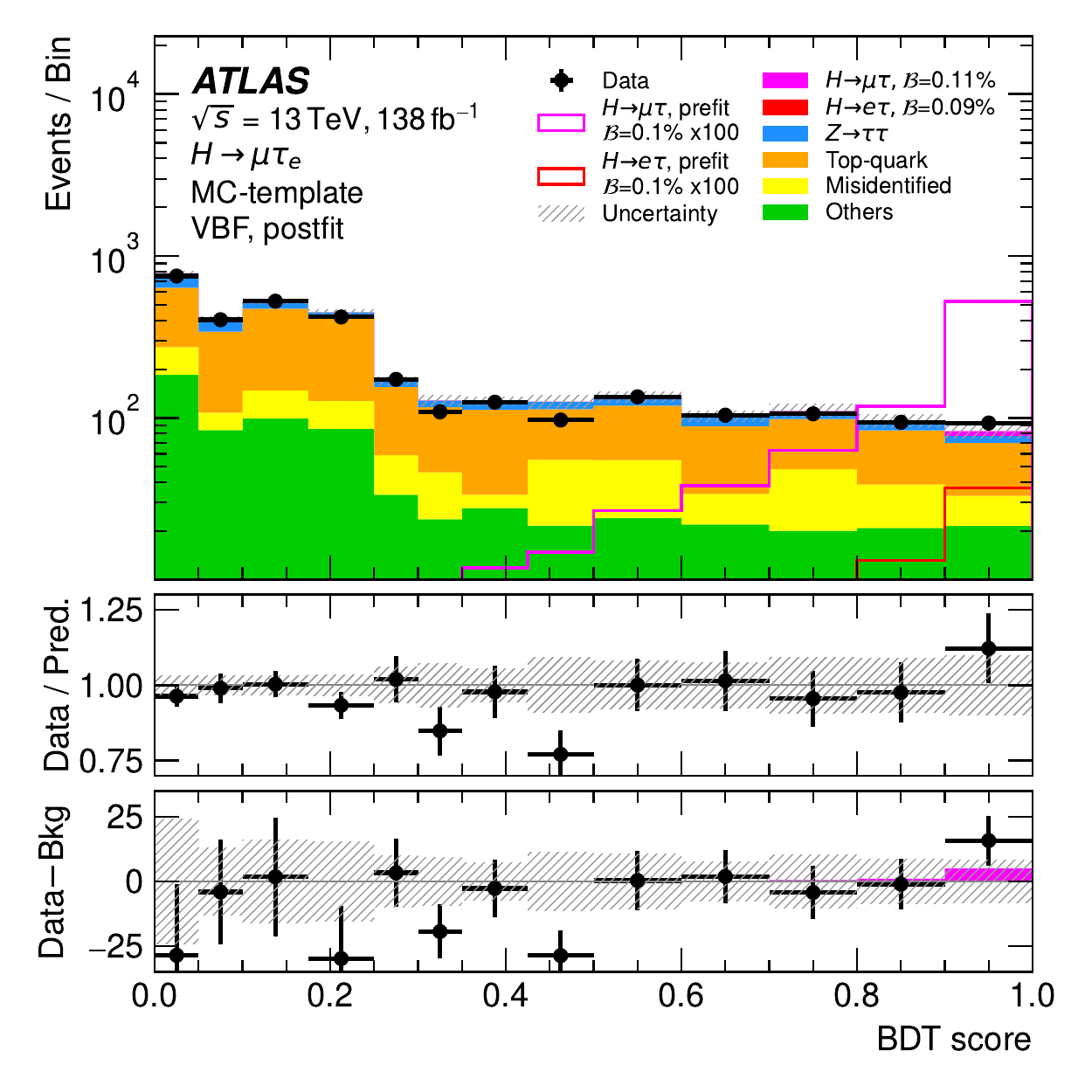}}
 
\caption{BDT score distributions for the $e\tau_\mu$ (left) and the
$\mu\tau_e$ (right) final states in the \nonVBF{} (top)
and \VBF{} (bottom) categories of the \MCbased{} \leplep{} channel.
The middle panel shows the ratio of data to prediction (background+signal).
The bottom panel shows the residuals between data and the background after the fit.
The hashed band indicates the total postfit uncertainty of the total predicted yields.
The prediction for each sample is determined from the likelihood fit performed to
measure \hte{} and \htm{} signals simultaneously, described in Section~\ref{sec:2poi}.
The binning is shown as used in the likelihood fit.
Overlaid prefit signal shapes assume $\BR(H \to \ell \tau) =0.1\%$
and are enhanced by a factor 100 for visibility. The postfit signal contributions are considered as part of the predictions.}
\label{fig:BDT_scores_leplep}
\end{figure}
 
\begin{figure}[!hptb]
\centering
\subfloat[]{\includegraphics[width=0.45\textwidth]{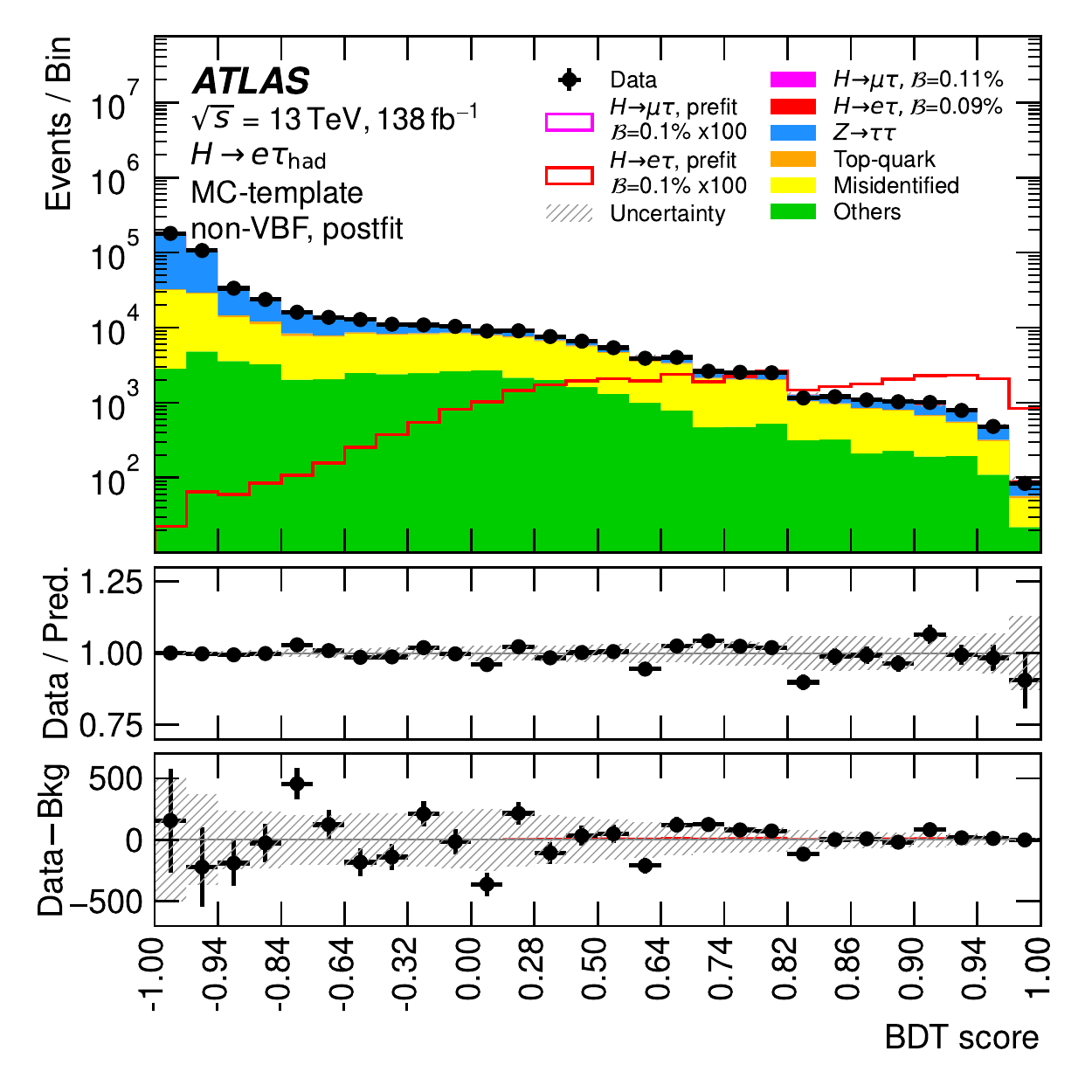}}
\subfloat[]{\includegraphics[width=0.45\textwidth]{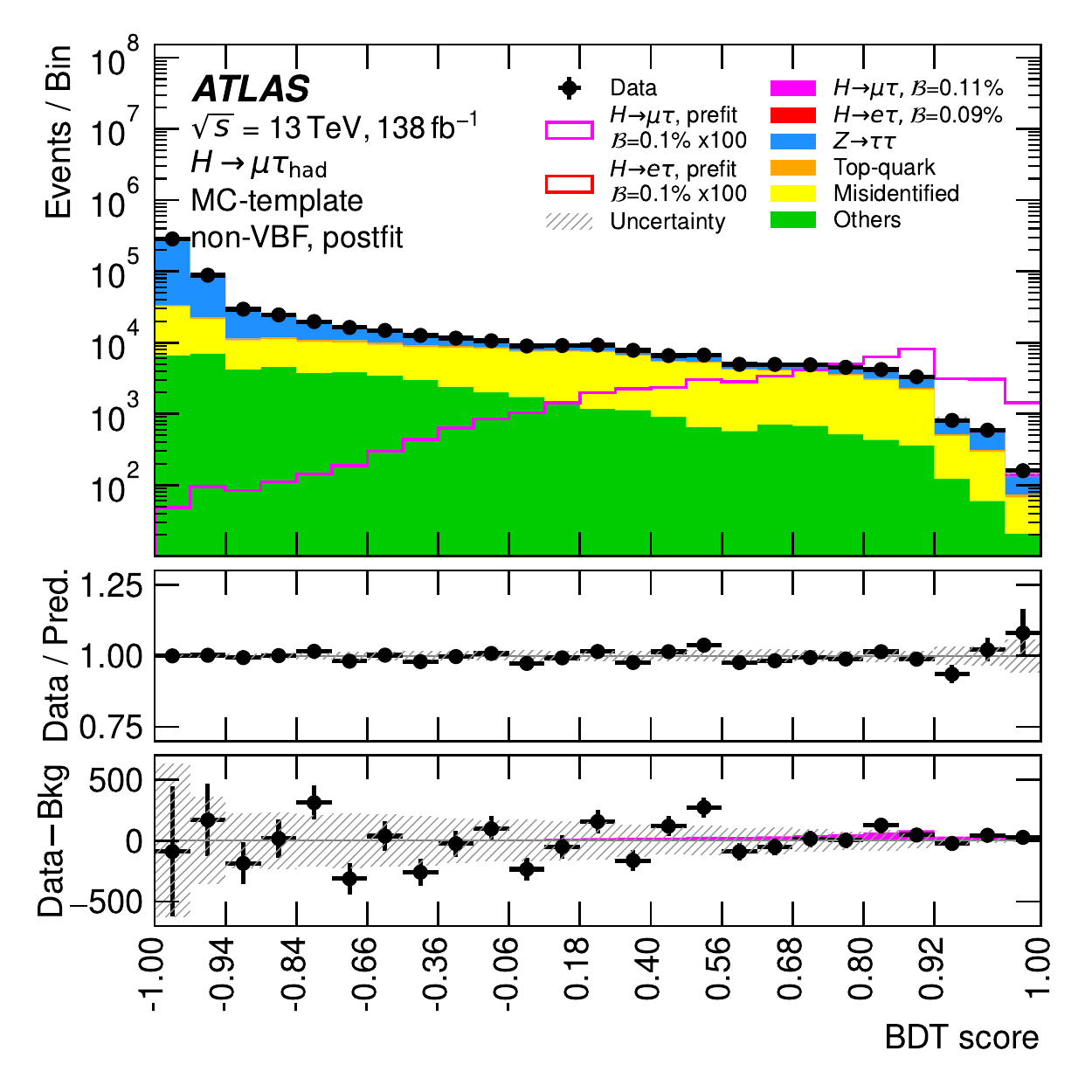}} \\
\subfloat[]{\includegraphics[width=0.45\textwidth]{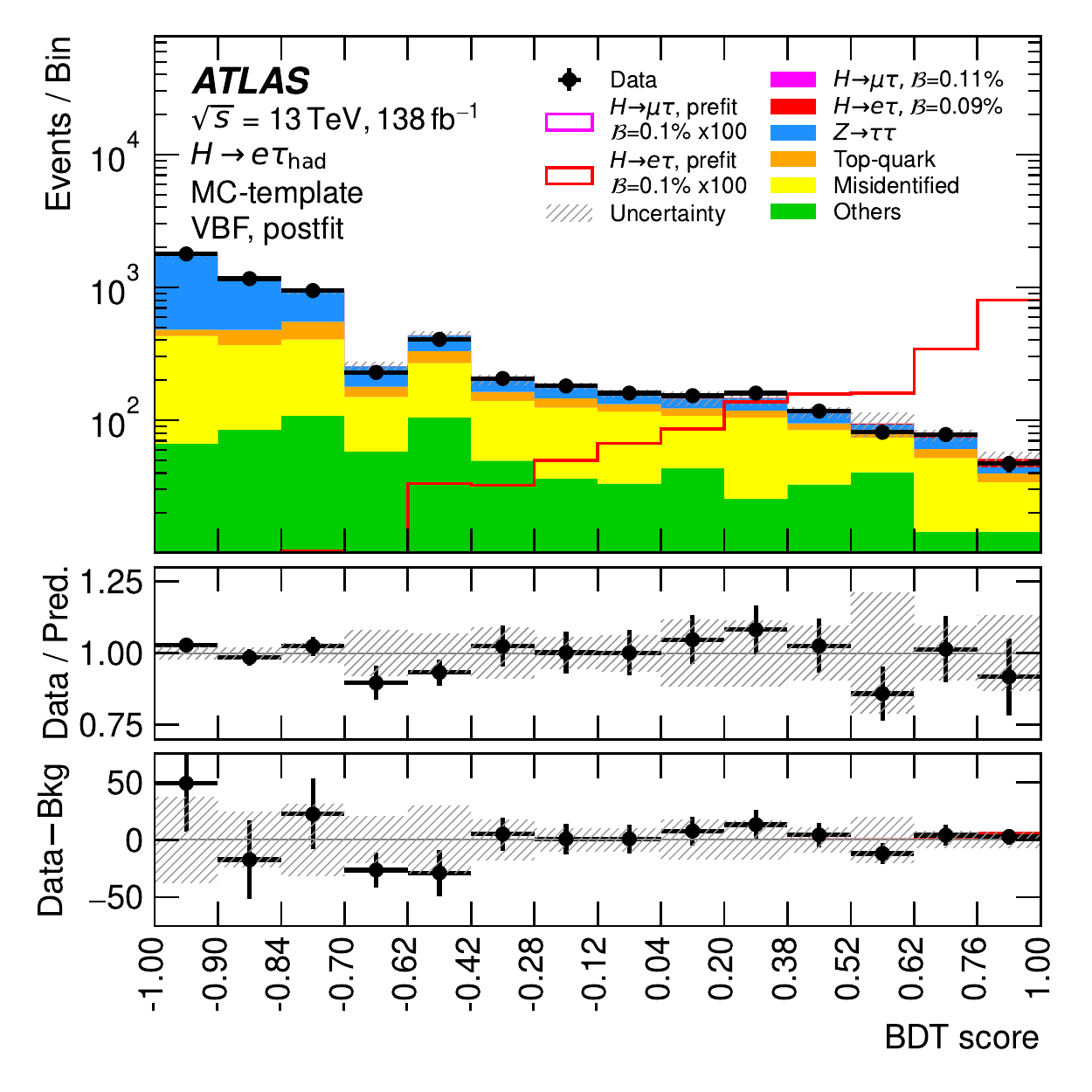}}
\subfloat[]{\includegraphics[width=0.45\textwidth]{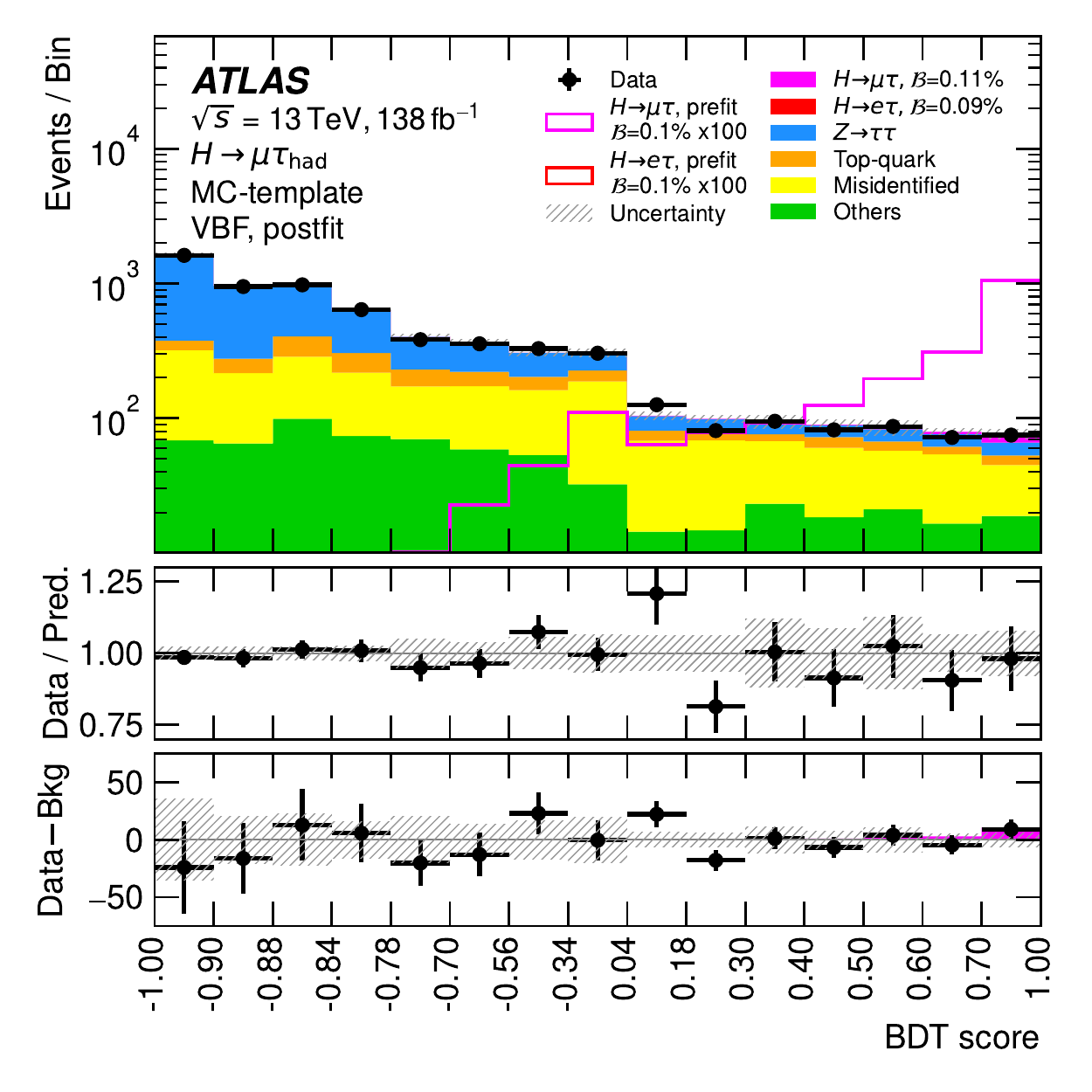}}
 
\caption{BDT score distributions for the $e\tauhad$ (left) and the
$\mu\tauhad$ (right) final states in the \nonVBF{} (top) and \VBF{}
(bottom) categories of the \MCbased{} \lephad{} channel.
The middle panel shows the ratio of data to prediction (background+signal).
The bottom panel shows the residuals between data and the background after the fit.
The hashed band indicates the total postfit uncertainty of the total predicted yields.
The prediction for each sample is determined from the likelihood fit performed to
measure \hte{} and \htm{} signals simultaneously, described in Section~\ref{sec:2poi}.
The binning is shown as used in the likelihood fit.
Overlaid prefit signal shapes assume $\BR(H \to \ell \tau) =0.1\%$
and are enhanced by a factor 100 for visibility. The postfit signal contributions are considered as part of the predictions.}
\label{fig:BDT_scores_lephad}
\end{figure}
 
\begin{figure}[!hptb]
\centering
\subfloat[]{\includegraphics[width=0.45\textwidth]{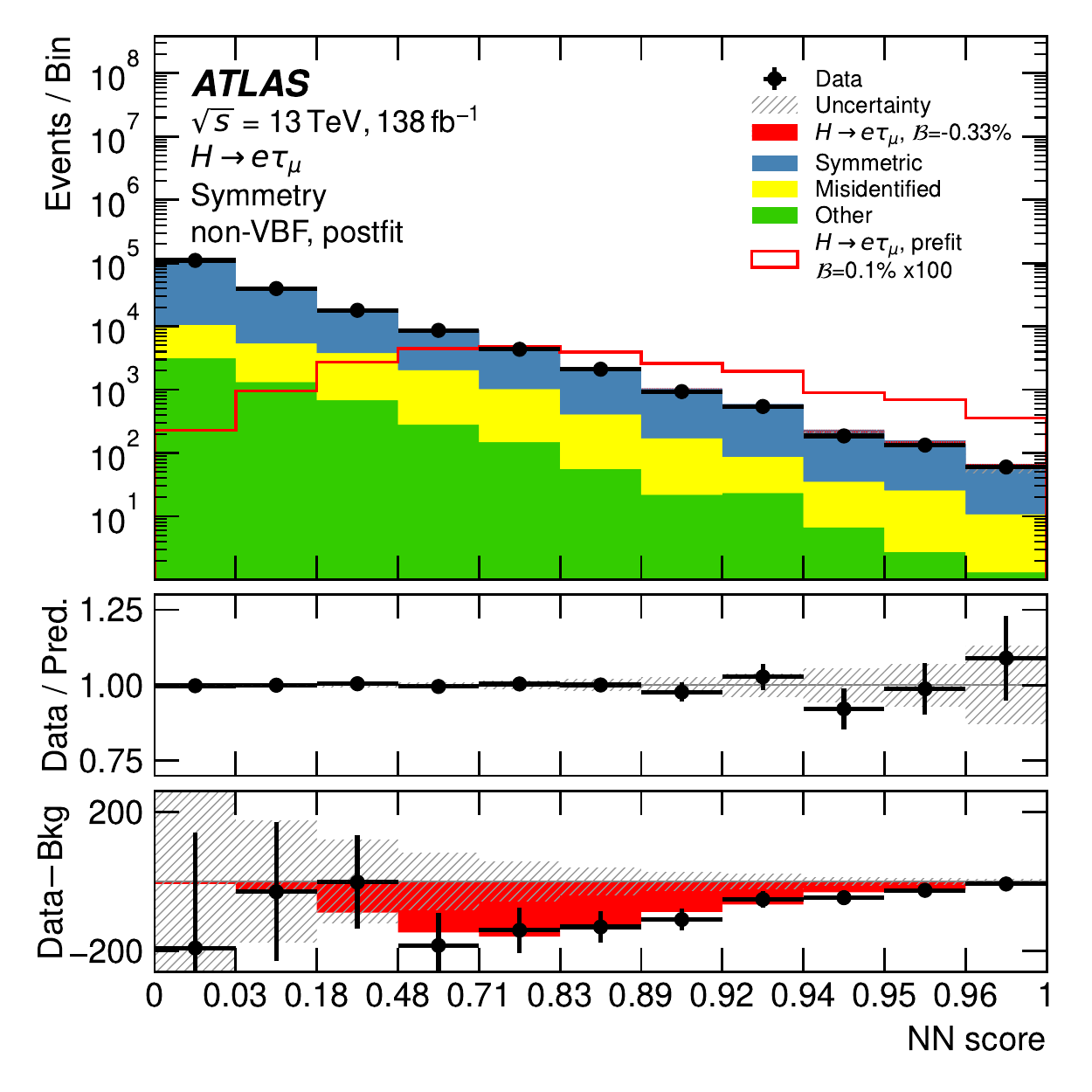}}
\subfloat[]{\includegraphics[width=0.45\textwidth]{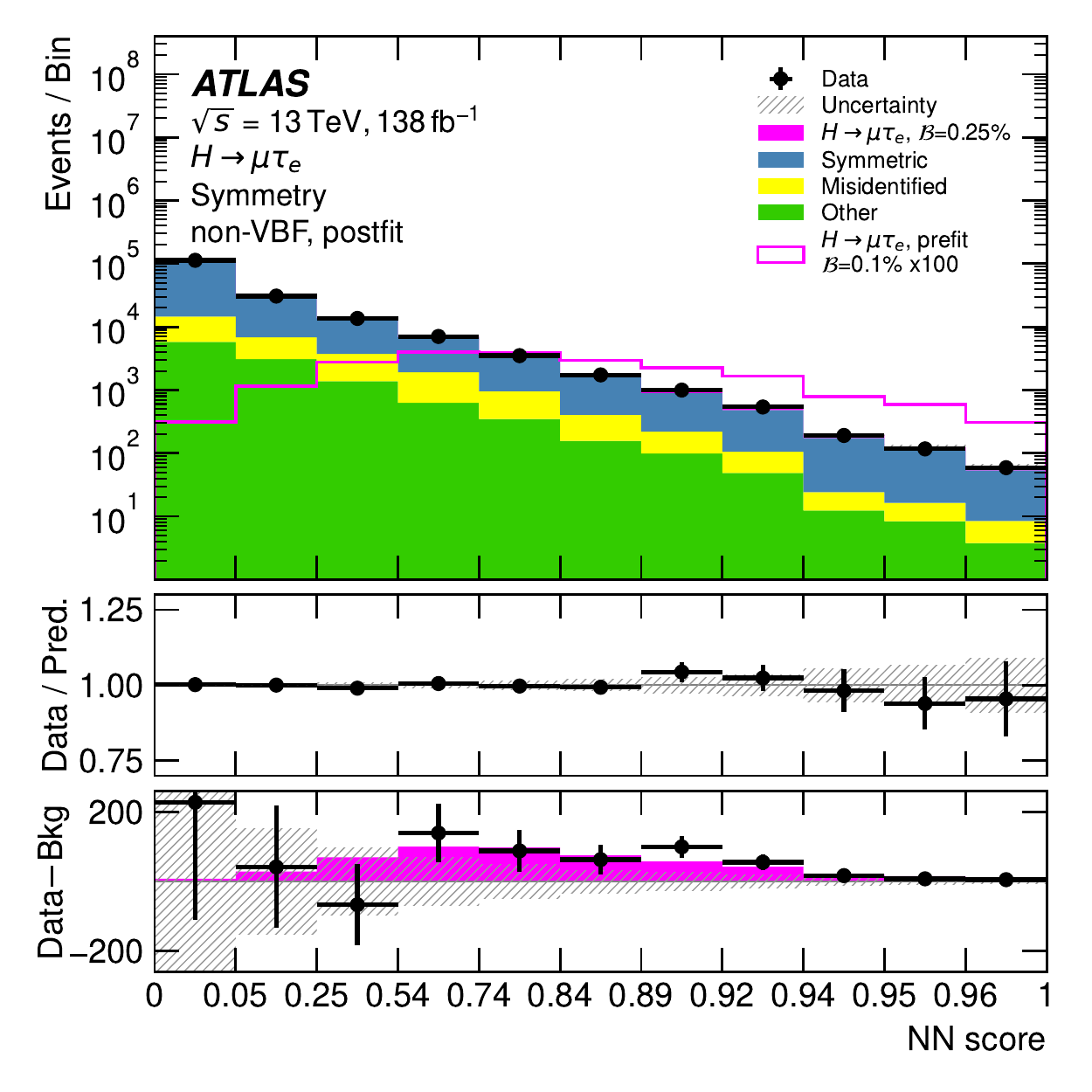}} \\
\subfloat[]{\includegraphics[width=0.45\textwidth]{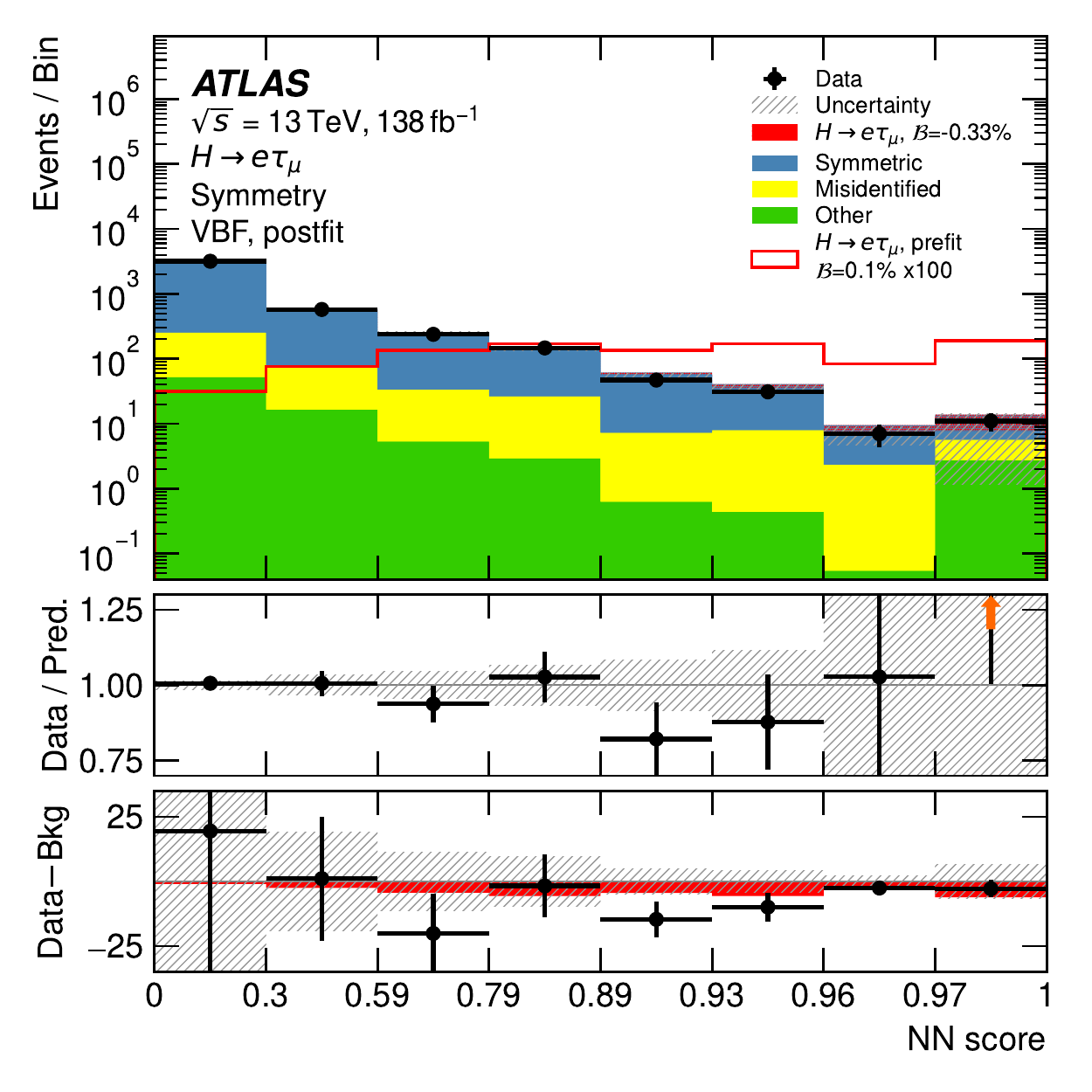}}
\subfloat[]{\includegraphics[width=0.45\textwidth]{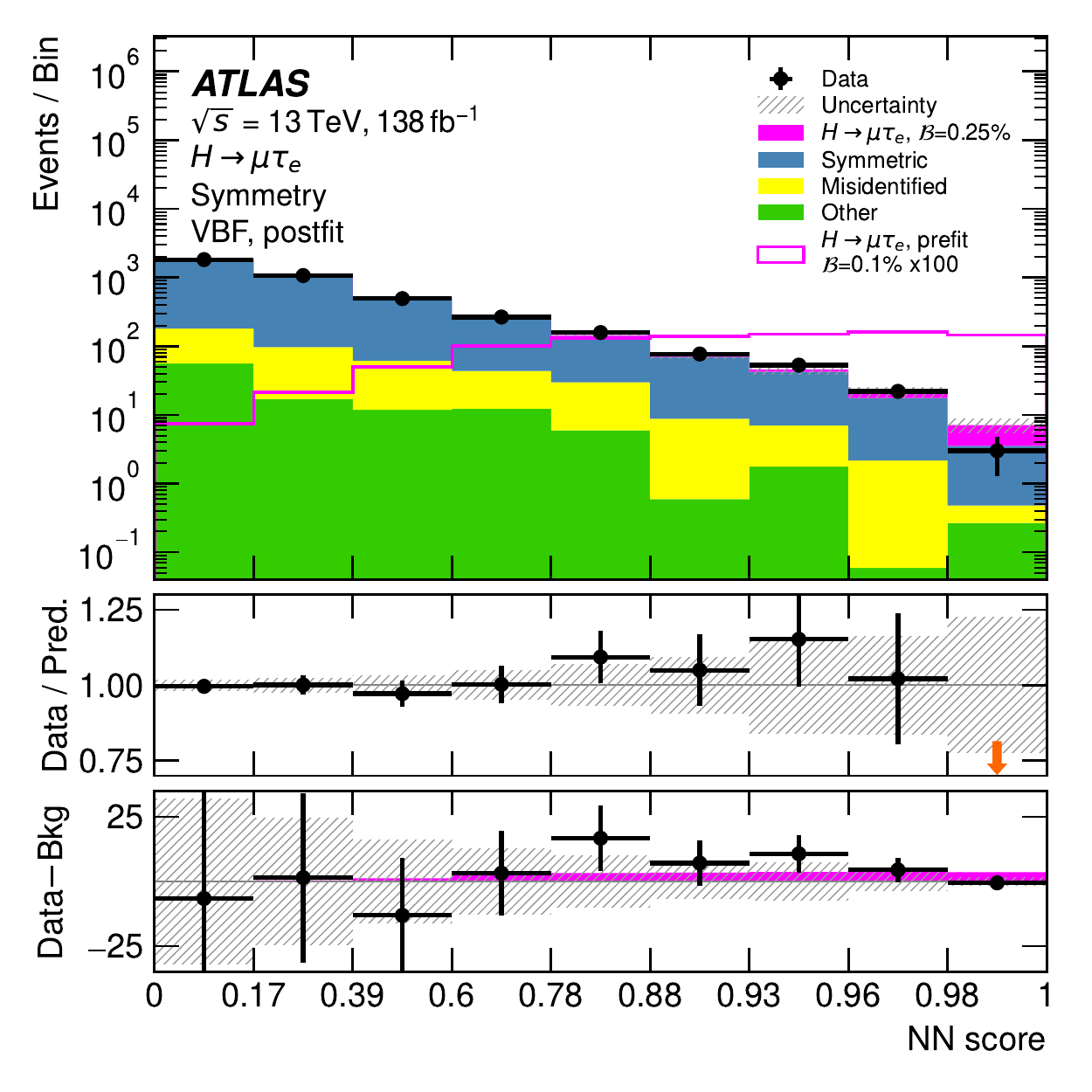}}
 
\caption{NN score distributions for the $e\tau_{\mu}$
(left) and the $\mu\tau_e$ (right) final states, in the \nonVBF{}
(top) and \VBF{} (bottom) categories of the \Symbased{} \leplep{} channel.
The middle panel shows the ratio of data to prediction (background+signal).
The bottom panel shows the residuals between data and the background after the fit.
The hashed band indicates the total postfit uncertainty on the total predicted yields.
The prediction for each sample is determined from the likelihood fit performed to
measure \hte{} (left) and \htm{} (right) signals,
described in Section~\ref{sec:1poi}.
The binning is shown as used in the likelihood fit
and points outside the displayed $y$-axis range are indicated by
arrows. Overlaid prefit signal shapes assume $\BR(H \to \ell \tau) =0.1\%$
and are enhanced by a factor 100 for visibility.
The postfit signal contributions are considered as part of the predictions.
As a result of the conversion of the fits of branching ratio differences to individual branching ratios,
the upward deviation in $\mu\tau_e$ appears as the downward deviation in $e\tau_{\mu}$.}
\label{fig:symm-background-postfit}
\end{figure}

\FloatBarrier


\section{Systematic uncertainties}
 
Systematic uncertainties affect the yields in the signal and control regions, as well as the shape of the MVA score distribution.
They can be separated into three groups: experimental uncertainties, theoretical uncertainties for the backgrounds, and theoretical uncertainties for the signal.
The systematic uncertainties related to the estimation of \fakebkg{} are discussed in
Section~\ref{sec:background} for the \MCbased{} \leplep{}
(Section~\ref{sec:background_leplep}), the \MCbased{} \lephad{}
(Section~\ref{sec:background_lephad}) and the \Symbased{} method
(Section~\ref{sec:lep_taulep_symm}).
 
\begin{table}[htb!p]
\caption{Summary of the different sources of uncertainty affecting the observed
$\BR(H \to \ell\tau)$ and their impact as computed
by the independent fits (1 POI) described in Section~\ref{sec:1poi} and
the simultaneous fit (2 POI) described in Section~\ref{sec:2poi}.
The values in the table are multiplied by a factor 100 to improve their readability.
Experimental uncertainties for reconstructed objects combine
efficiency and energy/momentum scale and resolution
uncertainties. \enquote{Background sample size} includes the
bin-by-bin statistical uncertainties in the simulated backgrounds as
well as statistical uncertainties in \fake{}
backgrounds, which are estimated using data.
}
\label{tab:impact_1POI}
\begin{center}
\begin{tabular}{l|c|c}
\toprule
1 POI             & \multicolumn{2}{c}{Impact on observed [$10^{-4}$]} \\
Source of uncertainty & $\hat{\BR}(\hte)$ & $\hat{\BR}(\htm)$  \\
\midrule
Flavour tagging                                   & 0.6 & 0.4 \\
Misidentified background (\lephad)                & 2.1 & 1.5 \\
Misidentified background (\leplep)                & 2.9 & 1.6 \\
Jet and \met                                      & 1.1 & 1.1 \\
Electrons and muons                               & 0.2 & 0.5 \\
Luminosity                                        & 0.6 & 0.5 \\
Hadronic $\tau$ decays                            & 0.9 & 1.0 \\
Theory (signal)                                   & 0.9 & 0.7 \\
Theory  (\Zjets{} processes)                      & 1.0 & 1.2 \\
Theory  (top-quark processes)                     & 0.3 & 0.3 \\
Theory  (diboson processes)                       & 0.4 & 0.7 \\
\Zll normalisation                                & 0.2 & 0.7 \\
Symmetric background estimate                     & 0.2 & 0.1 \\
Background sample size                            & 4.2 & 2.4 \\
\midrule
Total systematic uncertainty                      & 5.3 & 3.9 \\
Data sample size                                  & 2.9 & 2.7 \\
Total                                             & 6.1 & 4.7 \\
\bottomrule
\end{tabular}
\end{center}
\begin{center}
\begin{tabular}{l|c|c}
\toprule
2 POI             & \multicolumn{2}{c}{Impact on observed [$10^{-4}$]} \\
Source of uncertainty & $\hat{\BR}(\hte)$ & $\hat{\BR}(\htm)$  \\
\midrule
Flavour tagging                                  & 0.7 & 0.2 \\
Misidentified background ($e\tauhad$)            & 2.1 & 0.3 \\
Misidentified background ($e\tau_\mu$)           & 2.7 & 0.3 \\
Misidentified background ($\mu\tauhad$)          & 0.6 & 1.4 \\
Misidentified background ($\mu\tau_{e}$)         & 0.9 & 1.0 \\
Jet and \met                                     & 1.2 & 0.9 \\
Electrons and muons                              & 1.4 & 0.5 \\
Luminosity                                       & 0.6 & 0.4 \\
Hadronic $\tau$ decays                           & 0.9 & 0.9 \\
Theory (signal)                                  & 0.8 & 0.8 \\
Theory (\Zjets{} processes)                      & 0.8 & 1.0 \\
\Zll normalisation ($e\tau$)                     &<0.1\phantom{<} & <0.1\phantom{<} \\
\Zll normalisation ($\mu\tau$)                   & 0.2 & 0.9 \\
Background sample size                           & 3.7 & 2.3 \\
\midrule
Total systematic uncertainty                     & 5.1 & 3.6 \\
Data sample size                                 & 3.0 & 2.7 \\
Total                                            & 5.9 & 4.5 \\
\bottomrule
\end{tabular}
\end{center}
\end{table}
 
Experimental uncertainties include those originating from the trigger, reconstruction,
identification and isolation efficiencies and affecting final-state objects such as electrons~\cite{PERF-2017-03,PERF-2017-01}, muons~\cite{PERF-2015-10},
{\tauhadvis}~\cite{ATLAS-CONF-2017-029}, jets~\cite{PERF-2016-04,ATLAS-CONF-2014-018,PERF-2016-06,JETM-2018-05}
and $\MET$~\cite{PERF-2016-07}.
Their energy scale and resolution uncertainties are taken into account as well.
Experimental uncertainties affect the shape of the MVA score distribution, the background yields and the signal
cross-section through their effects on the acceptance and the migration between different analysis categories.
An additional uncertainty from the measurement of the luminosity~\cite{ATLAS-CONF-2019-021,LUCID2}, amounting to 1.7\%, is included.
 
In the \MCbased{} channels, theoretical uncertainties are considered for
the background processes estimated from simulation~\cite{HIGG-2019-09}.
Their effect on the normalisation and shape
of the MVA discriminant is considered in the statistical analysis.
For \Zjets{} events, systematic uncertainties include those due to renormalisation (\muR),
factorisation (\muF) and resummation scale (\qsf), the
jet-to-parton matching scheme (\ckkw)~\cite{CKKW}, and the choice of \alphas value and the PDFs.
For the top-quark background, uncertainties related to the choice of matrix element and parton shower generators~\cite{Bahr:2008pv,Bellm:2015jjp}, the initial- and final-state radiation model~\cite{ATL-PHYS-PUB-2017-007}, and the PDFs are considered~\cite{Harland-Lang:2014zoa}.
For the diboson production processes, an uncertainty of 6\% is assigned to the cross-section in the statistical analysis~\cite{Hoche:2010kg,Gleisberg:2005qq,Binoth:2009wk}.
 
With the \Symbased{} method, these background contributions are estimated from data, and it is not necessary to consider the aforementioned theoretical uncertainties.
 
The Higgs boson production cross-section uncertainties are obtained from Ref.~\cite{theo_sys_1}.
Effects on the signal expectations are treated
as uncorrelated between production modes.  Theoretical
uncertainties affecting the \ggF, VBF, $WH$, $ZH$ and \ttH production cross-sections are considered; more details can be found in Ref.~\cite{HIGG-2019-09}.
The uncertainties include components for those estimated by varying the PDF or \alphas value, or varying the choice of matrix element generator or parton shower and hadronisation model. For the matrix element variation, predictions by \POWHEGBOX[v2] are compared with those by \MGNLO~\cite{Alwall:2014hca}. The parton shower and hadronisation model variation replaces the nominal \PYTHIA[8] simulation with \HERWIG[7]~\cite{Bellm:2015jjp,Bahr:2008pv}.
For both the \MCbased{} and \Symbased{} methods, the Higgs boson production  cross-section uncertainties are considered for SM decays as well as for the \hte{} and \htm{} signals (\ggF, VBF, $WH$, $ZH$ production modes only).
 
Table~\ref{tab:impact_1POI} lists uncertainties for each measurement grouped by their respective sources.
In the independent measurements of the \hte{} and \htm{} signal (discussed in Section~\ref{sec:1poi}), labelled as 1~POI, the results based on the Symmetry method are included. Hence, statistical uncertainties in the background estimate which result from the symmetric component of the background estimate play an important role.
In the simultaneous measurement of the \hte{} and \htm{} signal (discussed in Section~\ref{sec:2poi}), labelled as 2~POI, the uncertainties are dominated by systematic sources, in particular the \fake{} background.


\section{Statistical analysis and results}
\label{sec:stat}
 
The statistical analysis uses a likelihood function
$\mathcal{L}(\mu,\theta)$, constructed as a product of Poisson
probability terms over all bins considered in the search. These
include the MVA score distributions of all the SRs and the
event yields in the  \leplep{} \MCbased{} method CRs, when included,
to constrain the normalisations of the major backgrounds estimated from simulation,
in particular the top-quark and \Ztt{}  background components.
The likelihood function depends on the parameters of interest (POIs),
$\mu_{\etau}$ and $\mu_{\mtau}$, defined as the branching ratios
$\BR(\hte)$ and $\BR(\htm)$, respectively, and a set of nuisance parameters $\theta$
that encode the effect of systematic uncertainties on the signal and
background expectations. All nuisance parameters are implemented in
the likelihood function as Gaussian or log-normal constraints.
The latter are used for normalisation factors to ensure that they
are always positive.
 
\begin{table}[bth]
\setlength{\tabcolsep}{10pt}
 
\caption{Combinations of channels, regions and categories used in the statistical analysis. In the independent searches, 1~POI (see Section~\ref{sec:1poi}), \hte{} and \htm{} signals are fit separately. When performing the simultaneous measurement, 2~POI (see Section~\ref{sec:2poi}), \hte{} and \htm{} signals are fit simultaneously.}
\label{tab:fit_setup}
\centering
\begin{tabular}[h]{llllcc}
\toprule
Method & Channel & Category & Region & 1 POI fit & 2 POI fit \\
\midrule
\multirow{6}{*}{\MCbased} & \multirow{6}{*}{\leplep{}}  & \multirow{3}{*}{\nonVBF{}} & SR       & \checkmark & \checkmark \\
&  &  & \Ztt{} CR  & \checkmark & \checkmark \\
&  &  & Top-quark CR   & \checkmark & \checkmark \\
\cmidrule{3-6}
& & \multirow{3}{*}{\VBF{}} & SR &  & \checkmark \\
&  &  & \Ztt{} CR     &  & \checkmark \\
&  & & Top-quark CR      &  & \checkmark \\
\midrule
\multirow{2}{*}{\MCbased} & \multirow{2}{*}{\lephad{}}  & \nonVBF{} & SR         & \checkmark & \checkmark \\
&  & \VBF{} & SR      & \checkmark & \checkmark \\
\midrule
\multirow{2}{*}{\Symbased} & \multirow{2}{*}{\leplep{}} & \nonVBF{} & SR      &  &  \\
&   & \VBF{} & SR          & \checkmark &  \\
\bottomrule
\end{tabular}
\end{table}
 
The likelihood function is fitted to the data
to test for the presence of a signal.
Estimates of the POIs are calculated with the
profile-likelihood-ratio test statistic $\tilde{q}_{\mu}$~\cite{Cowan:2010js},
and if no signal is found, the upper limits on the branching ratios are derived by using
$\tilde{q}_{\mu}$ and the CL$_{\mathrm{s}}$ method~\cite{Read:2002hq}.
 
Three different statistical analyses are performed, differing in the POI definitions and relying on different inputs from the two background estimation methods:
\begin{itemize}
\item An independent search for the \hte{} signal: a single POI, $\mu_{\etau}$, is estimated in the fit combining the \etm{} and \ethad{} final states
(discussed in \Sect{\ref{sec:1poi}}). \BR(\htm) is set to zero.
 
\item An independent search for the \htm{} signal: a single POI, $\mu_{\mtau}$, is estimated in the fit combining the \mte{} and \muthad{} final states
(discussed in \Sect{\ref{sec:1poi}}). \BR(\hte) is set to zero.
 
\item A simultaneous measurement of the \hte{} and \htm{} signals: two POIs ($\mu_{\etau}$ and $\mu_{\mtau}$) are estimated in the simultaneous fit of the \etm{}, \ethad{}, \mte{} and \muthad{} final states
(discussed in \Sect{\ref{sec:2poi}}).
\end{itemize}
The independent searches combine the \nonVBF{} and \VBF{} categories of the \lephad{} channel and the \nonVBF{} category of the \leplep{} channel from the \MCbased{} method with the \VBF{} category of the \leplep{} channel from the \Symbased{} method.
For the simultaneous measurements of \hte{} and \htm{} signals, the combined fit is performed with the regions from the \MCbased{} method for both \leplep{} and \lephad{} final states.
The SRs and CRs exploited in the different combined fits are specified in Table~\ref{tab:fit_setup}. In the fit combinations where a SR of the \leplep{} channel from the \MCbased{} method is considered, the corresponding top-quark and \Ztt{} CRs are also included.
 
Additionally, stand-alone fits of individual channels and categories
are performed employing either background estimation method.
Since the \Symbased{} method directly measures the branching ratio difference, the results of the \Symbased{} method can be converted into individual branching ratios with the assumption that the other signal is zero (employed in \Sect{\ref{sec:1poi}}). Alternatively, the results from the simultaneous fits with the \MCbased{} method can be utilised to calculate the branching ratio
difference to allow a direct comparison with the \Symbased{} method results.
The branching ratio difference determined from the stand-alone fits with either background estimation method for the \leplep{} final state, together with their compatibility, is discussed in Section~\ref{sec:comp-leplep}.
 
\subsection{Independent searches for \hte{} and \htm}
\label{sec:1poi}
 
A fit with a single POI is performed for each search separately by combining
the \leplep{} and \lephad{} final states.
The best-fit branching ratios and upper limits are evaluated assuming
$\BR(\hte) = 0$ for the \htm{} search and $\BR(\htm) = 0$ for the \hte{}
search.
Each search combines the \lephad{} channel and the \nonVBF{} category of the \leplep{}
channel from the \MCbased{} method with the \VBF{} category of the \leplep{} channel from the \Symbased{} method.
The choice of combining the \MCbased{} and \Symbased{} methods is based on the expected sensitivity~\cite{Cowan:2010js} of each individual category and selecting the method having higher significance.
For the \VBF{} category, the \Symbased{} method performs better than the \MCbased{} method by 13\% and 7\% for the \etau{} and \mtau{} final states, respectively.
For the \nonVBF{} category, the \MCbased{} method performs better than the \Symbased{} method by 28\% and 18\% for the \etau{} and \mtau{} final states, respectively.
For the \lephad{} channel, only the \MCbased{} method is available.
 
Three normalisation factors are used to normalise the \Ztt{} production cross-section in the \leplep{} and \lephad{} SRs estimated with the \MCbased{} method.
A common normalisation factor is used for the \nonVBF{} category SR of the \leplep{} channel and the  event yield of the
dedicated \Ztt{} CR.
In the \lephad{} channel, two independent normalisation factors are used for the \nonVBF{} and \VBF{} categories. They are constrained
by the data in the SRs.
In general, top-quark processes form a very small background contribution in
the \lephad{} channel and no top-quark CR is defined in the \lephad{} channel.
Therefore, in the \nonVBF{} category, where the \MCbased{} method is used for both final states,
a single normalisation factor is used to scale the top-quark processes' production cross-section jointly
in the \leplep{} and \lephad{} final state, making use of the \leplep{} top-quark CR.
Since the \Symbased{} method is exploited in
the \leplep{} channel in the \VBF{} category, it is not possible to constrain the normalisation of the
top-quark background in the \lephad{} final state by a CR and hence uncertainties
are applied to the top-quark cross-section.
 
\begin{figure}[tb]
\centering
\subfloat[]{\includegraphics[width=0.45\textwidth]{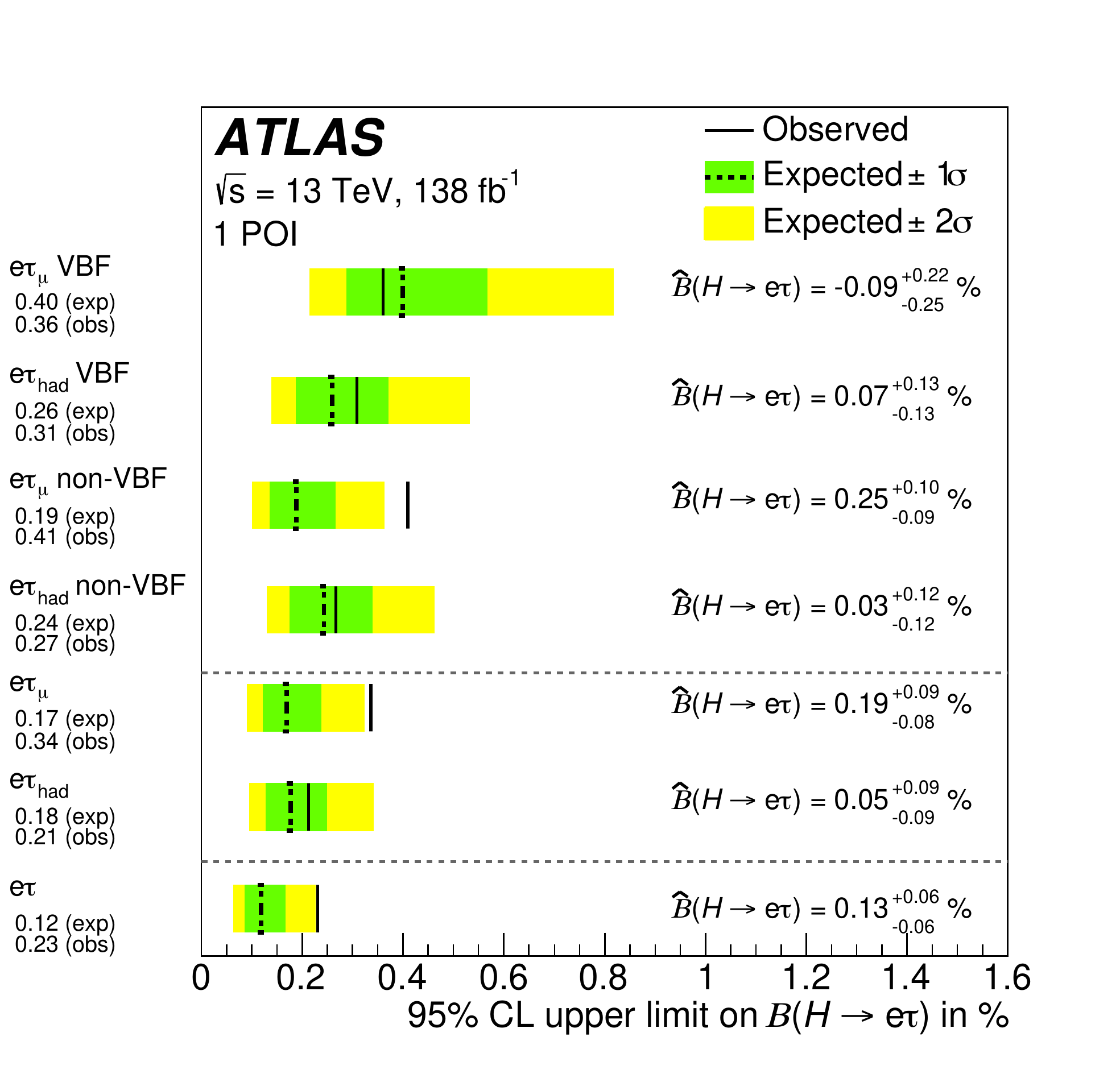}}
\subfloat[]{\includegraphics[width=0.45\textwidth]{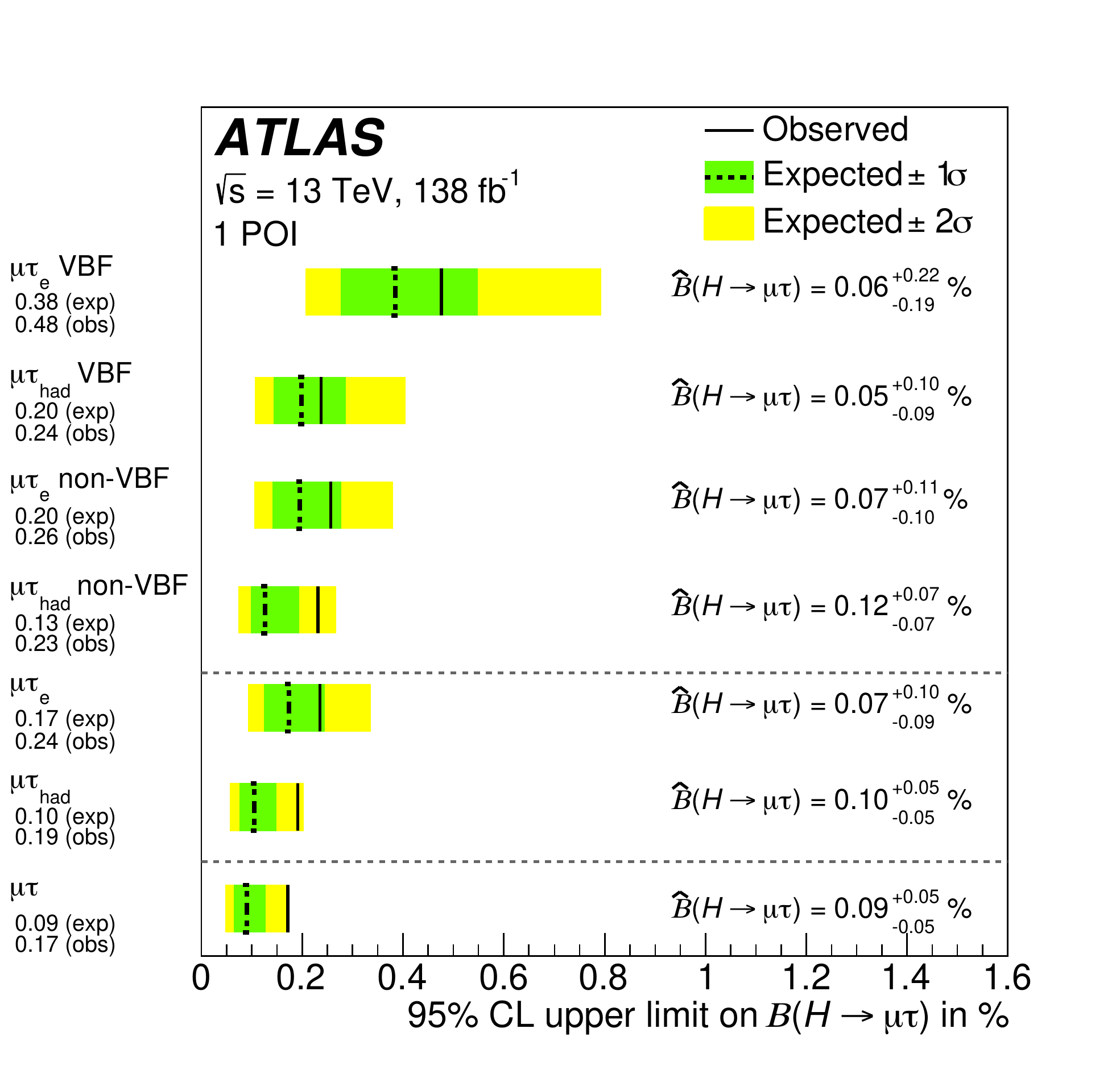}}\\
\caption{Fit results of the independent searches (1~POI) showing
95\% CL upper limits on the LFV branching ratios of the Higgs
boson, (a) $H\to e\tau$ and (b) $H\to \mu\tau$, indicated by
solid lines (observed results) or dashed lines (expected results).
Best-fit values of the branching ratios ($\hat{\BR}$) are also provided, in \%.
The limits are computed while assuming that either (a) $\BR(H\to \mu\tau)=0$ or (b) $\BR(H\to e\tau)=0$.
The channels and categories included in each likelihood fit
are shown on the $y$-axis, and the signal and control regions
from all other channels/categories are removed from the fit.
The results from stand-alone channels/categories fits shown at the top
are compared with the results of the combined fit displayed in the
last row.}
\label{fig:1POI_limits}
\end{figure}

Figure~\ref{fig:1POI_limits} shows the resulting  95\% CL upper
limits for the \hte{} and \htm{} searches, with the
breakdown of the stand-alone fits from each channel.  For the \htm{}
(\hte{}) signal, a $1.9\sigma$ ($2.2\sigma$) excess is observed.
In the case of the \hte{} signal, this is driven mainly by the \nonVBF{} category of
the \leplep{} final state based on the MC-template method. In the case of
the \htm{} signal, it is driven by the \nonVBF{} category of
the \lephad{} final state.
 
The branching ratio of the LFV Higgs boson decay is related to the
non-diagonal Yukawa coupling matrix elements~\cite{Harnik:2012pb} by
the formula
\begin{equation*}
|Y_{\ell\tau}|^2+|Y_{\tau\ell}|^2 = \frac{8\pi}{m_H}\,
\frac{\BR(H\to\ell\tau)}{1-\BR(H\to\ell\tau)}\,\Gamma_{H}(\mathrm{SM}),
\end{equation*}
where $\Gamma_{H}(\mathrm{SM}) =
4.07$~\MeV~\cite{LHCHiggsCrossSectionWorkingGroup:2013} is the
Higgs boson's width as predicted by the SM. The
observed 95\% CL upper limits on the branching ratio correspond to the following
limits on the coupling matrix elements: $\sqrt{|Y_{\tau e}|^2+|Y_{e
\tau}|^2} < 0.0014$ and $\sqrt{|Y_{\tau \mu}|^2+|Y_{\mu \tau}|^2}
< 0.0012$.
 
\begin{figure}[tb]
\centering
\subfloat[]{\includegraphics[width=0.45\textwidth]{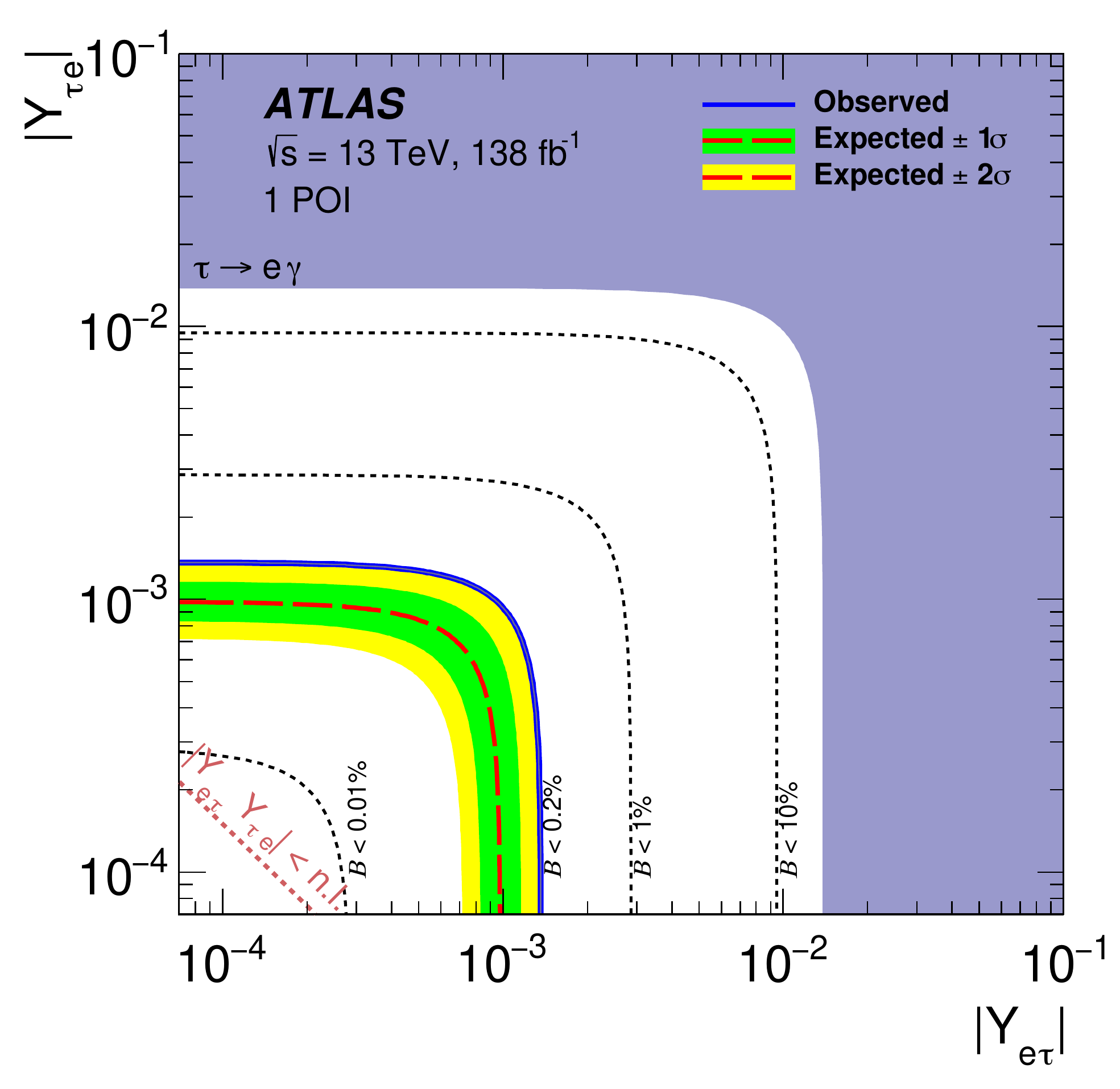}}
\subfloat[]{\includegraphics[width=0.45\textwidth]{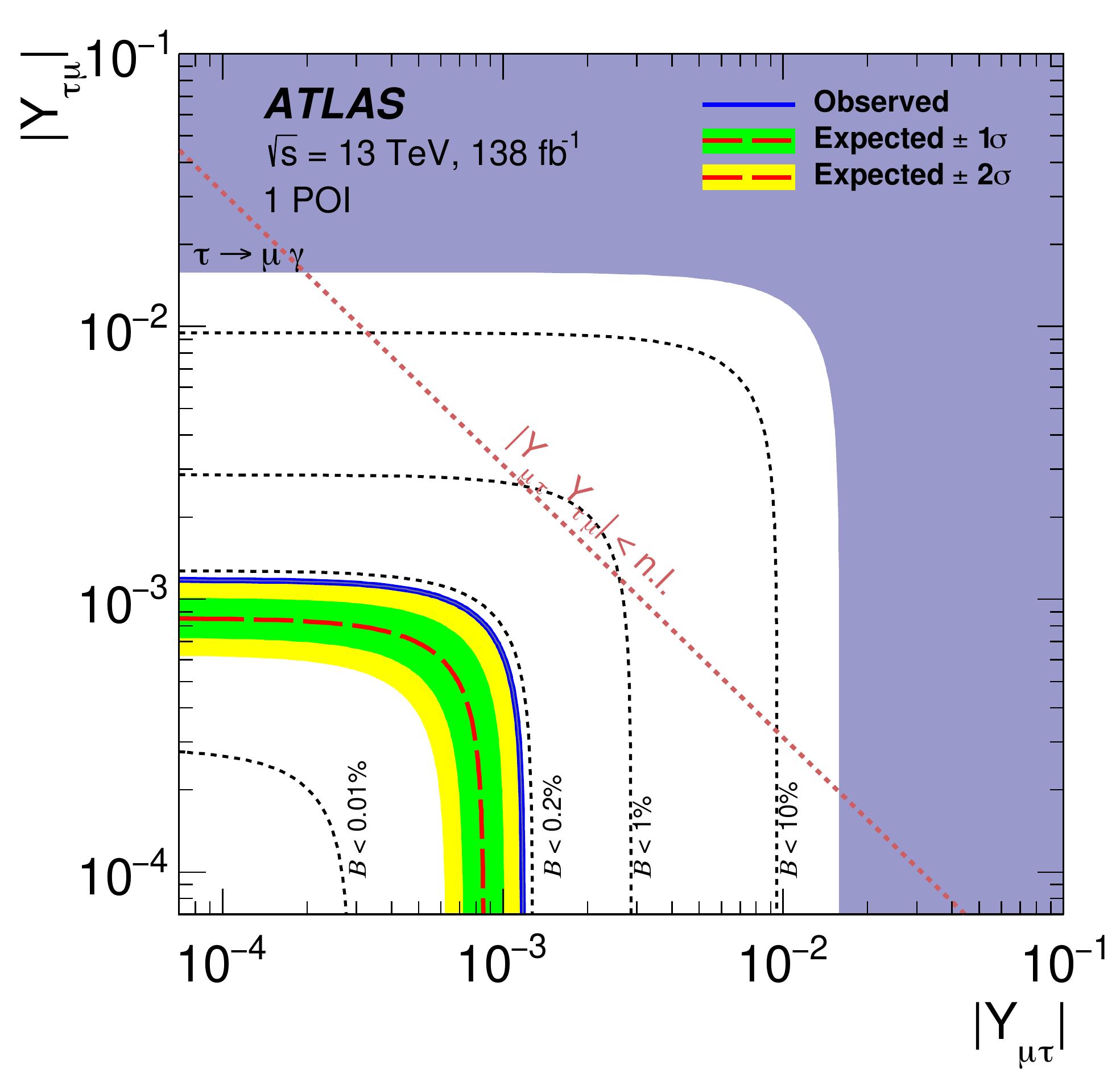}}\\
\caption{Expected (red long-dashed line) and observed (solid blue line)
95\% CL upper limits from the independent fits (1~POI) on the
absolute value of the couplings $Y_{\tau \ell}$ and $Y_{\ell \tau}$
together with the most stringent indirect limits from
$\tau \to \ell \gamma$ searches (dark purple region) for (a) $\ell=e$ or (b) $\ell=\mu$. The short-dashed lines represent the
limits corresponding to different branching ratios ($0.01\%$,
$0.2\%$, $1\%$ and $10\%$), while the dotted line indicates the
naturalness limit (denoted by n.l.). }
\label{fig:couplinglimits}
\end{figure}
 
Figure~\ref{fig:couplinglimits} shows the limits on the individual
coupling matrix elements $Y_{\tau \ell}$ and $Y_{\ell \tau}$ obtained
from the independent fits in the two searches.  The same figure also
shows the limits
from the $\tau\to\ell\gamma$
searches~\cite{Harnik:2012pb,Aubert:2009ag}.
Compared with the indirect limits from the $\tau\to\ell\gamma$
searches, the direct limits obtained  in this
search are  tighter by slightly less than one order of magnitude for $Y_{\tau e}$ and $Y_{e \tau}$, while for  $Y_{\tau \mu}$ and $Y_{\mu \tau}$ they are tighter by about a factor of ten. This clearly indicates the strength of direct searches at the LHC.
In the case of  \htm{}, the constraints are tighter than the naturalness limit
preventing a non-hierarchical mass spectrum from large off-diagonal terms in the Yukawa coupling matrix:
$| Y_{\tau \ell} Y_{\ell \tau} | \lesssim m_\tau m_\ell/v^2$,
where $v$ is the vacuum expectation value of the Higgs field~\cite{Harnik:2012pb};
in the case of \hte{}, the naturalness limit has not been reached yet.
This is in line with the previous ATLAS search~\cite{HIGG-2017-08} based on a partial dataset at $\sqrt{s}=13\,\TeV$, but with tighter limits.

\subsection{Simultaneous measurement of \hte{} and \htm{} signal}
\label{sec:2poi}
A simultaneous measurement of the \hte{} and \htm{} signals is performed using the \MCbased{} method,
where it is possible to remove the assumption about the absence of a \htm{} signal
in the fit of \BR(\hte) and vice versa.
The two POIs, corresponding to \BR(\hte) and \BR(\htm),
are estimated in the simultaneous fit of the \etm{}, \ethad{}, \mte{} and \muthad{} final states.
The analysis exploits eight SRs and eight CRs (one top-quark and one \Ztt{} CR for each \leplep{} SR).
The \Ztt{} and top-quark normalisation factors are correlated between the SRs and CRs, but not
between the \VBF{} and \nonVBF{} categories. The top-quark normalisation factors
are common to the \etm, \mte, \ethad{} and \muthad{} channels, while
\Ztt{} normalisation factors are decorrelated between \leplep{} and \lephad{} final states.
In total, two normalisation factors are used for the top-quark processes and four normalisation factors are used for the \Ztt{} processes.
 
\begin{figure}[tbp]
\centering
\subfloat[]{\includegraphics[width=0.45\textwidth]{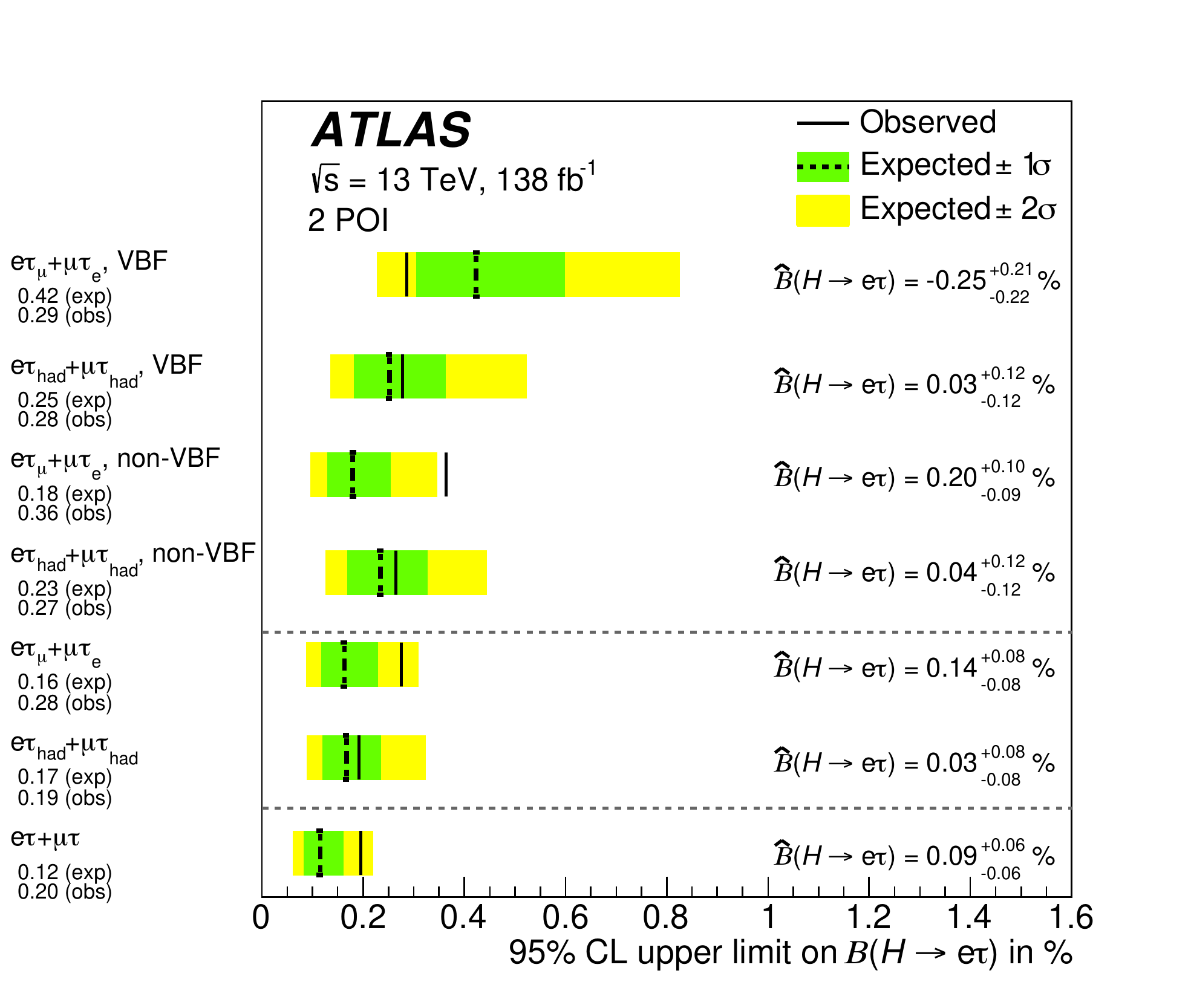}}
\subfloat[]{\includegraphics[width=0.45\textwidth]{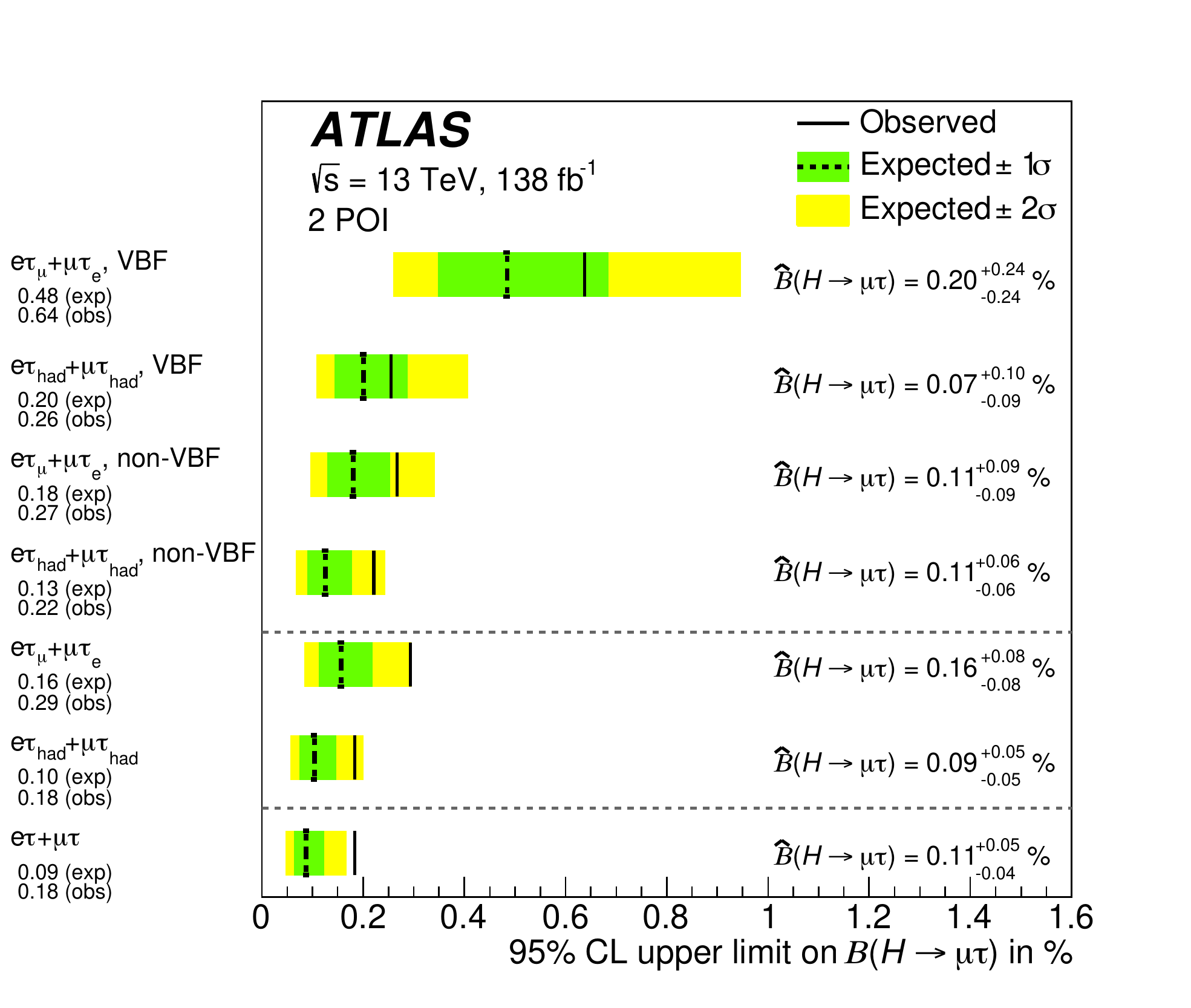}}\\
\caption{Fit results of the simultaneous measurement of the \hte{} and \htm{}
signals (2~POI) showing 95\% CL upper limits on the LFV
branching ratios of the Higgs boson, (a) $H\to e\tau$  and (b) $H\to \mu\tau$, indicated by
solid lines (observed results) or dashed lines (expected results).
Best-fit values of the branching ratios ($\hat{\BR}$) are also provided, in \%.
The channels and categories included in each likelihood fit
are shown on the $y$-axis, and the signal and control regions
from all other channels/categories are removed from the fit.
The results from stand-alone channels/categories fits shown at the top
are compared with the results of the combined fit displayed in the
last row.}
\label{fig:2POI_limits}
\end{figure}
 
\begin{figure}[p]
\centering
\includegraphics[width=0.70\textwidth]{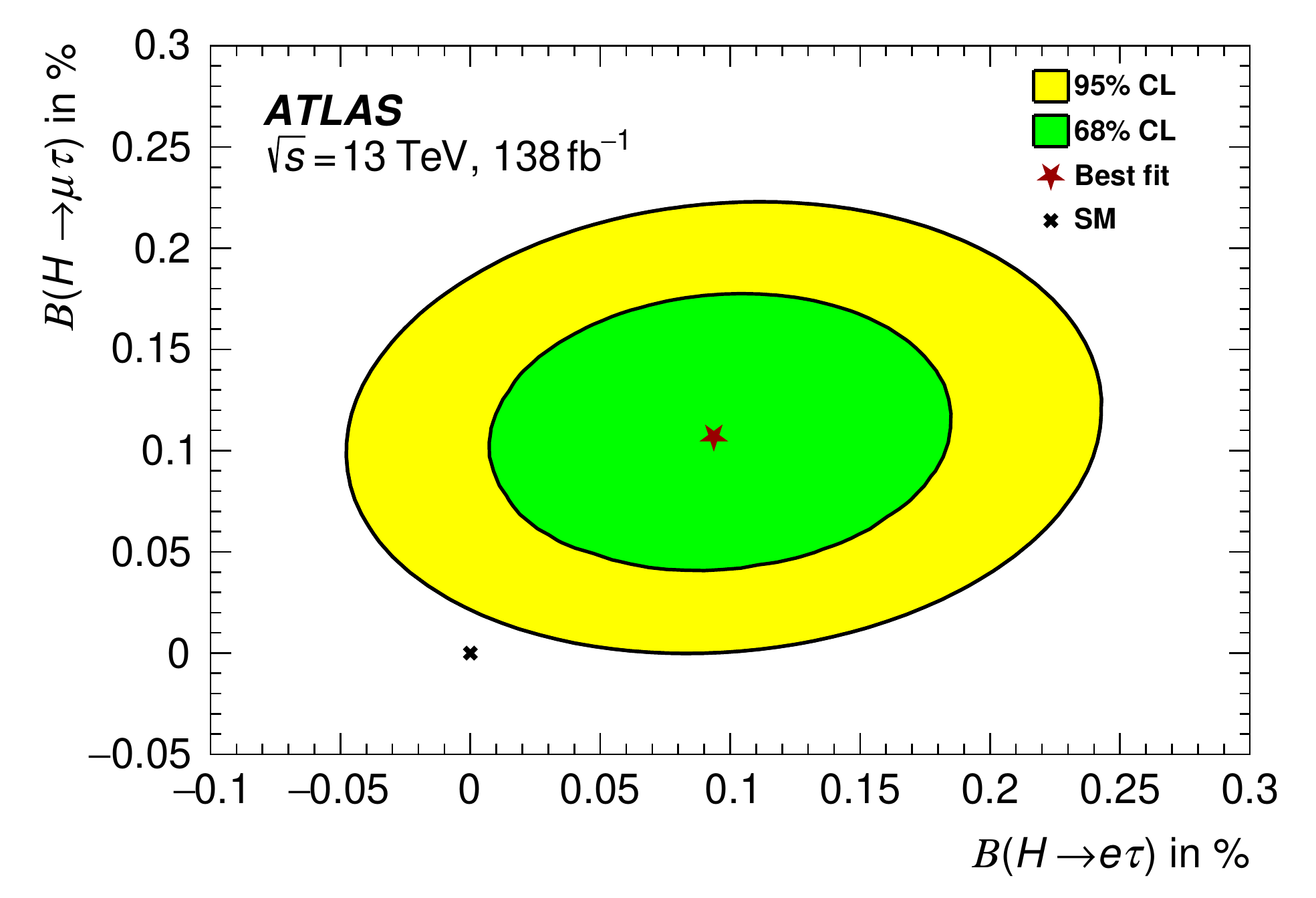}
\caption{Best-fit value (red star) of the branching ratios
$\mathcal{B}(H\to e\tau)$ and $\mathcal{B}(H\to \mu\tau)$, given in \%, and
likelihood contours at 68\% and 95\% CL obtained from the simultaneous
fit of \hte{} and \htm{} signals based on the \MCbased{} method,  compared with the SM expectation
(black cross).}
\label{fig:contour2POI}
\end{figure}
 
\begin{figure}[!htbp]
\centering
\subfloat[]{\includegraphics[width=0.45\textwidth]{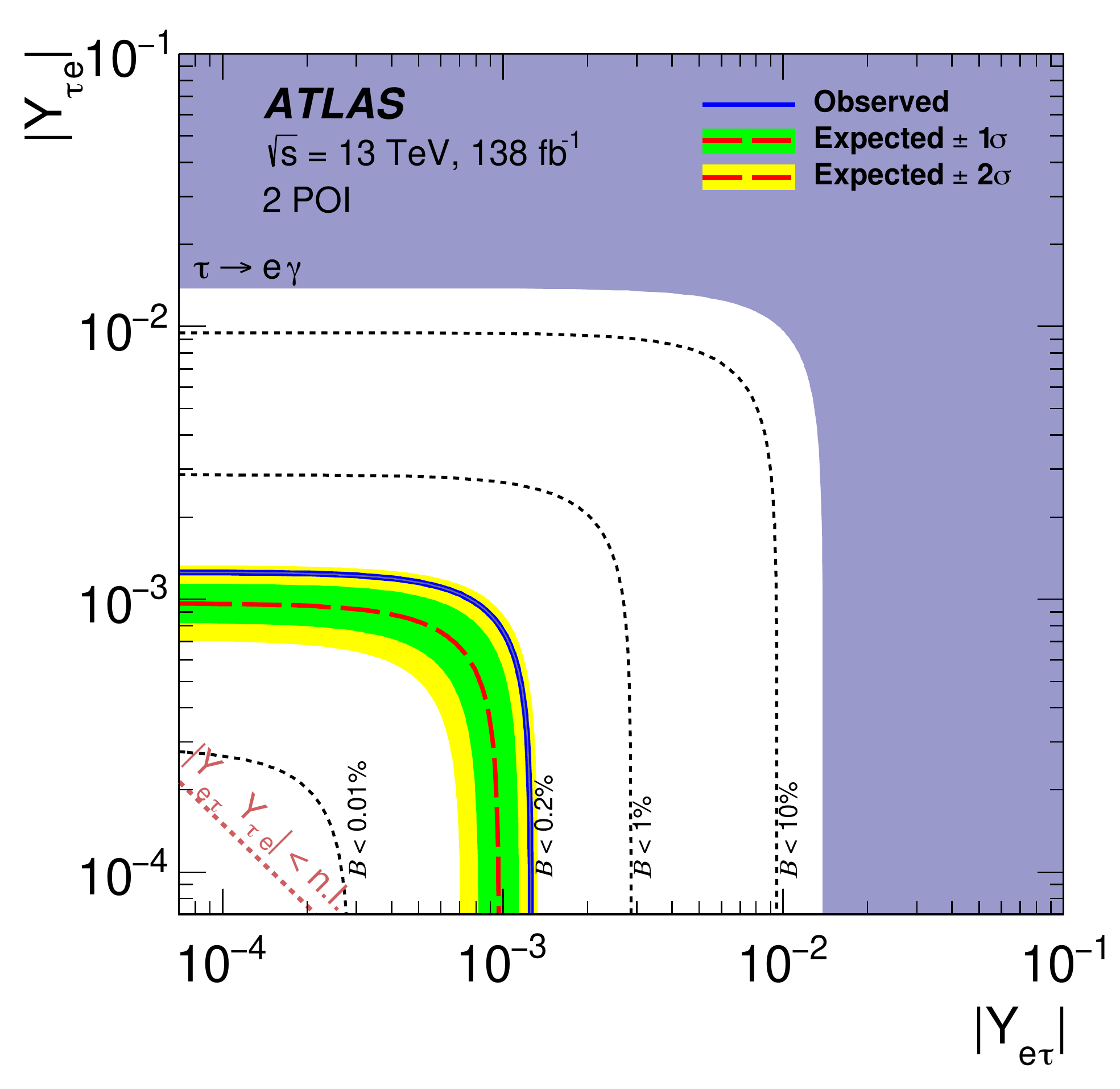}}
\subfloat[]{\includegraphics[width=0.45\textwidth]{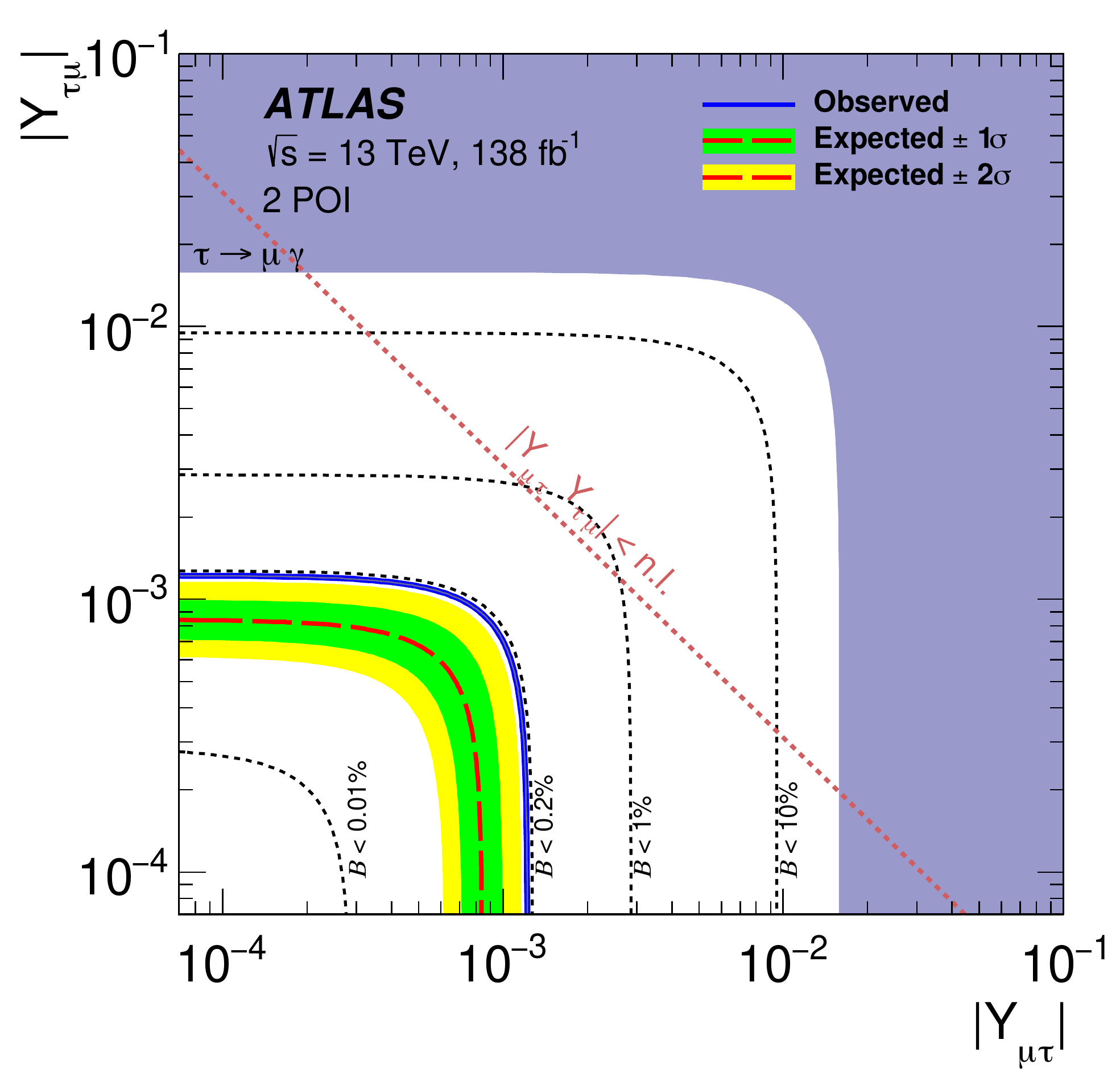}} \\
\caption{Expected (red long-dashed line) and observed (solid blue line)
95\% CL upper limits from the simultaneous fit (2~POI) of the
two searches on the absolute value of the couplings $Y_{\tau \ell}$
and $Y_{\ell \tau}$ together with the most stringent indirect limits
from $\tau \to \ell \gamma$ searches (dark purple region) for (a) $\ell=e$ or (b) $\ell=\mu$. The
short-dashed lines represent limits corresponding to
different branching ratios ($0.01\%$, $0.2\%$, $1\%$ and $10\%$), while the dotted line indicates the
naturalness limit (denoted by n.l.).}
\label{fig:couplinglimits_2POI}
\end{figure}
 
Figure~\ref{fig:2POI_limits} shows the resulting 95\% CL upper
limits for the \hte{} and \htm{} signals, together with the
contribution from each category. For
the \htm{} (\hte{}) signal, a $2.4\sigma$ ($1.6\sigma$) excess is observed.
In the case of the \hte{} signal, this is driven mainly by the \nonVBF{} category of
the \leplep{} final state based on the MC-template method. In the case of
the \htm{} signal, it is driven by the \nonVBF{} category of
the \lephad{} final state.
The result of the simultaneous fit of the \hte{} and \htm{}
signals, along with the 68\% and 95\% CL contours, is
shown in Figure~\ref{fig:contour2POI}.
The result is found to be compatible with the SM prediction within $2.1\sigma$.
 
Figure~\ref{fig:couplinglimits_2POI} shows the limits on the
individual coupling matrix elements $Y_{\tau \ell}$ and $Y_{\ell
\tau}$ obtained from the simultaneous fit of the two signals.
The observed 95\% CL upper limits on the branching ratios correspond to the following
limits on the coupling matrix elements: $\sqrt{|Y_{\tau e}|^2+|Y_{e \tau}|^2} < 0.0013$, and $\sqrt{|Y_{\tau \mu}|^2+|Y_{\mu \tau}|^2}
< 0.0012$.
Both the indirect limits from  $\tau\to\ell\gamma$
searches~\cite{Harnik:2012pb,Aubert:2009ag} and the naturalness limit
$| Y_{\tau \ell} Y_{\ell \tau} | \lesssim  m_\tau m_\ell/v^2$~\cite{Harnik:2012pb} are shown as well.

\subsection{Measurement of the branching ratio difference in the \leplep{} final state}
\label{sec:comp-leplep}
 
The \Symbased{} method uses the \etm{} dataset to estimate the
background in the \htm{} search, and the \mte{} dataset to estimate the
background in the \hte{} search. Therefore, this method measures the
difference between the two branching ratios, $\BR(\hte)$ and $\BR({\htm})$.
In the search for \htm{}, the difference $\BR(\htm)-\BR({\hte})$ is measured and in
the search for \hte{}, the difference $\BR(\hte)-\BR({\htm})$ is measured.
The two measurements of the branching ratio difference with the \Symbased{} method are found to be compatible, taking into account the anti-correlation between the two measurements.
A more detailed description can be found in Appendix~\ref{app:compatability}.
The $\BR(\htm)-\BR({\hte})$ measurement has the smaller expected uncertainty and is chosen as main result for the branching ratio difference from the \Symbased{} method.
For the combination of the \nonVBF{} and \VBF{} categories, $\BR({\htm})-\BR(\hte)$ measured by the \Symbased{} method is $(0.25 \pm 0.10)\%$, compatible with zero within $2.5\sigma$.
 
The branching ratio difference for the \MCbased{} method is determined from the simultaneous fit of $\BR(\hte)$ and $\BR({\htm})$ following the approach discussed in Section~\ref{sec:2poi}, but based only on data in the \leplep{} final state. The obtained difference $\BR({\htm})-\BR(\hte)$ takes into account the correlation between the two branching ratio parameters from the fit. The difference $\BR({\htm})-\BR(\hte)$ is measured to be $(0.02 \pm 0.12)\%$ for the combination of the \nonVBF{} and \VBF{} categories, compatible with zero well within $1\sigma$.
 
\begin{figure}[htb!]
\centering
\includegraphics[width=0.5\textwidth]{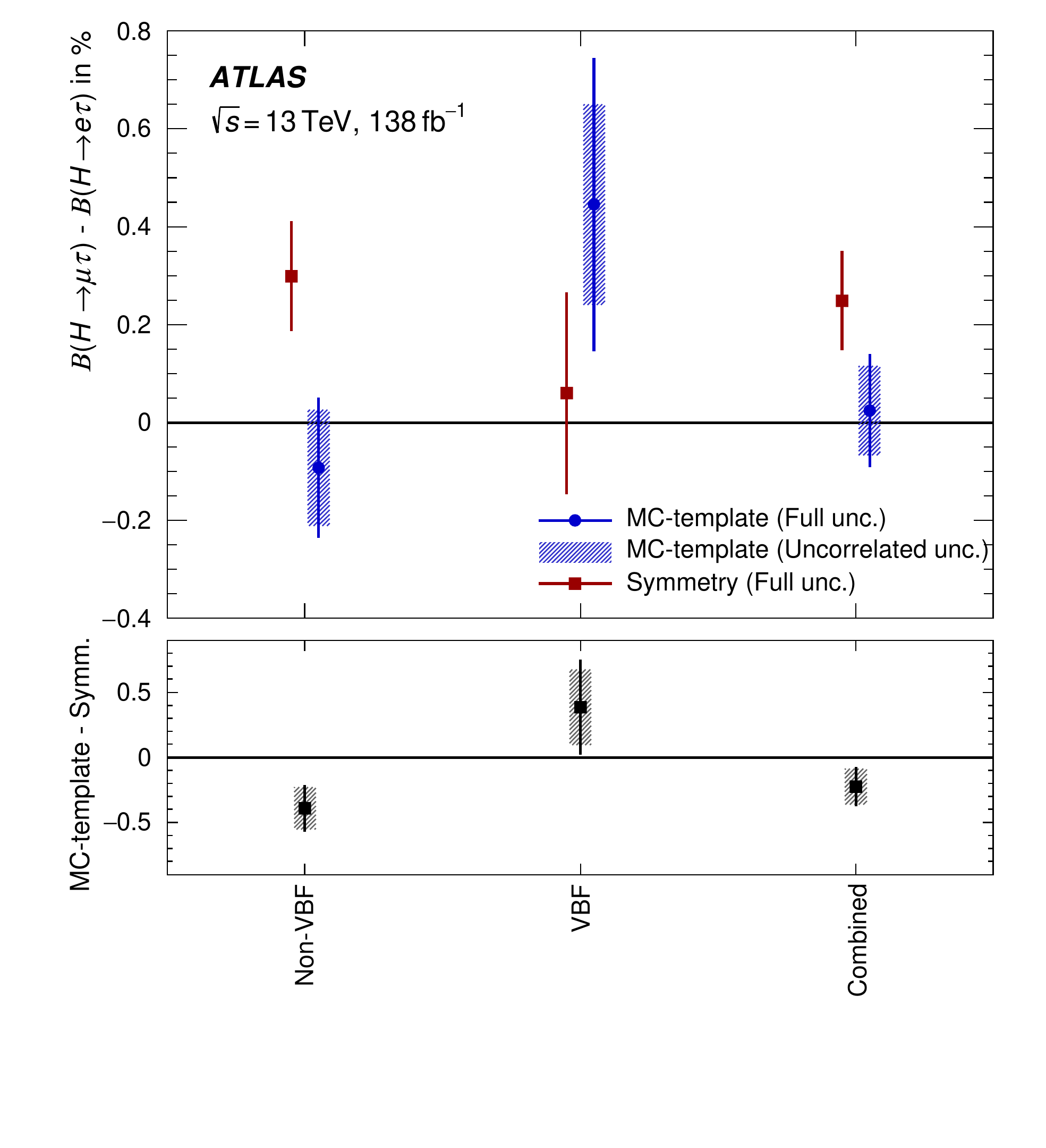}
\caption{Best-fit $\BR(\htm)-\BR({\hte})$ values, given in \%,
obtained from the stand-alone fits of the \leplep{} final state
with either background estimation method.  The
uncertainty of the difference displayed in the lower panel is
calculated taking into account either only the sources of uncertainty that
are uncorrelated between the two analyses (grey bars) or all uncertainties (black lines).} \label{fig:compatibility}
\end{figure}
 
The results of the branching ratio difference measurement are shown in Figure~\ref{fig:compatibility}, displaying the results for \nonVBF{} and \VBF{} categories separately as well as combined. The \Symbased{} method favours a larger branching ratio for the \htm{} signal than for the \hte{} signal, with a significance of $2.5\sigma$ for the combination of the \nonVBF{} and \VBF{} categories, driven mainly by the \nonVBF{} category.
The \MCbased{} method observes a branching ratio difference compatible with zero for the combination of the \nonVBF{} and \VBF{} categories, driven also by the \nonVBF{} category.
 
The measurement of the branching ratio difference with two independent background estimation methods in the \leplep{} final state allows the results of the two methods to be checked for compatibility.
The compatibility of the two background estimation methods is tested, taking into account the correlations between
the \Symbased{} and \MCbased{} methods.
Due to the large overlap between the data that are used by the two methods,
the data statistical uncertainties are correlated.
Likewise, the two methods use the same simulated samples for signal,
so the signal uncertainties are correlated.
All other uncertainties are considered to be uncorrelated.
For the \MCbased{} method, the correlated uncertainties are fixed to their best-fit
values and the uncertainties are re-evaluated considering only the uncorrelated sources.
For the \Symbased{} method, the full uncertainty is considered.
The results obtained with the \MCbased{} and \Symbased{} methods are shown in Figure~\ref{fig:compatibility} and are found to be compatible within 2$\sigma$.

\FloatBarrier


\section{Conclusion}
Two direct searches for lepton-flavour-violating Higgs boson decays, \hte{} and \htm, are presented.
They are based on $\sqrt{s}=13$~\TeV{}  proton--proton collision data collected by the ATLAS experiment at
the LHC, corresponding to an integrated luminosity of \ilumi.
Two complementary background estimation techniques are exploited.
The first relies on simulation and is used in both the \leplep{} and \lephad{} channels.
The second one relies on the \Symbased{} method and is applied in the \leplep{} channel.
 
A simultaneous fit of possible \hte{} and \htm{} signals was performed
combining the \leplep{} and \lephad{} channels, relying on inputs from
the \MCbased{} method only. In line with the independent fits,
small excesses with respect to the SM background are observed, but below the threshold for evidence of a new signal.
The observed (expected) 95\% CL upper limits on the branching
ratios of $H\to e\tau$ and $H\to\mu\tau$ are $0.20\%$ ($0.12\%$) and $0.18\%$ ($0.09\%$), respectively.
The result
of the simultaneous fit of \hte{} and \htm{} signals
is found to be compatible with a branching ratio value of zero for both processes within $2.1\sigma$.
 
The measurement of the branching ratio difference with the \Symbased{} method, based only on data in the \leplep{} final state, favours a larger branching ratio for the \htm{} signal than for the \hte{} signal, though the difference is not statistically significant.
The best-fit value for the difference $\BR({\htm})-\BR(\hte)$ obtained with the \Symbased{} method is $(0.25 \pm 0.10)\%$, which is compatible with a value of zero within $2.5\sigma$. It is also compatible within 2$\sigma$ with the best-fit value of $(0.02 \pm 0.12)\%$, obtained with the \MCbased{} method in the \leplep{} final state.
 
The observed limits based on the full dataset recorded by ATLAS at $\sqrt{s}=13$~\TeV\ are more stringent by factors of up to 2.5 (1.6)
than the corresponding previous limits for the $H\to e\tau$ ($H\to\mu\tau$) decay based on a partial dataset at $\sqrt{s}=13$~\TeV,
while the expected sensitivity  for the  $H\to e\tau$ ($H\to\mu\tau$) signal improves by a factor of about 3.1 (4.1).
In addition to the approximately four times larger dataset, the main sources of improvement are significantly more sophisticated features of the analysis method. These include lepton assignment in
the approximate Higgs boson rest frame for the $\leplep$ channel, use of
the symmetry method for the $\leplep$ $\VBF$ category in the fit with one parameter of interest, introduction of a simultaneous fit of
$H\to e\tau$  and $H\to\mu\tau$ signals (two parameters of interest),
more advanced multivariate classifiers for signal extraction, and improved object reconstruction, in particular the $\tau$
identification algorithm.


\section*{Acknowledgements}


We thank CERN for the very successful operation of the LHC, as well as the
support staff from our institutions without whom ATLAS could not be
operated efficiently.
 
We acknowledge the support of
ANPCyT, Argentina;
YerPhI, Armenia;
ARC, Australia;
BMWFW and FWF, Austria;
ANAS, Azerbaijan;
CNPq and FAPESP, Brazil;
NSERC, NRC and CFI, Canada;
CERN;
ANID, Chile;
CAS, MOST and NSFC, China;
Minciencias, Colombia;
MEYS CR, Czech Republic;
DNRF and DNSRC, Denmark;
IN2P3-CNRS and CEA-DRF/IRFU, France;
SRNSFG, Georgia;
BMBF, HGF and MPG, Germany;
GSRI, Greece;
RGC and Hong Kong SAR, China;
ISF and Benoziyo Center, Israel;
INFN, Italy;
MEXT and JSPS, Japan;
CNRST, Morocco;
NWO, Netherlands;
RCN, Norway;
MEiN, Poland;
FCT, Portugal;
MNE/IFA, Romania;
MESTD, Serbia;
MSSR, Slovakia;
ARRS and MIZ\v{S}, Slovenia;
DSI/NRF, South Africa;
MICINN, Spain;
SRC and Wallenberg Foundation, Sweden;
SERI, SNSF and Cantons of Bern and Geneva, Switzerland;
MOST, Taiwan;
TENMAK, T\"urkiye;
STFC, United Kingdom;
DOE and NSF, United States of America.
In addition, individual groups and members have received support from
BCKDF, CANARIE, Compute Canada and CRC, Canada;
PRIMUS 21/SCI/017 and UNCE SCI/013, Czech Republic;
COST, ERC, ERDF, Horizon 2020 and Marie Sk{\l}odowska-Curie Actions, European Union;
Investissements d'Avenir Labex, Investissements d'Avenir Idex and ANR, France;
DFG and AvH Foundation, Germany;
Herakleitos, Thales and Aristeia programmes co-financed by EU-ESF and the Greek NSRF, Greece;
BSF-NSF and MINERVA, Israel;
Norwegian Financial Mechanism 2014-2021, Norway;
NCN and NAWA, Poland;
La Caixa Banking Foundation, CERCA Programme Generalitat de Catalunya and PROMETEO and GenT Programmes Generalitat Valenciana, Spain;
G\"{o}ran Gustafssons Stiftelse, Sweden;
The Royal Society and Leverhulme Trust, United Kingdom.
 
The crucial computing support from all WLCG partners is acknowledged gratefully, in particular from CERN, the ATLAS Tier-1 facilities at TRIUMF (Canada), NDGF (Denmark, Norway, Sweden), CC-IN2P3 (France), KIT/GridKA (Germany), INFN-CNAF (Italy), NL-T1 (Netherlands), PIC (Spain), ASGC (Taiwan), RAL (UK) and BNL (USA), the Tier-2 facilities worldwide and large non-WLCG resource providers. Major contributors of computing resources are listed in Ref.~\cite{ATL-SOFT-PUB-2021-003}.


\clearpage
\part*{Appendix}
\addcontentsline{toc}{part}{Appendix}
\appendix

\section{Control and validation regions summary}
 
Table~\ref{tab:control_regions} summarizes the control and validation
regions (CR and VR respectively) used in the analysis to characterize the various background
sources.
 
\begin{table}[htbp]
\caption{Control and validation regions selection in the {\leplep} and {\lephad}
channels. The detailed description of the \fakebkg{} CR can be found in
Sections~\ref{sec:background_leplep} and~\ref{sec:background_lephad}.}\centering 
\begin{tabular}{lcc}
\toprule
Selection & \leplep & \lephad \\
\midrule
\multirow{2}{*}{\fakebkg{} CR} & \multicolumn{2}{c}{\nonVBF{} (or \VBF) category with } \\
& \multicolumn{2}{c}{statistically independent lepton ($\ell$ or $\thadvis$) selection, see text} \\
\midrule
\multirow{4}{*}{\Zmumu{} CR/VR (\leplep/\lephad)}   & \Baseline{} with $\SI{35}{\GeV}<\pt^{\ell_1} < \SI{45}{\GeV}$  & \Baseline{} \\
& $\SI{75}{\GeV} < m_{\ell_1\ell_2} < \SI{100}{\GeV}$                & $|\eta(\tau)|<0.1$ \\
& $|\Delta \phi(\ell_2, \met)|<1.5$                                & $\SI{90}{\GeV} <\mcol(\mu,\tau) < \SI{110}{\GeV}$ \\
& $1.25<\pt^{\mathrm{track}}(\ell_2)/\pt^{\mathrm{cluster}}(\ell_2)<3$  & \\
\midrule
\multirow{2}{*}{top-quark CR} & \nonVBF{} (or \VBF) selection  & \multirow{2}{*}{$-$} \\
& with inverted $b$-veto requirement & \\
\midrule
\multirow{2}{*}{\Ztt{} CR}    &  \nonVBF{} (or \VBF) selection                      & \multirow{2}{*}{$-$}  \\
&  with $\SI{35}{\GeV} < \pt^{\ell_1} < \SI{45}{\GeV}$ &  \\
\midrule
\multirow{5}{*}{Diboson VR}   & \Baseline{}                                        & \multirow{5}{*}{$-$}  \\
& $\pt^{\ell_2} > \SI{30}{\GeV}$                      &  \\
& $\SI{100}{\GeV} < m_{\ell_1\ell_2} < \SI{150}{\GeV}$ & \\
& $m_{\mathrm{T}} > \SI{30}{\GeV}$                     & \\
& veto events with jets with $\pT >30$~{\GeV}        & \\
\bottomrule
\end{tabular}
\label{tab:control_regions}
\end{table}


\section{Background and signal yields}
 
\subsection{\MCbased{}{ \leplep} channel}
\label{app:yieldsMCleplep}
 
The signal and background event yields in all SRs and CRs obtained after
performing the simultaneous fit of the \hte{} and \htm{} signals (Section~\ref{sec:2poi}) using only data from the \leplep{}
final state are detailed in Tables~\ref{tab:MC_yields_leplep_etau_2poi}
and~\ref{tab:MC_yields_leplep_mutau_2poi}.
 
\begin{table}[htbp]
\small
\begin{center}
\caption{Observed event yields and predictions as computed by the
fit in the signal region (SR) and control regions (CRs) of
the \etm{} channel.  The prediction for each sample is
determined from the likelihood fit performed to measure the
\hte{} and \htm{} signals using the \leplep{} data only
(Section~\ref{sec:2poi}).  The signal event yields are given for
$\BR(\hte)=0.13\%$ and $\BR(\htm)=0.17\%$, which correspond to
the best-fit values obtained from the fit mentioned above.
Uncertainties include statistical and systematic
components. Uncertainties in the total background prediction
include the effect of correlations between individual
uncertainty sources as determined by the fit.  }
\label{tab:MC_yields_leplep_etau_2poi}
\resizebox{\columnwidth}{!}{
\begin{tabular}{lr|r @{${}\pm{}$} l|r @{${}\pm{}$} l|r @{${}\pm{}$} l|r @{${}\pm{}$} l|r @{${}\pm{}$} l|r @{${}\pm{}$} l}
\toprule
& Channel & \multicolumn{12}{c}{\etm} \\
& Region & \multicolumn{4}{c}{SR} & \multicolumn{4}{c}{\Ztt{} CR} & \multicolumn{4}{c}{Top-quark CR}\\
& Category& \multicolumn{2}{c}{\nonVBF{}} & \multicolumn{2}{c}{\VBF{}} & \multicolumn{2}{c}{\nonVBF{}} & \multicolumn{2}{c}{\VBF{}} & \multicolumn{2}{c}{\nonVBF{}} & \multicolumn{2}{c}{\VBF{}} \\
\midrule
\midrule
\multirow{3}{*}{Signal, \hte} & \ggF             & 270  & 160  & 4.3  & 2.6 & 37   & 22  & 0.55   & 0.33  & 14    & 8    & 0.57  & 0.34  \\
& VBF              & 15   & 9    & 8    & 5   & 2.0  & 1.2 & 0.8    & 0.5   & 1.6   & 1.0  & 0.8   & 0.5  \\
& $VH$             & 12   & 7    & 0.22 &0.13 & 1.4  & 0.9 & 0.027  & 0.023 & 2.9   & 1.7  & 0.053 & 0.033 \\
\midrule
\multirow{3}{*}{Signal, \htm} & \ggF             & 19   & 9    & 0.57 & 0.30& 3.6  & 1.7 & 0.045  & 0.023 & 1.4  & 0.7   & 0.07  & 0.05  \\
& VBF              & 1.7  & 0.8  & 0.7  & 0.4 & 0.18 & 0.09 & 0.045 & 0.024 & 0.17 & 0.09  & 0.09  & 0.05 \\
& $VH$             & 2.4  & 1.2  & 0.06 & 0.04& 0.32 & 0.16 & 0.004 & 0.002 & 0.51 & 0.26  & 0.011 & 0.013  \\
\midrule
\multirow{9}{*}{Background}   & \Ztt             & 23000 & 400 & 450  & 80  & 57500 & 900  & 250  & 40    & 2620 & 180   & 133   & 15  \\
& \Zll             & 510  & 150  & 9    &  4  & 340   & 120  & 5    &  5    & 53   & 19    & 4.3   & 2.5 \\
& Top-quark        & 34600& 1300 & 1850 &  70 & 6270  & 220  & 319  & 22    &298800& 1700  & 15470 & 120  \\
& SM \ggF          & 1200 & 100   & 37  & 7   & 1090  & 60   & 12.1 & 2.3   & 92   & 13    & 5.6   & 2.1  \\
& SM VBF           & 109  & 4    & 74.9 & 3.2 & 46.6  & 2.4  & 19.5 & 1.1   & 13.1 & 0.8   & 8.0   & 0.6  \\
& SM $VH$                & 34   & 5    & 0.6  & 0.4 & 12.0  & 1.9  & 0.19 & 0.12  & 7.6  & 1.2   & 0.19  & 0.07  \\
& SM $ttH$               & 8.5  & 1.4  & 0.45 & 0.08& 1.28  & 0.21 & 0.080& 0.018 & 214  &   34  & 8.8   & 1.5  \\
& $VV$             & 31800& 1500 & 640  &   70& 11800 &  600 & 100  &    12 & 2020 &  180  & 106   & 12  \\
& Misidentified    & 9200 & 900  & 370  &   50&  6000 & 1000 &  98  &    25 & 9200 & 1600  & 590   & 100  \\
\midrule
\midrule
Total Background              &              & 100800& 400     & 3450& 60   & 82700 & 400  & 810  & 31    &313100& 600  &16340& 120  \\
Data                         &               & \multicolumn{2}{c|}{100769} & \multicolumn{2}{c|}{3383} & \multicolumn{2}{c|}{82902} & \multicolumn{2}{c|}{878} & \multicolumn{2}{c|}{312902} & \multicolumn{2}{c}{16382} \\
\bottomrule
\end{tabular}
}
\end{center}
\end{table}

\begin{table}[htb!]
\begin{center}
\caption{Observed event yields and predictions as computed by the
fit in the signal region (SR) and control regions (CRs) of
the \mte{} channel.  The prediction for each sample is
determined from the likelihood fit performed to measure the
\hte{} and \htm{} signals using the \leplep{} data only
(Section~\ref{sec:2poi}).  The signal event yields are given for
$\BR(\hte)=0.13\%$ and $\BR(\htm)=0.17\%$, which correspond to
the best-fit values obtained from the fit mentioned above.
Uncertainties include statistical and systematic
components. Uncertainties in the total background prediction
include the effect of correlations between individual
uncertainty sources as determined by the fit.
} \label{tab:MC_yields_leplep_mutau_2poi}
\resizebox{\columnwidth}{!}{
\begin{tabular}{lr|r @{${}\pm{}$} l|r @{${}\pm{}$} l|r @{${}\pm{}$} l|r @{${}\pm{}$} l|r @{${}\pm{}$} l|r @{${}\pm{}$} l}
\toprule
& Channel & \multicolumn{12}{c}{\mte} \\
& Region & \multicolumn{4}{c}{SR} & \multicolumn{4}{c}{\Ztt{} CR} & \multicolumn{4}{c}{Top-quark CR}\\
& Category & \multicolumn{2}{c}{\nonVBF{}} & \multicolumn{2}{c}{\VBF{}} & \multicolumn{2}{c}{\nonVBF{}} & \multicolumn{2}{c}{\VBF{}} & \multicolumn{2}{c}{\nonVBF{}} & \multicolumn{2}{c}{\VBF{}} \\
\midrule
\midrule
\multirow{3}{*}{Signal, \hte}& \ggF             &  15   & 9   & 0.44  & 0.27 & 2.7   & 1.6  & 0.041 & 0.026 & 1.1  & 0.6  & 0.08  & 0.06 \\
& VBF              &  1.2  & 0.8 & 0.6   & 0.4  & 0.14  & 0.09 & 0.042 & 0.026 & 0.15 & 0.09 & 0.07  & 0.04 \\
&$VH$              &  1.9  & 1.1 & 0.06  & 0.04 & 0.26  & 0.15 & 0.005 & 0.004 & 0.41 & 0.25 & 0.012 & 0.008 \\
\midrule
\multirow{3}{*}{Signal, \htm} &\ggF             &  270  & 130 & 4.5   & 2.3  & 38    & 18  & 0.55   & 0.28  & 14   & 7   & 0.64  & 0.33 \\
&VBF              &  15   & 7   & 8     & 4    & 2.1   & 1.1 & 0.9    & 0.5   & 1.6  & 0.8 & 0.8   & 0.4 \\
&$VH$             &  12   & 6   & 0.2   & 0.1 & 1.5    & 0.8 & 0.022  & 0.011 & 3.0  & 1.5 & 0.059 & 0.031 \\
\midrule
\multirow{9}{*}{Background}   & \Ztt            & 18200 & 400  & 360  & 70   & 46400 & 600  & 213   & 32    & 2000 & 160 & 113   & 13  \\
&\Zll             & 610   & 190  & 5.1  & 2.9  & 640   & 180  & 2.1   & 1.4   & 48   & 22  & 1.5   & 1.1 \\
& Top-quark             & 30900 & 1200 & 1750 & 60   & 5770  & 200  & 306   & 22    &274500& 900 & 14500 & 100 \\
& SM \ggF         & 942   & 70   & 32   & 6    & 910   & 50   & 10.5  & 1.9   & 79   & 11  & 4.4  & 1.6 \\
& SM VBF          & 92    & 4    & 64.7 & 2.8  & 40.9  & 2.1  & 17.4  & 1.0   & 11.1 & 0.7 & 7.0  & 0.5 \\
& SM $VH$         & 27    & 4    & 0.6  & 0.4  & 9.7   & 1.4  & 0.17  & 0.09  & 6.7  & 1.3 & 0.07 & 0.10 \\
& SM $ttH$        & 7.7   & 1.3  & 0.43 & 0.08 & 1.25  & 0.21 & 0.061 & 0.023 & 200  & 110 & 8.0  & 1.3   \\
& $VV$            & 28000 & 1300 & 610  & 60   & 10600 & 500  & 100   & 12    & 1820 & 160 & 100  & 10  \\
& Misidentified   & 11400 & 700  & 390  & 40   & 8200  & 600  & 81    & 18    & 8900 & 600 & 460  & 50 \\
\midrule
\midrule
Total Background            &                   & 90500 & 400  & 3230 & 60 & 72680 & 330 & 733 & 28 & 287600 & 500 & 15210 & 110 \\
Data                        &             & \multicolumn{2}{c|}{90531} & \multicolumn{2}{c|}{3138} & \multicolumn{2}{c|}{72511} & \multicolumn{2}{c|}{712} & \multicolumn{2}{c|}{287734} & \multicolumn{2}{c}{15228} \\
\bottomrule
\end{tabular}
}
\end{center}
\end{table}

\clearpage
\subsection{\MCbased{}{ \lephad} channel}
\label{app:yieldsMClephad}
 
The signal and background event yields in all SRs obtained after
performing the simultaneous fit of the \hte{} and \htm{} signals (Section~\ref{sec:2poi}) using only data of the \lephad{}
final state are detailed in Table~\ref{tab:MC_yields_lephad_2poi}.

\begin{table}[htb!]
\begin{center}
\caption{Observed event yields and predictions as
computed by the fit in the signal regions (SRs)
of the \ethad{} and \muthad{} channels.
The prediction for each sample
is determined from the likelihood fit performed to
measure the \hte{} and \htm{} signals using the
\lephad{} data only (Section~\ref{sec:2poi}).  The
signal event yields are given for $\BR(\hte)=0.03\%$
and $\BR(\htm)=0.09\%$, which correspond to the
best-fit values obtained from the fit mentioned
above.  Uncertainties include statistical and
systematic components. Uncertainties in the total
background prediction include the effect of
correlations between individual uncertainty sources
as determined by the fit.  }
\label{tab:MC_yields_lephad_2poi}
\resizebox{\columnwidth}{!}{
\begin{tabular}{lr|r @{${}\pm{}$} l|r @{${}\pm{}$} l|r @{${}\pm{}$} l|r @{${}\pm{}$} l}
\toprule
& Channel & \multicolumn{4}{c}{\ethad} & \multicolumn{4}{c}{\muthad}\\
& Region & \multicolumn{4}{c}{SR} & \multicolumn{4}{c}{SR}\\
& Category & \multicolumn{2}{c}{\nonVBF{}} & \multicolumn{2}{c}{\VBF{}} & \multicolumn{2}{c}{\nonVBF{}} & \multicolumn{2}{c}{\VBF{}}\\
\midrule
\midrule
\multirow{3}{*}{Signal, \hte} & \ggF            &  110    & 270   & 2  & 5  & \multicolumn{2}{c|}{<0.001} & \multicolumn{2}{c}{<0.001} \\
& VBF             &  5      & 14    & 4  & 10 & \multicolumn{2}{c|}{<0.001} & \multicolumn{2}{c}{<0.001} \\
& $VH$            &  4      & 9     & 0.06    & 0.16 & 0.07  & 0.17 & 0.003 & 0.006 \\
\midrule
\multirow{3}{*}{Signal, \htm} & \ggF            &  0.019  & 0.013 & 0.0027 & 0.0029 & 440  & 250 & 5.7  & 3.4 \\
& VBF             &  0.0014 & 0.0010& \multicolumn{2}{c|}{<0.001} & 22   & 13  & 12 & 7 \\
& $VH$            &  0.08   & 0.05  & 0.0024 & 0.0019 & 14 & 8 & 0.21 & 0.13 \\
\midrule
\multirow{9}{*}{Background}   & \Ztt            &  286800 & 1800  & 2700  & 100   & 389100 & 2600 & 3440  & 130 \\
&\Zll             &  32100  & 2000  & 280   & 80    & 33200  & 2900 & 180   & 40 \\
& Top-quark       &  7400   & 2000  & 520   & 150   & 9500   & 2500 & 590   & 160 \\
& SM \ggF         &  2110   & 130   & 49    & 10    & 3220   & 190  & 50    & 10 \\
& SM VBF          &  135    & 13    & 103   & 7     & 196    & 16   & 121   & 8 \\
& SM $VH$         &  100    & 13    & 2.1   & 0.6   & 131    & 17   & 2.5   & 0.5 \\
& SM $ttH$        &  4.3    & 0.6   & 0.28  & 0.05  & 4.9    & 0.7  & 0.32  & 0.06 \\
& $VV,W$          &  9000   & 500   & 250   & 40    & 13100  & 700  & 280   & 30 \\
& Misidentified   &  140200 & 2100  & 1810  & 130   & 140800 & 1900 & 1520  & 90 \\
\midrule
\midrule
Total Background              &                 & 478000  & 1200  & 5700  & 110   & 589700 & 1200 & 6200 & 100 \\
Data                          &            & \multicolumn{2}{c|}{477927} & \multicolumn{2}{c|}{5713} & \multicolumn{2}{c|}{589198} & \multicolumn{2}{c}{6178} \\
\bottomrule
\end{tabular}
}
\end{center}
\end{table}

\subsection{\Symbased-based \leplep{} channel}
\label{app:yieldsSymm}
 
The signal and background event yields in the SRs obtained after
performing the independent fit (Section~\ref{sec:1poi}) using only data passing the
symmetry-method specific \leplep{} event selection are detailed in Table~\ref{tab:symmYields}.
The background contribution
is divided into the symmetric, \mcfakes{} and \fffakes{} components.
Contributions from the other lepton channel are marked with the \enquote{tilde} sign.

\begin{table}
\caption{Observed event yields in the \VBF{} and \nonVBF{}
categories for the \hte{} and \htm{} searches following the
Symmetry-method, after the statistical analysis based on
events passing the \leplep{} Symmetry-method specific
selection.  The signal event yields are given for the
best-fit branching ratio values based on the \leplep{} data,
$\BR(\hte)=-0.33\%$ and $\BR(\htm)=0.25\%$, where the fit of
the difference is translated into individual branching
ratios, by setting the other branching ratio explicitly to
zero. This results in a negative shift for $\BR(\hte)$, and
a positive shift for $\BR(\htm)$ (cf.\ Section~\ref{sec:lep_taulep_symm}).
Contributions in the search for \hte{} coming from the \mte{} selection are marked
with the \enquote{tilde} sign, those from the \etm{} selection are
listed without. For the \htm{} search, the notation is the
same with interchanged regions.  Uncertainties include
statistical and systematic components. Uncertainties in the
total background prediction include the effect of
correlations between individual uncertainty sources as
determined by the fit.  }
\label{tab:symmYields}
\resizebox{\textwidth}{!}{
\begin{tabular}{lr|r @{${}\pm{}$} l|r @{${}\pm{}$} l|r @{${}\pm{}$} l|r @{${}\pm{}$} l}
\toprule
& & \multicolumn{4}{c}{\hte{} (no \htm{})} & \multicolumn{4}{c}{\htm{} (no \hte{})} \\
& Process & \multicolumn{2}{c}{\nonVBF{}} & \multicolumn{2}{c}{\VBF{}} & \multicolumn{2}{c}{\nonVBF{}} & \multicolumn{2}{c}{\VBF{}} \\
\midrule
\midrule
\multirow{6}{*}{Signal} & \ggF               &   $-$720 & 230  &  $-$11   & 4    &  480    & 190  &  8       & 4  \\
& VBF                &   $-$38  & 12   &  $-$20   & 7    &  26     & 11   &  14      & 6   \\
& $VH$               &   $-$31  & 11   &  $-$0.55 & 0.21 &  21     & 9    &  0.33    & 0.16   \\
& $\widetilde{\ggF}$ &   48     & 17   &  1.3     & 0.5  &  $-$35  & 15   &  $-$1.0  & 0.5   \\
& $\widetilde{VBF}$  &   3.9    & 1.3  &  1.7     & 0.6  &  $-$2.8 & 1.2  &  $-$1.2  & 0.5   \\
& $\widetilde{VH}$   &   5.7    & 2    &  0.14    & 0.06 &  $-$4.2 & 1.8  &  $-$0.10 & 0.05   \\
\midrule
\multirow{3}{*}{Symmetric background} & $\widetilde{\text{Data}}$ &  193700   & 1200  &  4300   & 50  &  162900   & 900  &  3890   & 50   \\
& $\widetilde{\fffakes{}}$  &  $-$19000 & 2000  &  $-$340 & 40  &  $-$15100 & 1500 &  $-$300 & 34   \\
& $\widetilde{\mcfakes{}}$  &  $-$12600 & 1800  &  $-$112 & 30  &  $-$5400  & 700  &  $-$72  & 17   \\
\midrule
\multirow{2}{*}{Misidentified background} & \fffakes{}   & 17900 & 1800  &  330 & 40  &  17200 & 1700  &  320 & 40   \\
& \mcfakes{}   & 5700  & 800   &  81  & 19  &  11500 & 1600  &  106 & 28   \\
\midrule
\midrule
Total background                         &               &   185700 & 500  &  4260 & 50  &  171000 & 500  &  3950 & 50   \\
Data & &  \multicolumn{2}{c|}{184887}  &  \multicolumn{2}{c|}{4230}  &  \multicolumn{2}{c|}{171675}  &  \multicolumn{2}{c}{3967}  \\
\bottomrule
\end{tabular}
}
\end{table}
 
\clearpage


\section{MVA optimisation}
\label{app:MVA}

The MVAs are trained using the four-momenta of the selected reconstructed particles as well as other
derived observables, such as the invariant masses, angular separations and the $\met$.
In the \VBF{} category, additional jet-related variables are used to exploit the characteristic VBF topology.
The complete list of the variables used is shown in Table~\ref{tab:input_variables}.
For the \MCbased{} method, the list of input variables is optimised by maximising the expected signal significance based on fits to Asimov datasets~\cite{Cowan:2010js} considering statistical uncertainties only.
For the \Symbased{} method, the hyperparameters and the input variables are optimised with respect to the expected signal
significance by performing the statistical analysis taking into account all systematic uncertainties and using only \leplep{} data (as described in \Sect{\ref{sec:stat}}).
For each category, the lowest-ranked variables making marginal contributions to the expected sensitivity are removed.
 
The collinear mass (\mcol)~\cite{Ellis:1987xu}, the Higgs boson mass
obtained with the Missing Mass Calculator (MMC) technique
(\mmmc)~\cite{MMCpaper} and \mvis{} show the highest separation power
together with transverse mass calculated from the Higgs decay products
and the $\met$ ($\mT(\tau,\met), \ \mT(\ell,\met)$).  The MMC
algorithm is tuned specifically to reconstruct the mass of the LFV
Higgs boson.
In the collinear approximation, the azimuthal angle
between the prompt lepton from the Higgs boson and the $\met$,
$\Delta\phi (\ell_H, \met)$, is expected to be large for signal
events, while the azimuthal angle
between the visible products of the $\tau$-lepton decay
and the $\met$, $\Delta\phi(\tau, \met)$, is expected to be small for signal
events.
Other angular separations are included as well:
$\Delta R(\ell_H, \tau)$, $\Delta \eta(\ell_H, \tau)$,
$\Delta\phi (\ell_H, \tau)$, where $\tau$ refers to $\tau_\ell$ or
$\tauhadvis$ in the corresponding channel. The $\Delta\phi$
angular separations involving $\ell_H$ are evaluated in the
approximate Higgs boson rest frame.
 
In the \leplep{} channel, two vertex-related variables are also included because
they provide additional discrimination power: the difference of the
transverse impact parameters of the leptons (\deltad0) and
the $d_0$ significance of the lepton from the $\tau$:
$\sigmad0LTau$. The $\Delta\alpha$ discriminant~\cite{Davidson:2012wn}
exploited in the \leplep{} channels is expected to be close to zero if
the decay products of $\tau$-lepton are collinear and the transverse
momentum of the Higgs boson can be neglected, while for the background
events, this value deviates from zero.
Additionally, in the \Symbased{} \leplep{} channel, the $\eta$-centrality, defined as
\begin{equation*}
\eta\mhyphen\mathrm{centrality}(\ell)=\exp(-4/(\eta_{\mathrm{j_1}}-\eta_{\mathrm{j_2}})^2\cdot (\eta_\ell -
0.5(\eta_{\mathrm{j_1}}+\eta_{\mathrm{j_2}}))^2),
\end{equation*}
is included in the \VBF{} category.
 
\begin{table}[h!]
\caption{ List of input variables used in the \nonVBF{} and \VBF{} categories.
Those labelled with \enquote{rest} are evaluated in the approximate
Higgs boson rest frame as described in
Section~\ref{sec:evt_sel}. The variable $\pt^\mathrm{tot}$,
used only in the \VBF{} category of the \MCbased{} \leplep{}
channel, corresponds to the absolute value of the vectorial
sum of the \pt{} of the two leptons, the two jets and
the \met, and it is used to veto the third jet.
When a variable is used in both the \leplep{} and \lephad{}
channels, the $\tau$ symbol refers to either $\tau_\ell$ or
$\tauhadvis$ in the corresponding channel.}
\label{tab:input_variables}
\centering
\begin{tabular}{c|c c|c c|c c}
\hline
\multirow{2}{2cm}{ {Variable} }     & \multicolumn{2}{c}{\lephad{}}  & \multicolumn{2}{c}{\leplep{} \MCbased{}} & \multicolumn{2}{c}{\leplep{} \Symbased{}}    \\
& \nonVBF       & \VBF             & \nonVBF       & \VBF                     & \nonVBF       & \VBF        \\
\hline
$\mcol$                             & \checkmark & \checkmark        & \checkmark   & \checkmark              & \checkmark   & \checkmark \\
$\mvis$                             & \checkmark & \checkmark        & \checkmark   & \checkmark              & \checkmark   & \checkmark \\
$\mmmc$                             &              &                 &  \checkmark  & \checkmark              & \checkmark   & \checkmark \\
$\mT(\tau,\met)$                    & \checkmark   & \checkmark      &              &                         & \checkmark   & \checkmark \\
$\mT(\ell_H,\met)$                     & \checkmark   & \checkmark      &              &                         & \checkmark   & \checkmark \\
$\mT(\ell_{1},\met)$                &              &                 & \checkmark   & \checkmark              &              &            \\
$\mT(\ell_{2},\met)$                &              &                 & \checkmark   & \checkmark              &              &            \\
$\met$                              & \checkmark   & \checkmark      & \checkmark   & \checkmark              & \checkmark   & \checkmark \\
$\pT(\ell_H)$                       & \checkmark   & \checkmark      &              &                         &              &            \\
$\pT(\thadvis)$                     & \checkmark   & \checkmark      &              &                         &              &            \\
$\pT(\ell_2+\met)/\pT(\ell_1)$      &              &                 & \checkmark   & \checkmark              &              &            \\
$\pT^{\mathrm{rest}}(\ell_H) $        &              &                 & \checkmark   & \checkmark              & \checkmark   & \checkmark \\
$\pT^{\mathrm{rest}}(\ell_\tau) $     &              &                 & \checkmark   &  \checkmark             & \checkmark   & \checkmark \\
$\pt^\mathrm{tot}$                   &              &                 &              & \checkmark              & \checkmark   & \checkmark \\
$\pT(\ell_H)/\met$                  &              &                 &              &                         & \checkmark   & \checkmark \\
$\pT(\ell_H)/\pT(\ell_\tau) $       &              &                 &              &                         & \checkmark   & \checkmark \\
$\pT(\ell_\tau+\met)/\pT(\ell_H)$       &              &                 &              &                         & \checkmark   & \checkmark \\
$\sum \pt $                         &              &                 &              &                         & \checkmark   & \checkmark \\
$\Delta R (\ell_H,\tau)$               & \checkmark   & \checkmark      & \checkmark   & \checkmark              & \checkmark   & \checkmark \\
$\Delta \eta (\ell_H,\tau)$            & \checkmark   & \checkmark      &              &                         & \checkmark   & \checkmark \\
$\Delta \phi (\ell_H,\tau)$            & \checkmark   &                 &              &                         & \checkmark   &            \\
$\Delta \phi (\ell_\tau,\met)$         &              &                 &  \checkmark  & \checkmark              & \checkmark   & \checkmark \\
$|\Delta \phi (\ell_H, \met)| - |\Delta \phi (\thadvis, \met)|$         & \checkmark   &                 &              &                         &              &             \\
$\dalpha$                           &              &                 &  \checkmark  & \checkmark              & \checkmark   & \checkmark \\
$\dphiLHiggs$                       &              & \checkmark      & \checkmark   &  \checkmark             & \checkmark   & \checkmark \\
$\deltad0$                          &              &                 & \checkmark   &  \checkmark             & \checkmark   &            \\
$\sigmad0LTau$                      &              &                 & \checkmark   &  \checkmark             &              &            \\
$\eta (\tauhadvis)$                 & \checkmark   & \checkmark      &              &                         &              &  \\
$\mjj$                              &              & \checkmark      &              & \checkmark              &              & \checkmark \\
$N_{\mathrm{jets} (\pT > 30 \GeV)}$ &              & \checkmark      &              &                         &              &  \\
$|\detajj|$                         &              & \checkmark      &              &                         &              & \checkmark \\
$\Delta R(\mathrm{j},\mathrm{j})$   &              &                 &              & \checkmark              &              & \checkmark \\
$|\detajj|\cdot\eta_{\mathrm{j_1}}\cdot \eta_{\mathrm{j_2}}$     &              &                 &              & \checkmark              &              &  \\
$\pT(\mathrm{j_1})$                 &              &                 &              &                         &               & \checkmark \\
$\pT(\mathrm{j_2})$                 &              &                 &              &                         &               & \checkmark \\
$\Delta\phi(\mathrm{j_1},\met)$     &              &                 &              &  \checkmark             &               & \checkmark \\
$\Delta\phi(\mathrm{j_2},\met)$     &              &                 &              &  \checkmark             &               & \checkmark \\
$\eta\mhyphen\mathrm{centrality}(\ell_H)$      &              &                 &              &  \checkmark             &               & \checkmark \\
$\eta\mhyphen\mathrm{centrality}(\ell_\tau)$   &              &                 &              &  \checkmark             &               & \checkmark \\
\hline
\end{tabular}
\end{table}

Different MVA techniques are used for the \nonVBF{} and \VBF{} categories to exploit the VBF topology, taking into account for the different limited training sample sizes.
 
For the \MCbased{} method, the parameters used to configure the BDT
are summarised in Table~\ref{tab:bdtparameters}.
The BDTs are trained using the Toolkit for Multivariate Data Analysis~\cite{TMVA}.
A $k$-fold cross validation procedure~\cite{k_fold} is implemented for the BDT training to avoid overtraining and to enable efficient usage of the available events in the training process.
In both the \leplep{} and \lephad{} final states, 5-folds are used.
 
\begin{table}[h!]
\caption{Optimised BDT parameters, for all the regions and
channels used in the \MCbased{} method. The Boost type and Number-of-cuts are set
to Gradient and 20 respectively.}
\label{tab:bdtparameters}
\centering
\begin{tabular}{c|c|c|c|c|c}
\hline
Region & Channel & \texttt{NTrees} & \texttt{MaxDepth} & \texttt{MinNodeSize} & \texttt{Shrinkage}\\ 
\hline
\nonVBF & $\ethad$,$\muthad$         & 500 & 7 & 1  & 0.05\phantom{00}   \\
\hline
\VBF & $\ethad$                   & 300 & 10\phantom{0} & 1 & 0.01\phantom{00}   \\
& $\muthad$ $\mathrm{BDT}_1$ & 300 & 8  & 1 & 0.009\phantom{0}  \\
& $\muthad$ $\mathrm{BDT}_2$ & 300 & 6  & 1 & 0.0095 \\
\hline
\nonVBF,\VBF & \leplep{} & 750 & 8  & \phantom{.5}2.5 & 0.1\phantom{000} \\
\hline
\end{tabular}
\end{table}

In the \leplep{} channel of the \Symbased{} method,
the fully connected deep NN training with Keras~\cite{chollet2015keras} and Tensorflow~\cite{abadi2016tensorflow} as the backend uses supervised learning with  categorical cross-entropy as the cost function and L2 weight regularisation to prevent overtraining, as well as 10-fold cross validation.
The hyperparameter optimisation is carried out using the \emph{Optuna} framework~\cite{optuna_2019}, where the L2 weight regularisation parameter is included in the hyperparameter optimisation.
The full set of hyperparameters used for the NNs in the \Symbased{}
\leplep{} channel is summarised in Table~\ref{tab:NNHyperp}.
\begin{table}[h!]
\caption{Hyperparameter configuration for the NNs
used in the \Symbased{} \leplep\ analysis.}
\label{tab:NNHyperp}
\centering
\begin{tabular}{lcccc}
\toprule
Hyperparameter & \multicolumn{4}{c}{Value}\\
& \nonVBF{} NN 	& \VBF$_{\Ztt}$ NN	& \VBF$_\text{Top-quark}$ NN & \VBF$_\text{misID}$ NN \\
\midrule
\# nodes in 1st layer    		& 512      			& 128      			& 128      			& 128      \\
\# hidden layers         		& \phantom{00}2	& \phantom{00}4	& \phantom{00}3	& \phantom{00}4        \\
\# output layers         			& \phantom{00}3	& \phantom{00}2	& \phantom{00}2	& \phantom{00}2        \\
L2 weight reg. param.    		& 0.000048 		& 0.000292 		& 0.000094		& 0.000356 \\
Leaky ReLU slope below 0 	& 0.080537 		& 0.019614 		& 0.062515		& 0.084219 \\
Optimiser                			& SGD      		& Adam     		& Adam     		& Adam     \\
Learning rate            			& 0.025810 		& 0.000142 		& 0.000215 		& 0.003507 \\
Batch size               			& 128      			& 128      			& 512      			& 1024\phantom{0}     \\
Epochs                   			& 100      			& 100      			& 100      			& 100      \\
\bottomrule
\end{tabular}
\end{table}

\clearpage


\section{Compatibility of branching ratio differences}
\label{app:compatability}
 
The compatibility of the two measured branching ratio
differences, $\Delta_{e\tau}= \BR(\hte)-\BR({\htm})$ and $\Delta_{\mu\tau} = \BR(\htm)-\BR({\hte})$, in
the \Symbased{} method is tested. Measuring $\Delta_{e\tau}$
($\Delta_{\mu\tau}$) is equivalent to measuring $\mu_{e\tau}$
($\mu_{\mu\tau}$) when assuming $\mu_{\mu\tau}$ ($\mu_{e\tau}$) = 0.
This implies that the sum of the two fitted values for $\mu_{e\tau}$ and
$\mu_{\mu\tau}$ should be consistent with zero.  This was quantified
using a $\chi^2$ test, accounting for the error of the
mean value of the upward and downward systematic variations of $\mu_{e\tau}$ and
$\mu_{\mu\tau}$, and the correlation coefficient $\rho$ between
the \etm{} and \mte{} datasets.
The latter is found to be $\rho = -0.80$ using the bootstrap
methodology~\cite{ATL-PHYS-PUB-2021-011} when considering the full
systematic uncertainties in the fit.
The results obtained in the \nonVBF{} and \VBF{} categories, as well as in the combined fit,
indicate that the $\Delta_{e\tau}$ and $\Delta_{\mu\tau}$ measurements are compatible within $1.3\sigma$.
 
The branching ratio difference, $\Delta_{\mu\tau}$, from the \Symbased{} method can be compared with the value from the \MCbased{} method, as  discussed in \Sect{\ref{sec:comp-leplep}}. The result is displayed in \Tab{\ref{tab:compatibility}}.
 
\begingroup
\renewcommand{\arraystretch}{1.3}
\begin{table}[htb!]
\centering
\caption{Best-fit $\BR(\htm)-\BR({\hte})$ values, given in \%,
obtained from the stand-alone fits of the \leplep{} final state
with either background estimation method. The results for \nonVBF{} and \VBF{} category
are provided, as well as their combination.
The last row shows the difference between the \MCbased{} method and the \Symbased{} method.
The results for the \MCbased{} method are quoted once with the full uncertainty and once considering only the sources of uncertainty that
are uncorrelated between the two analyses.
The uncertainty in the difference between the two methods is calculated taking into account only the uncorrelated uncertainty sources for the \MCbased{} method and the full uncertainty for \Symbased{} method.}
\label{tab:compatibility}
\begin{tabular}{lccc}
\toprule
& \multicolumn{3}{c}{$\BRhat(\htm)-\BRhat({\hte})$ [\%] } \\
Method   & \nonVBF{}  & \VBF{} & Combined \\
\midrule
\MCbased{} (full uncertainties)               & $-0.10 \pm 0.14$  & $0.45 \pm 0.30$ & $ 0.02 \pm 0.12$ \\
\MCbased{} (uncorrelated uncertainties)               & $-0.10 \pm 0.12$  & $0.45 \pm 0.21$ & $ 0.02 \pm 0.09$ \\
\Symbased{} (full uncertainties)         	 & \phantom{$-$}$0.30 \pm 0.11$  & $0.06 \pm 0.21$ & $ 0.25  \pm 0.10$ \\
\midrule
\MCbased{} (uncorr. uncertainties) $-$ \Symbased{} & $-0.39 \pm 0.16$  & $0.39 \pm 0.29$ & $-0.22  \pm 0.14$\phantom{$-$} \\
 
\bottomrule
\end{tabular}
\end{table}
\endgroup

\clearpage


\printbibliography

@inproceedings{k_fold, 
    author = {Blum, Avrim and Kalai, Adam and Langford, John}, 
    title = {Beating the Hold-out: Bounds for K-Fold and Progressive Cross-Validation}, 
    year = {1999}, 
    booktitle = {Proceedings of the Twelfth Annual Conference on Computational Learning Theory}, 
    pages = {203–208},
    url = {https://doi.org/10.1145/307400.307439}, 
    doi = {10.1145/307400.307439}, 
}

@article{Davidson:2012wn,
    author = "Davidson, Sacha and Lacroix, Sylvain and Verdier, Patrice",
    title = "{LHC sensitivity to lepton flavour violating Z boson decays}",
    eprint = "1207.4894",
    archivePrefix = "arXiv",
    primaryClass = "hep-ph",
    doi = "10.1007/JHEP09(2012)092",
    journal = "JHEP",
    volume = "09",
    pages = "092",
    year = "2012"
}

@article{Ellis:1987xu,
	author = "Ellis, R. Keith and Hinchliffe, I. and Soldate, M. and van der Bij, J. J.",
	title = "{Higgs decay to $\tau^+$ $\tau^-$ A possible signature of intermediate mass Higgs bosons at high energy hadron colliders}",
	reportNumber = "FERMILAB-PUB-87-100-T, LBL-23504",
	doi = "10.1016/0550-3213(88)90019-3",
	journal = "Nucl. Phys. B",
	volume = "297",
	pages = "221--243",
	year = "1988"
}

@article{Bressler:2014jta,
	author = "Bressler, Shikma and Dery, Avital and Efrati, Aielet",
	title = "{Asymmetric lepton-flavor violating Higgs boson decays}",
	eprint = "1405.4545",
	archivePrefix = "arXiv",
	primaryClass = "hep-ph",
	doi = "10.1103/PhysRevD.90.015025",
	journal = "Phys. Rev. D",
	volume = "90",
	number = "1",
	pages = "015025",
	year = "2014"
}

@article{MEG:2013oxv,
    author = "Adam, J. and others",
    collaboration = "MEG",
    title = "{New Constraint on the Existence of the $\mu^+ \to e^+\gamma$ Decay}",
    eprint = "1303.0754",
    archivePrefix = "arXiv",
    primaryClass = "hep-ex",
    doi = "10.1103/PhysRevLett.110.201801",
    journal = "Phys. Rev. Lett.",
    volume = "110",
    pages = "201801",
    year = "2013"
}

@article{Guralnik:1964eu,
        author                  = "Guralnik, G.S. and Hagen, C.R. and Kibble, T.W.B.",
        title                           = "{Global Conservation Laws and Massless Particles}",
        journal                 = "Phys. Rev. Lett.",
        volume                  = "13",
        pages                   = "585-587",
        doi                             = "10.1103/PhysRevLett.13.585",
        year                    = "1964",
        SLACcitation    = "%%CITATION = PRLTA,13,585;%%",
}

@article{HIGGS3,
        author                  = "Higgs, Peter W.",
        title                           = "{Spontaneous Symmetry Breakdown without Massless Bosons}",
        journal                 = "Phys. Rev.",
        volume                  = "145",
        pages                   = "1156-1163",
        doi                             = "10.1103/PhysRev.145.1156",
        year                    = "1966",
        SLACcitation    = "%%CITATION = PHRVA,145,1156;%%",
}

@article{Kibble:1967sv,
        author                  = "Kibble, T.W.B.",
        title                           = "{Symmetry Breaking in Non-Abelian Gauge Theories}",
        journal                 = "Phys. Rev.",
        volume                  = "155",
        pages                   = "1554-1561",
        doi                             = "10.1103/PhysRev.155.1554",
        year                    = "1967",
        SLACcitation    = "%%CITATION = PHRVA,155,1554;%%",
}

@article{Arana-Catania:2013xma,
      author         = "Arana-Catania, M. and Arganda, E. and Herrero, M. J.",
      title          = "{Non-decoupling SUSY in LFV Higgs decays: a window to new
                        physics at the LHC}",
      journal        = "JHEP",
      volume         = "09",
      year           = "2013",
      pages          = "160",
      doi            = "10.1007/JHEP09(2013)160",
      note           = "{(Erratum:
                  \href{http://dx.doi.org/10.1007/JHEP10(2015)192}{JHEP \textbf{10} (2015) 192})}",
%%		  ,\texttt{\href{http://arxiv.org/abs/1304.3371}{arXiv:1304.3371 [hep-ph]}}}

@article{Arhrib:2012ax,
      author         = "Arhrib, Abdesslam and Cheng, Yifan and Kong, Otto C. W.",
      title          = "{Comprehensive analysis on lepton flavor violating Higgs
                        boson to $\mu^\mp \tau^\pm$ decay in supersymmetry without $R$ parity}",
      journal        = "Phys.\ Rev. D",
      volume         = "87",
      pages          = "015025",
      doi            = "10.1103/PhysRevD.87.015025",
      year           = "2013",
      eprint         = "1210.8241",
      archivePrefix  = "arXiv",
      primaryClass   = "hep-ph",
      reportNumber   = "NCU-HEP-K055",
      SLACcitation   = "%%CITATION = ARXIV:1210.8241;%%",
}

@article{Ko:2017lzd,
    author = "Ko, P. and Omura, Yuji and Shigekami, Yoshihiro and Yu, Chaehyun",
    title = "{LHCb anomaly and B physics in flavored Z' models with flavored Higgs doublets}",
    eprint = "1702.08666",
    archivePrefix = "arXiv",
    primaryClass = "hep-ph",
    doi = "10.1103/PhysRevD.95.115040",
    journal = "Phys. Rev. D",
    volume = "95",
    number = "11",
    pages = "115040",
    year = "2017"
}

@article{Carmona:2015ena,
    author = "Carmona, Adrian and Goertz, Florian",
    title = "{Lepton Flavor and Nonuniversality from Minimal Composite Higgs Setups}",
    eprint = "1510.07658",
    archivePrefix = "arXiv",
    primaryClass = "hep-ph",
    reportNumber = "CERN-PH-TH-2015-251",
    doi = "10.1103/PhysRevLett.116.251801",
    journal = "Phys. Rev. Lett.",
    volume = "116",
    number = "25",
    pages = "251801",
    year = "2016"
}

@article{Botella:2015hoa,
    author = "Botella, F. J. and Branco, G. C. and Nebot, M. and Rebelo, M. N.",
    title = "{Flavour-changing Higgs couplings in a class of two Higgs doublet models}",
    eprint = "1508.05101",
    archivePrefix = "arXiv",
    primaryClass = "hep-ph",
    reportNumber = "IFIC-15-62",
    doi = "10.1140/epjc/s10052-016-3993-0",
    journal = "Eur. Phys. J. C",
    volume = "76",
    number = "3",
    pages = "161",
    year = "2016"
}

@article{PhysRevLett.115.111803,
  title = {Measurement of the Ratio of Branching Fractions $\mathcal{B}({\overline{B}}^{0}\ensuremath{\rightarrow}{D}^{*+}{\ensuremath{\tau}}^{\ensuremath{-}}{\overline{\ensuremath{\nu}}}_{\ensuremath{\tau}})/\mathcal{B}({\overline{B}}^{0}\ensuremath{\rightarrow}{D}^{*+}{\ensuremath{\mu}}^{\ensuremath{-}}{\overline{\ensuremath{\nu}}}_{\ensuremath{\mu}})$},
  %author = "Aaij, R. and others",
  author = "{LHCb Collaboration}",
  collaboration = "LHCb",
  journal = {Phys. Rev. Lett.},
  volume = {115},
  number = {11},
  pages = {111803},
  numpages = {10},
  year = {2015},
  month = {Sep},
  publisher = {American Physical Society},
  doi = {10.1103/PhysRevLett.115.111803},
 %%% url = {https://link.aps.org/doi/10.1103/PhysRevLett.115.111803},
  eprint = "1506.08614",
  archivePrefix  = "arXiv",
  primaryClass   = "hep-ex",
}

@article{LHCb:2017smo,
    %author = "Aaij, R. and others",
    author = "{LHCb Collaboration}",
    collaboration = "LHCb",
    title = "{Measurement of the Ratio of the $B^0 \to D^{*-} \tau^+ \nu_{\tau}$ and $B^0 \to D^{*-} \mu^+ \nu_{\mu}$ Branching Fractions Using Three-Prong $\tau$-Lepton Decays}",
    eprint = "1708.08856",
    archivePrefix = "arXiv",
    primaryClass = "hep-ex",
    reportNumber = "LHCB-PAPER-2017-017, CERN-EP-2017-212",
    doi = "10.1103/PhysRevLett.120.171802",
    journal = "Phys. Rev. Lett.",
    volume = "120",
    number = "17",
    pages = "171802",
    year = "2018"
}

@article{LHCb:2015tgy,
    %author = "Aaij, Roel and others",
    author = "{LHCb Collaboration}",
    collaboration = "LHCb",
    title = "{Differential branching fraction and angular analysis of $\Lambda^{0}_{b} \rightarrow \Lambda \mu^+\mu^-$ decays}",
    eprint = "1503.07138",
    archivePrefix = "arXiv",
    primaryClass = "hep-ex",
    reportNumber = "LHCB-PAPER-2015-009, CERN-PH-EP-2015-078",
    doi = "10.1007/JHEP06(2015)115",
    journal = "JHEP",
    volume = "06",
    pages = "115",
    year = "2015",
    note = "{(Erratum: \href{https://doi.org/10.1007/JHEP09(2018)145}{JHEP \textbf{09} (2018) 145})}"
}

@article{Bjorken:1977vt,
      author         = "Bjorken, J. D. and Weinberg, Steven",
      title          = "{Mechanism for Nonconservation of Muon Number}",
      journal        = "Phys.\ Rev.\ Lett.",
      volume         = "38",
      pages          = "622",
      doi            = "10.1103/PhysRevLett.38.622",
      year           = "1977",
      reportNumber   = "SLAC-PUB-1868",
      SLACcitation   = "%%CITATION = PRLTA,38,622;%%",
}

@article{DiazCruz:1999xe,
      author         = "Diaz-Cruz, J. Lorenzo and Toscano, J. J.",
      title          = "{Lepton flavor violating decays of Higgs bosons beyond
                        the standard model}",
      journal        = "Phys.\ Rev. D",
      volume         = "62",
      pages          = "116005",
      doi            = "10.1103/PhysRevD.62.116005",
      year           = "2000",
%%      eprint         = "hep-ph/9910233",
%%      archivePrefix  = "arXiv",
      reportNumber   = "HEP-IFUAP-99-2",
      SLACcitation   = "%%CITATION = HEP-PH/9910233;%%",
}

@article{Agashe:2009di,
      author         = "Agashe, Kaustubh and Contino, Roberto",
%%      title          = "{Composite Higgs-Mediated FCNC}",
      title          = "{Composite Higgs-mediated flavor-changing neutral current}",
      journal        = "Phys.\ Rev. D",
      volume         = "80",
      pages          = "075016",
      doi            = "10.1103/PhysRevD.80.075016",
      year           = "2009",
      eprint         = "0906.1542",
      archivePrefix  = "arXiv",
      primaryClass   = "hep-ph",
      reportNumber   = "UMD-PP-09-038, CERN-PH-TH-2009-085",
      SLACcitation   = "%%CITATION = ARXIV:0906.1542;%%",
}

@article{Ishimori:2010au,
      author         = "Ishimori, Hajime and others",
      title          = "{Non-Abelian Discrete Symmetries in Particle Physics}",
      journal        = "Prog.\ Theor.\ Phys.\ Suppl.",
      volume         = "183",
      pages          = "1",
      doi            = "10.1143/PTPS.183.1",
      year           = "2010",
      eprint         = "1003.3552",
      archivePrefix  = "arXiv",
      primaryClass   = "hep-th",
      reportNumber   = "KUNS-2260",
      SLACcitation   = "%%CITATION = ARXIV:1003.3552;%%",
}

@article{Perez:2008ee,
      author         = "Perez, Gilad and Randall, Lisa",
      title          = "{Natural neutrino masses and mixings from warped
                        geometry}",
      journal        = "JHEP",
      volume         = "01",
      pages          = "077",
      doi            = "10.1088/1126-6708/2009/01/077",
      year           = "2009",
      eprint         = "0805.4652",
      archivePrefix  = "arXiv",
      primaryClass   = "hep-ph",
      SLACcitation   = "%%CITATION = ARXIV:0805.4652;%%",
}

@article{Azatov:2009na,
      author         = "Azatov, Aleksandr and Toharia, Manuel and Zhu, Lijun",
      title          = "{Higgs mediated flavor changing neutral currents in warped extra dimensions}",
      journal        = "Phys.\ Rev. D",
      volume         = "80",
      pages          = "035016",
      doi            = "10.1103/PhysRevD.80.035016",
      year           = "2009",
      eprint         = "0906.1990",
      archivePrefix  = "arXiv",
      primaryClass   = "hep-ph",
      reportNumber   = "UMD-PP-09-039",
      SLACcitation   = "%%CITATION = ARXIV:0906.1990;%%",
}

@inproceedings{abadi2016tensorflow,
  title={Tensorflow: A system for large-scale machine learning},
  author={Abadi, Mart{\'\i}n and Barham, Paul and Chen, Jianmin and Chen, Zhifeng and Davis, Andy and Dean, Jeffrey and Devin, Matthieu and Ghemawat, Sanjay and Irving, Geoffrey and Isard, Michael and others},
  booktitle={12th USENIX Symposium on Operating Systems Design and Implementation (OSDI'16)},
  pages={265--283},
  year={2016}
}

@Article{CKKW,
   author         = "L{\"o}nnblad, L. and Prestel, S.",
   title          = "{Matching tree-level matrix elements with interleaved showers}",
   journal        = "JHEP",
   volume         = "03",
   pages          = "019",
   year           = "2012",
   eprint         = "1109.4829",
   archivePrefix  = "arXiv",
   primaryClass   = "hep-ph",
   doi            = "10.1007/JHEP03(2012)019",
}

@Article{LHCXS,
%   author         = "{LHC Higgs Cross Section Working Group} and de Florian, D. and Grojean, C. and Maltoni, F. and Nikitenko, A. and others",
   author         = "{D. de Florian et al., LHC Higgs Cross Section Working Group}",
   title          = "{Handbook of LHC Higgs cross sections: 4. Deciphering the nature of the Higgs sector}",
   year           = "2017",
   eprint         = "1610.07922",
   archivePrefix  = "arXiv",
   primaryClass   = "hep-ph",
   report         = "CERN Yellow Reports: Monographs Volume 2/2017 (CERN--2017--002-M)",
   doi            = "10.23731/CYRM-2017-002",
}

@Article{SM,
  AUTHOR  = "A. Djouadi",
  title   = "{The anatomy of electroweak symmetry breaking: Tome I: The Higgs boson in the Standard Model}",
  journal = "Phys. Rep.",
  volume  = "457",
  year    = "2008",
  pages   = "1, and references therein"
}

@Article{HIGGS,
      author         = "Higgs, Peter W.",
      title          = "{Broken symmetries, massless particles and gauge fields}",
      journal        = "Phys. Lett. ",
      volume         = "12",
      pages          = "132-133",
      doi            = "10.1016/0031-9163(64)91136-9",
      year           = "1964",
      SLACcitation   = "%%CITATION = PHLTA,12,132;%%",
}

@Article{HIGGS2,
      author         = "Higgs, Peter W.",
      title          = "{Broken Symmetries and the Masses of Gauge Bosons}",
      journal        = "Phys. Rev. Lett. ",
      volume         = "13",
      pages          = "508-509",
      doi            = "10.1103/PhysRevLett.13.508",
      year           = "1964",
      SLACcitation   = "%%CITATION = PRLTA,13,508;%%",
}

@Article{ENGLERT,
      author         = "Englert, F. and Brout, R.",
      title          = "{Broken Symmetry and the Mass of Gauge Vector Mesons}",
      journal        = "Phys. Rev. Lett. ",
      volume         = "13",
      pages          = "321-323",
      doi            = "10.1103/PhysRevLett.13.321",
      year           = "1964",
      SLACcitation   = "%%CITATION = PRLTA,13,321;%%",
}

@article{Pythia8,
      author         = "{T. Sj\"ostrand, S. Mrenna and P. Skands}",
      title          = "{A Brief Introduction to PYTHIA 8.1}",
      journal        = "Comput. Phys. Commun.",
      volume         = "178",
      pages          = "852-867",
      doi            = "10.1016/j.cpc.2008.01.036",
      year           = "2008",
      eprint         = "0710.3820",
      archivePrefix  = "arXiv",
      primaryClass   = "hep-ph",
      reportNumber   = "CERN-LCGAPP-2007-04, LU-TP-07-28,
                        FERMILAB-PUB-07-512-CD-T",
      SLACcitation   = "%%CITATION = ARXIV:0710.3820;%%",
}

@Article{Fastjet,
      author         = "Cacciari, Matteo and Salam, Gavin P. and Soyez, Gregory",
      title          = "{FastJet user manual}",
      journal        = "Eur. Phys. J. C",
      volume         = "72",
      year           = "2012",
      pages          = "1896",
      doi            = "10.1140/epjc/s10052-012-1896-2",
      eprint         = "1111.6097",
      archivePrefix  = "arXiv",
      primaryClass   = "hep-ph",
      reportNumber   = "CERN-PH-TH-2011-297",
}

@Article{Geant4,
     hideauthor = "{GEANT4 Collaboration (S. Agostinelli et al.)}",
     author = "{S. Agostinelli et al., GEANT4 Collaboration}",
     title = "{GEANT4---a simulation toolkit}",
      journal        = "Nucl. Instrum. Meth. A",
      volume         = "506",
      pages          = "250-303",
      doi            = "10.1016/S0168-9002(03)01368-8",
      year           = "2003",
      reportNumber   = "SLAC-PUB-9350, FERMILAB-PUB-03-339",
      SLACcitation   = "%%CITATION = NUIMA,A506,250;%%",
}

@Article{MMCpaper,
     author         = "{A. Elagin, P. Murat, A. Pranko and A. Safonov}",
     title          = "{A new mass reconstruction technique for resonances
                        decaying to $\tau\tau$}",
      journal        = "Nucl. Instrum. Meth. A",
      volume         = "654",
      pages          = "481-489",
      doi            = "10.1016/j.nima.2011.07.009",
      year           = "2011",
      eprint         = "1012.4686",
      archivePrefix  = "arXiv",
      primaryClass   = "hep-ex",
      SLACcitation   = "%%CITATION = ARXIV:1012.4686;%%",
}

@Article{ATLASHiggsJuly2012,
      hideauthor = "{ATLAS Collaboration (G. Aad et al.)}",
      author = "{ATLAS Collaboration}",
      title  = "{Observation of a new particle in the search for the Standard Model Higgs boson with the ATLAS detector at the LHC}",
      journal        = "Phys. Lett. ",
      volume         = "B 716",
      pages          = "1-29",
      doi            = "10.1016/j.physletb.2012.08.020",
      year           = "2012",
      eprint         = "1207.7214",
      archivePrefix  = "arXiv",
      primaryClass   = "hep-ex",
      reportNumber   = "CERN-PH-EP-2012-218",
      SLACcitation   = "%%CITATION = ARXIV:1207.7214;%%",
}

@Article{CMSHiggsJuly2012,
      hideauthor = "{CMS Collaboration (S. Chatrchyan et al.)}",
      author = "{CMS Collaboration}",
      title  = "{Observation of a new boson at a mass of 125 GeV with the CMS experiment at the LHC}",
      journal        = "Phys. Lett. ",
      volume         = "B 716",
      pages          = "30-61",
      doi            = "10.1016/j.physletb.2012.08.021",
      year           = "2012",
      eprint         = "1207.7235",
      archivePrefix  = "arXiv",
      primaryClass   = "hep-ex",
      reportNumber   = "CMS-HIG-12-028, CERN-PH-EP-2012-220",
      SLACcitation   = "%%CITATION = ARXIV:1207.7235;%%",
}

@article{LHCHiggsCrossSectionWorkingGroup:2013,
      title          = "{Handbook of LHC Higgs Cross Sections: 3. Higgs
                        Properties}",
      author         = "{LHC Higgs Cross Section Working Group}",
      doi            = "10.5170/CERN-2013-004",
      year           = "2013",
      eprint         = "1307.1347",
      archivePrefix  = "arXiv",
      primaryClass   = "hep-ph",
      reportNumber   = "CERN-2013-004",
      SLACcitation   = "%%CITATION = ARXIV:1307.1347;%%",
}

@Misc{FTAG-2019-07,
    author         = "{ATLAS Collaboration}",
    title          = "{ATLAS flavour-tagging algorithms for the LHC Run 2 $pp$ collision dataset}",
    year           = "2022",
    doi            = "10.48550/ARXIV.2211.16345",
    eprint         = "2211.16345",
    archivePrefix  = "arXiv",
    primaryClass   = "hep-ex",
}

@article{Blankenburg:2012ex,
      author         = "Blankenburg, Gianluca and Ellis, John and Isidori, Gino",
      title          = "{Flavour-changing decays of a 125 GeV Higgs-like
                        particle}",
      journal        = "Phys.\ Lett. B",
      volume         = "712",
      pages          = "386",
      doi            = "10.1016/j.physletb.2012.05.007",
      year           = "2012",
      eprint         = "1202.5704",
      archivePrefix  = "arXiv",
      primaryClass   = "hep-ph",
      reportNumber   = "CERN-PH-TH-2012-049, KCL-PH-TH-2012-09, LCTS-2012-05",
      SLACcitation   = "%%CITATION = ARXIV:1202.5704;%%",
}

@article{Harnik:2012pb,
      author         = "Harnik, Roni and Kopp, Joachim and Zupan, Jure",
      title          = "{Flavor violating Higgs decays}",
      journal        = "JHEP",
      volume         = "03",
      pages          = "026",
      doi            = "10.1007/JHEP03(2013)026",
      year           = "2013",
      eprint         = "1209.1397",
      archivePrefix  = "arXiv",
      primaryClass   = "hep-ph",
      reportNumber   = "FERMILAB-PUB-12-498-T",
      SLACcitation   = "%%CITATION = ARXIV:1209.1397;%%",
}

@article{Albrecht:2009xr,
      author         = "Albrecht, Michaela E. and Blanke, Monika and Buras,
                        Andrzej J. and Duling, Bjorn and Gemmler, Katrin",
      title          = "{Electroweak and flavour structure of a warped extra
                        dimension with custodial protection}",
      journal        = "JHEP",
      volume         = "09",
      pages          = "064",
      doi            = "10.1088/1126-6708/2009/09/064",
      year           = "2009",
      eprint         = "0903.2415",
      archivePrefix  = "arXiv",
      primaryClass   = "hep-ph",
      reportNumber   = "TUM-TEP-711-09, MPP-2009-17, TUM-HEP-711-09",
      SLACcitation   = "%%CITATION = ARXIV:0903.2415;%%",
}

@article{Goudelis:2011un,
      author         = "Goudelis, Andreas and Lebedev, Oleg and Park, Jae-hyeon",
      title          = "{Higgs-induced lepton flavor violation}",
      journal        = "Phys.\ Lett. B",
      volume         = "707",
      pages          = "369",
      doi            = "10.1016/j.physletb.2011.12.059",
      year           = "2012",
      eprint         = "1111.1715",
      archivePrefix  = "arXiv",
      primaryClass   = "hep-ph",
      reportNumber   = "DESY-11-194",
      SLACcitation   = "%%CITATION = ARXIV:1111.1715;%%",
}

@article{McKeen:2012av,
      author         = "McKeen, David and Pospelov, Maxim and Ritz, Adam",
      title          = "{Modified Higgs branching ratios versus CP and lepton
                        flavor violation}",
      journal        = "Phys.\ Rev. D",
      volume         = "86",
      pages          = "113004",
      doi            = "10.1103/PhysRevD.86.113004",
      year           = "2012",
      eprint         = "1208.4597",
      archivePrefix  = "arXiv",
      primaryClass   = "hep-ph",
      SLACcitation   = "%%CITATION = ARXIV:1208.4597;%%",
}

@article{Aubert:2009ag,
      author         = "{BaBar Collaboration}",
      title          = "{Searches for Lepton Flavor Violation in the Decays
                        $\tau^{\pm}\to e^{\pm}\gamma$ and $\tau^{\pm}\to\mu^{\pm}\gamma$}",
      collaboration  = "BaBar",
      journal        = "Phys.\ Rev.\ Lett.",
      volume         = "104",
      pages          = "021802",
      doi            = "10.1103/PhysRevLett.104.021802",
      year           = "2010",
      eprint         = "0908.2381",
      archivePrefix  = "arXiv",
      primaryClass   = "hep-ex",
      reportNumber   = "SLAC-PUB-13753, BABAR-PUB-09-026",
      SLACcitation   = "%%CITATION = ARXIV:0908.2381;%%",
}

@BOOK{bdt,
  author        = {L. Breiman and J. Friedman and R. Olshen and C. Stone},
  title         = {{Classification and Regression Trees}},
  publisher     = {Chapman \& Hall},
  address       = {New York},
  year          = {1984},
  pages         = {358},
}

@Article{theo_sys_1,
    author         = "{S. Dittmaier et al., LHC Higgs Cross Section Working Group}",
    title          = "{Handbook of LHC Higgs Cross Sections: 1. Inclusive Observables}",
    howpublished   = "{CERN-2011-002 (CERN, Geneva, 2011)}",
    eprint         = "1101.0593",
    archivePrefix  = "arXiv",
    primaryClass   = "hep-ph",
    year           = "2011",
    doi            = "10.5170/CERN-2011-002"
}

@article{LUCID2,
  author={G. Avoni and others},
  title={The new LUCID-2 detector for luminosity measurement and monitoring in ATLAS},
  journal={JINST},
  volume={13},
  number={07},
  pages={P07017},
  doi="10.1088/1748-0221/13/07/P07017",
  year={2018},
}

@inproceedings{optuna_2019,
	author = {Akiba, Takuya and Sano, Shotaro and Yanase, Toshihiko and Ohta, Takeru and Koyama, Masanori},
	booktitle = {Proceedings of the 25rd {ACM} {SIGKDD} International Conference on Knowledge Discovery and Data Mining},
	title = {Optuna: A Next-generation Hyperparameter Optimization Framework},
	year = {2019},
	doi = "10.1145/3292500.3330701"
}

@article{Belle1,
    author = "{Belle Collaboration}",
    %author = "Caria, G. and others",
    %collaboration = "Belle",
    title = "{Measurement of $\mathcal{R}(D)$ and $\mathcal{R}(D^*)$ with a Semileptonic Tagging Method}",
    eprint = "1910.05864",
    archivePrefix = "arXiv",
    primaryClass = "hep-ex",
    reportNumber = "Belle-2019-18, KEK-2019-40",
    doi = "10.1103/PhysRevLett.124.161803",
    journal = "Phys. Rev. Lett.",
    volume = "124",
    number = "16",
    pages = "161803",
    year = "2020"
}

@article{BaBar1,
    author = "{BaBar Collaboration}",
    %author = "Lees, J. P. and others",
    %collaboration = "BaBar",
    title = "{Evidence for an Excess of $\bar{B} \to D^{(*)} \tau^-\bar{\nu}_\tau$ Decays}",
    eprint = "1205.5442",
    archivePrefix = "arXiv",
    primaryClass = "hep-ex",
    reportNumber = "BABAR-PUB-12-012, SLAC-PUB-15028",
    doi = "10.1103/PhysRevLett.109.101802",
    journal = "Phys. Rev. Lett.",
    volume = "109",
    pages = "101802",
    year = "2012"
}

@article{Hoche:2010kg,
    author = "Hoche, Stefan and Krauss, Frank and Schonherr, Marek and Siegert, Frank",
    title = "{NLO matrix elements and truncated showers}",
    eprint = "1009.1127",
    archivePrefix = "arXiv",
    primaryClass = "hep-ph",
    reportNumber = "ZU-TH-13-10, IPPP-10-73, DCPT-10-146, CERN-PH-TH-2010-195, MCNET-10-17",
    doi = "10.1007/JHEP08(2011)123",
    journal = "JHEP",
    volume = "08",
    pages = "123",
    year = "2011"
}

@article{Gleisberg:2005qq,
    author = "Gleisberg, Tanju and Krauss, Frank and Schalicke, Andreas and Schumann, Steffen and Winter, Jan-Christopher",
    title = "{Studying $W^{+} W^{-}$ production at the Fermilab Tevatron with SHERPA}",
    eprint = "hep-ph/0504032",
    archivePrefix = "arXiv",
    doi = "10.1103/PhysRevD.72.034028",
    journal = "Phys. Rev. D",
    volume = "72",
    pages = "034028",
    year = "2005"
}

@article{Binoth:2009wk,
    author = "Binoth, T. and Gleisberg, T. and Karg, S. and Kauer, N. and Sanguinetti, G.",
    title = "{NLO QCD corrections to ZZ+ jet production at hadron colliders}",
    eprint = "0911.3181",
    archivePrefix = "arXiv",
    primaryClass = "hep-ph",
    reportNumber = "EDINBURGH-2009-18, LAPTH-1362-09, PITHA-09-30, SFB-CPP-09-113, SLAC-PUB-13833",
    doi = "10.1016/j.physletb.2009.12.013",
    journal = "Phys. Lett. B",
    volume = "683",
    pages = "154--159",
    year = "2010"
}

@article{Aloni:2015wvn,
    author = "Aloni, Daniel and Nir, Yosef and Stamou, Emmanuel",
    title = "{Large BR($h \to \tau \mu$) in the MSSM?}",
    eprint = "1511.00979",
    archivePrefix = "arXiv",
    primaryClass = "hep-ph",
    doi = "10.1007/JHEP04(2016)162",
    journal = "JHEP",
    volume = "04",
    pages = "162",
    year = "2016"
}

@Article{Cacciari:2008gp,
     author    = "Cacciari, Matteo and Salam, Gavin P. and Soyez, Gregory",
     title     = "{The anti-\(k_{t}\) jet clustering algorithm}",
     journal   = "JHEP",
     volume    = "04",
     year      = "2008",
     pages     = "063",
     eprint    = "0802.1189",
     archivePrefix = "arXiv",
     primaryClass  =  "hep-ph",
     doi       = "10.1088/1126-6708/2008/04/063",
     SLACcitation  = "%%CITATION = 0802.1189;%%"
}

@Article{Butterworth:2015oua,
      author         = "Butterworth, Jon and others",
      title          = "{PDF4LHC recommendations for LHC Run II}",
      journal        = "J. Phys. G",
      volume         = "43",
      year           = "2016",
      pages          = "023001",
      doi            = "10.1088/0954-3899/43/2/023001",
      eprint         = "1510.03865",
      archivePrefix  = "arXiv",
      primaryClass   = "hep-ph",
      reportNumber   = "OUTP-15-17P, SMU-HEP-15-12, TIF-UNIMI-2015-14,
                        LCTS-2015-27, CERN-PH-TH-2015-249",
      SLACcitation   = "%%CITATION = ARXIV:1510.03865;%%"
}

@Article{Pumplin:2002vw,
     author    = "Pumplin, J. and others",
     title     = "{New Generation of Parton Distributions with Uncertainties from Global QCD Analysis}",
     journal   = "JHEP",
     volume    = "07",
     year      = "2002",
     pages     = "012",
     doi       = "10.1088/1126-6708/2002/07/012",
     eprint    = "hep-ph/0201195",
     archivePrefix = "arXiv",
     SLACcitation  = "%%CITATION = HEP-PH/0201195;%%"
}

@article{Lai:2010vv,
      author         = "Lai, H.-L. and others",
      title          = "{New parton distributions for collider physics}",
      journal        = "Phys. Rev. D",
      volume         = "82",
      pages          = "074024",
      doi            = "10.1103/PhysRevD.82.074024",
      year           = "2010",
      eprint         = "1007.2241",
      archivePrefix  = "arXiv",
      primaryClass   = "hep-ph",
      reportNumber   = "MSUHEP-100707, SMU-HEP-10-10",
      SLACcitation   = "%%CITATION = ARXIV:1007.2241;%%",
}

@article{Harland-Lang:2014zoa,
      author         = "Harland-Lang, L. A. and Martin, A. D. and Motylinski, P.
                        and Thorne, R. S.",
      title          = "{Parton distributions in the LHC era: MMHT 2014 PDFs}",
      journal        = "Eur. Phys. J. C",
      number         = "5",
      volume         = "75",
      pages          = "204",
      doi            = "10.1140/epjc/s10052-015-3397-6",
      year           = "2015",
      eprint         = "1412.3989",
      archivePrefix  = "arXiv",
      primaryClass   = "hep-ph",
      reportNumber   = "LCTS-2014-47, IPPP-14-97, DCPT-14-194",
      SLACcitation   = "%%CITATION = ARXIV:1412.3989;%%",
}

@article{Ball:2012cx,
      author         = "{NNPDF Collaboration} and Ball, Richard D. and others",
      title          = "{Parton distributions with LHC data}",
      journal        = "Nucl. Phys. B",
      volume         = "867",
      year           = "2013",
      pages          = "244",
      doi            = "10.1016/j.nuclphysb.2012.10.003",
      eprint         = "1207.1303",
      archivePrefix  = "arXiv",
      primaryClass   = "hep-ph",
      reportNumber   = "EDINBURGH-2012-08, IFUM-FT-997, FR-PHENO-2012-014,
                        RWTH-TTK-12-25, CERN-PH-TH-2012-037, SFB-CPP-12-47\,
                        --CERN-PH-TH-2012-037",
      SLACcitation   = "%%CITATION = ARXIV:1207.1303;%%"
}

@article{Ball:2014uwa,
      author         = "{The NNPDF Collaboration} and Ball, Richard D. and others",
      title          = "{Parton distributions for the LHC run II}",
      journal        = "JHEP",
      volume         = "04",
      year           = "2015",
      pages          = "040",
      doi            = "10.1007/JHEP04(2015)040",
      eprint         = "1410.8849",
      archivePrefix  = "arXiv",
      primaryClass   = "hep-ph",
      reportNumber   = "EDINBURGH-2014-15, IFUM-1034-FT, CERN-PH-TH-2013-253,
                        OUTP-14-11P, CAVENDISH-HEP-14-11",
      SLACcitation   = "%%CITATION = ARXIV:1410.8849;%%"
}

@article{Sjostrand:2014zea,
      author         = "Sj{\"o}strand, Torbj{\"o}rn and Ask, Stefan and Christiansen,
                        Jesper R. and Corke, Richard and Desai, Nishita and Ilten,
                        Philip and Mrenna, Stephen and Prestel, Stefan and
                        Rasmussen, Christine O. and Skands, Peter Z.",
      title          = "{An introduction to PYTHIA 8.2}",
      journal        = "Comput. Phys. Commun.",
      volume         = "191",
      year           = "2015",
      pages          = "159",
      doi            = "10.1016/j.cpc.2015.01.024",
      eprint         = "1410.3012",
      archivePrefix  = "arXiv",
      primaryClass   = "hep-ph",
      reportNumber   = "LU-TP-14-36, MCNET-14-22, CERN-PH-TH-2014-190,
                        FERMILAB-PUB-14-316-CD, DESY-14-178, SLAC-PUB-16122,
                        --FERMILAB-PUB-14-316-CD",
      SLACcitation   = "%%CITATION = ARXIV:1410.3012;%%"
}

@Article{Alwall:2014hca,
      author         = "Alwall, J. and Frederix, R. and Frixione, S. and Hirschi,
                        V. and Maltoni, F. and Mattelaer, O. and Shao, H. -S. and
                        Stelzer, T. and Torrielli, P. and Zaro, M.",
      title          = "{The automated computation of tree-level and
                        next-to-leading order differential cross sections, and
                        their matching to parton shower simulations}",
      journal        = "JHEP",
      volume         = "07",
      year           = "2014",
      pages          = "079",
      doi            = "10.1007/JHEP07(2014)079",
      eprint         = "1405.0301",
      archivePrefix  = "arXiv",
      primaryClass   = "hep-ph",
      reportNumber   = "CERN-PH-TH-2014-064, CP3-14-18, LPN14-066, MCNET-14-09,
                        ZU-TH-14-14",
      SLACcitation   = "%%CITATION = ARXIV:1405.0301;%%"
}

@Article{Bahr:2008pv,
      author         = "B{\"a}hr, M. and others",
      title          = "{Herwig++ physics and manual}",
      journal        = "Eur. Phys. J. C",
      volume         = "58",
      year           = "2008",
      pages          = "639",
      doi            = "10.1140/epjc/s10052-008-0798-9",
      eprint         = "0803.0883",
      archivePrefix  = "arXiv",
      primaryClass   = "hep-ph",
      reportNumber   = "CERN-PH-TH-2008-038, CAVENDISH-HEP-08-03, KA-TP-05-2008,
                        DCPT-08-22, IPPP-08-11, CP3-08-05",
      SLACcitation   = "%%CITATION = ARXIV:0803.0883;%%"
}

@Article{Bellm:2015jjp,
      author         = "Bellm, Johannes and others",
      title          = "{Herwig 7.0/Herwig++ 3.0 release note}",
      journal        = "Eur. Phys. J. C",
      volume         = "76",
      year           = "2016",
      number         = "4",
      pages          = "196",
      doi            = "10.1140/epjc/s10052-016-4018-8",
      eprint         = "1512.01178",
      archivePrefix  = "arXiv",
      primaryClass   = "hep-ph",
      reportNumber   = "CERN-PH-TH-2015-289, MAN-HEP-2015-15, IFJPAN-IV-2015-13,
                        HERWIG-2015-01, KA-TP-18-2015, DCPT-15-142, MCNET-15-28,
                        IPPP-15-71, --HERWIG-2015-01",
      SLACcitation   = "%%CITATION = ARXIV:1512.01178;%%"
}

@Article{Nason:2004rx,
      author         = "Nason, Paolo",
      title          = "{A new method for combining NLO QCD with shower Monte Carlo algorithms}",
      journal        = "JHEP",
      volume         = "11",
      pages          = "040",
      doi            = "10.1088/1126-6708/2004/11/040",
      year           = "2004",
      eprint         = "hep-ph/0409146",
      archivePrefix  = "arXiv",
}

@Article{Frixione:2007vw,
      author         = "Frixione, Stefano and Nason, Paolo and Oleari, Carlo",
      title          = "{Matching NLO QCD computations with parton shower
                        simulations: the POWHEG method}",
      journal        = "JHEP",
      volume         = "11",
      pages          = "070",
      doi            = "10.1088/1126-6708/2007/11/070",
      year           = "2007",
      eprint         = "0709.2092",
      archivePrefix  = "arXiv",
      primaryClass   = "hep-ph",
}

@Article{Alioli:2010xd,
      author         = "Alioli, Simone and Nason, Paolo and Oleari, Carlo and Re,
                        Emanuele",
      title          = "{A general framework for implementing NLO calculations in
                        shower Monte Carlo programs: the POWHEG BOX}",
      journal        = "JHEP",
      volume         = "06",
      pages          = "043",
      doi            = "10.1007/JHEP06(2010)043",
      year           = "2010",
      eprint         = "1002.2581",
      archivePrefix  = "arXiv",
      primaryClass   = "hep-ph",
}

@article{Hartanto:2015uka,
      author         = "Hartanto, Heribertus B. and J{\"a}ger, Barbara and Reina,
                        Laura and Wackeroth, Doreen",
      title          = "{Higgs boson production in association with top quarks in
                        the POWHEG BOX}",
      journal        = "Phys. Rev. D",
      volume         = "91",
      year           = "2015",
      number         = "9",
      pages          = "094003",
      doi            = "10.1103/PhysRevD.91.094003",
      eprint         = "1501.04498",
      archivePrefix  = "arXiv",
      primaryClass   = "hep-ph",
      SLACcitation   = "%%CITATION = ARXIV:1501.04498;%%"
}

@article{Frixione:2007nw,
      author         = "Frixione, Stefano and Ridolfi, Giovanni and Nason, Paolo",
      title          = "{A positive-weight next-to-leading-order Monte Carlo for
                        heavy flavour hadroproduction}",
      journal        = "JHEP",
      volume         = "09",
      pages          = "126",
      doi            = "10.1088/1126-6708/2007/09/126",
      year           = "2007",
      eprint         = "0707.3088",
      archivePrefix  = "arXiv",
      primaryClass   = "hep-ph",
      SLACcitation   = "%%CITATION = ARXIV:0707.3088;%%",
}

@Article{Schumann:2007mg,
      author         = "Schumann, Steffen and Krauss, Frank",
      title          = "{A parton shower algorithm based on Catani--Seymour dipole factorisation}",
      journal        = "JHEP",
      volume         = "03",
      pages          = "038",
      doi            = "10.1088/1126-6708/2008/03/038",
      year           = "2008",
      eprint         = "0709.1027",
      archivePrefix  = "arXiv",
      primaryClass   = "hep-ph",
}

@article{Bothmann:2019yzt,
      author = "Bothmann, Enrico and others",
      title = "{Event generation with Sherpa 2.2}",
      journal = "SciPost Phys.",
      volume = "7",
      year = "2019",
      number = "3",
      pages = "034",
      doi = "10.21468/SciPostPhys.7.3.034",
      reportNumber = "FERMILAB-PUB-19-218-T, SLAC-PUB-17433, IPPP/19/42, MCNET-19-11",
      eprint = "1905.09127",
      archivePrefix = "arXiv",
      primaryClass = "hep-ph",
}

@Article{Read:2002hq,
    Author = {Read, Alexander L.},
    Journal = {J. Phys. G},
    Pages = {2693},
    doi = "10.1088/0954-3899/28/10/313",
    Title = "{Presentation of search results: the \(CL_S\) technique}",
    Volume = {28},
    Year = {2002}
}

@Article{Cowan:2010js,
   Author = {Cowan, Glen and Cranmer, Kyle and Gross, Eilam and Vitells, Ofer},
   Title = {{Asymptotic formulae for likelihood-based tests of new physics}},
   Journal = {Eur. Phys. J. C},
   Volume = {71},
   Year = {2011},
   Pages = {1554},
   doi = "10.1140/epjc/s10052-011-1554-0",
   Eprint = "1007.1727",
   Archiveprefix = {arXiv},
   Primaryclass = {physics.data-an},
   related = "Cowan:2010js-err",
   relatedstring  = "Erratum:",
}

@Booklet{TMVA,
      author         = "Hoecker, Andreas and others",
      title          = "{TMVA - Toolkit for Multivariate Data Analysis}",
      year           = "2007",
      eprint         = "physics/0703039",
      archivePrefix  = "arXiv",
      primaryClass   = "physics.data-an",
}

@online{chollet2015keras,
      author         = {Chollet, Fran\c{c}ois and others},
      title          = {Keras},
      year           = {2015},
      url            = {https://keras.io},
}

@Article{PERF-2007-01,
    author         = "{ATLAS Collaboration}",
    title          = "{The ATLAS Experiment at the CERN Large Hadron Collider}",
    journal        = "JINST",
    volume         = "3",
    year           = "2008",
    pages          = "S08003",
    doi            = "10.1088/1748-0221/3/08/S08003",
    primaryClass   = "hep-ex",
}

@Article{SOFT-2010-01,
    author         = "{ATLAS Collaboration}",
    title          = "{The ATLAS Simulation Infrastructure}",
    journal        = "Eur. Phys. J. C",
    volume         = "70",
    year           = "2010",
    pages          = "823",
    doi            = "10.1140/epjc/s10052-010-1429-9",
    eprint         = "1005.4568",
    archivePrefix  = "arXiv",
    primaryClass   = "physics.ins-det",
}

@Article{STDM-2012-23,
    author         = "{ATLAS Collaboration}",
    title          = "{Measurement of the \(Z/\gamma^*\) boson transverse momentum distribution in \(pp\) collisions at \(\sqrt{s} = 7\,\text{TeV}\) with the ATLAS detector}",
    journal        = "JHEP",
    volume         = "09",
    year           = "2014",
    pages          = "145",
    doi            = "10.1007/JHEP09(2014)145",
    reportNumber   = "CERN-PH-EP-2014-075",
    eprint         = "1406.3660",
    archivePrefix  = "arXiv",
    primaryClass   = "hep-ex",
}

@Article{HIGG-2013-21,
    author         = "{ATLAS Collaboration}",
    title          = "{Measurements of Higgs boson production and couplings in the four-lepton channel in \(pp\) collisions at center-of-mass energies of \(7\) and \(8\,\text{TeV}\) with the ATLAS detector}",
    journal        = "Phys. Rev. D",
    volume         = "91",
    year           = "2015",
    pages          = "012006",
    doi            = "10.1103/PhysRevD.91.012006",
    reportNumber   = "CERN-PH-EP-2014-170",
    eprint         = "1408.5191",
    archivePrefix  = "arXiv",
    primaryClass   = "hep-ex",
}

@Article{HIGG-2013-32,
    author         = "{ATLAS Collaboration}",
    title          = "{Evidence for the Higgs-boson Yukawa coupling to tau leptons with the ATLAS detector}",
    journal        = "JHEP",
    volume         = "04",
    year           = "2015",
    pages          = "117",
    doi            = "10.1007/JHEP04(2015)117",
    reportNumber   = "CERN-PH-EP-2014-262",
    eprint         = "1501.04943",
    archivePrefix  = "arXiv",
    primaryClass   = "hep-ex",
}

@Article{PERF-2013-06,
    author         = "{ATLAS Collaboration}",
    title          = "{Identification and energy calibration of hadronically decaying tau leptons with the ATLAS experiment in \(pp\) collisions at \(\sqrt{s} = 8\,\text{TeV}\)}",
    journal        = "Eur. Phys. J. C",
    volume         = "75",
    year           = "2015",
    pages          = "303",
    doi            = "10.1140/epjc/s10052-015-3500-z",
    reportNumber   = "CERN-PH-EP-2014-227",
    eprint         = "1412.7086",
    archivePrefix  = "arXiv",
    primaryClass   = "hep-ex",
}

@Article{HIGG-2014-14,
    author         = "{ATLAS and CMS Collaborations}",
    title          = "{Combined Measurement of the Higgs Boson Mass in \(pp\) Collisions at \(\sqrt{s} = 7\) and \(8\,\text{TeV}\) with the ATLAS and CMS Experiments}",
    journal        = "Phys. Rev. Lett.",
    volume         = "114",
    year           = "2015",
    pages          = "191803",
    doi            = "10.1103/PhysRevLett.114.191803",
    eprint         = "1503.07589",
    archivePrefix  = "arXiv",
    primaryClass   = "hep-ex",
}

@Article{PERF-2014-03,
    author         = "{ATLAS Collaboration}",
    title          = "{Performance of pile-up mitigation techniques for jets in \(pp\) collisions at \(\sqrt{s} = 8\,\text{TeV}\) using the ATLAS detector}",
    journal        = "Eur. Phys. J. C",
    volume         = "76",
    year           = "2016",
    pages          = "581",
    doi            = "10.1140/epjc/s10052-016-4395-z",
    reportNumber   = "CERN-PH-EP-2015-206",
    eprint         = "1510.03823",
    archivePrefix  = "arXiv",
    primaryClass   = "hep-ex",
}

@Article{PERF-2014-07,
    author         = "{ATLAS Collaboration}",
    title          = "{Topological cell clustering in the ATLAS calorimeters and its performance in LHC Run~1}",
    journal        = "Eur. Phys. J. C",
    volume         = "77",
    year           = "2017",
    pages          = "490",
    doi            = "10.1140/epjc/s10052-017-5004-5",
    reportNumber   = "CERN-PH-EP-2015-304",
    eprint         = "1603.02934",
    archivePrefix  = "arXiv",
    primaryClass   = "hep-ex",
}

@Article{HIGG-2015-09,
    author         = "{ATLAS Collaboration}",
    title          = "{Search for lepton-flavour-violating decays of the Higgs and \(Z\) bosons with the ATLAS detector}",
    journal        = "Eur. Phys. J. C",
    volume         = "77",
    year           = "2017",
    pages          = "70",
    doi            = "10.1140/epjc/s10052-017-4624-0",
    reportNumber   = "CERN-EP-2016-055",
    eprint         = "1604.07730",
    archivePrefix  = "arXiv",
    primaryClass   = "hep-ex",
}

@Article{PERF-2015-01,
    author         = "{ATLAS Collaboration}",
    title          = "{Reconstruction of primary vertices at the ATLAS experiment in Run~1 proton--proton collisions at the LHC}",
    journal        = "Eur. Phys. J. C",
    volume         = "77",
    year           = "2017",
    pages          = "332",
    doi            = "10.1140/epjc/s10052-017-4887-5",
    reportNumber   = "CERN-EP-2016-150",
    eprint         = "1611.10235",
    archivePrefix  = "arXiv",
    primaryClass   = "hep-ex",
}

@Article{PERF-2015-09,
    author         = "{ATLAS Collaboration}",
    title          = "{Jet reconstruction and performance using particle flow with the ATLAS Detector}",
    journal        = "Eur. Phys. J. C",
    volume         = "77",
    year           = "2017",
    pages          = "466",
    doi            = "10.1140/epjc/s10052-017-5031-2",
    reportNumber   = "CERN-EP-2017-024",
    eprint         = "1703.10485",
    archivePrefix  = "arXiv",
    primaryClass   = "hep-ex",
}

@Article{PERF-2015-10,
    author         = "{ATLAS Collaboration}",
    title          = "{Muon reconstruction performance of the ATLAS detector in proton--proton collision data at \(\sqrt{s} = 13\,\text{TeV}\)}",
    journal        = "Eur. Phys. J. C",
    volume         = "76",
    year           = "2016",
    pages          = "292",
    doi            = "10.1140/epjc/s10052-016-4120-y",
    reportNumber   = "CERN-EP-2016-033",
    eprint         = "1603.05598",
    archivePrefix  = "arXiv",
    primaryClass   = "hep-ex",
}

@Article{PERF-2016-04,
    author         = "{ATLAS Collaboration}",
    title          = "{Jet energy scale measurements and their systematic uncertainties in proton--proton collisions at \(\sqrt{s} = 13\,\text{TeV}\) with the ATLAS detector}",
    journal        = "Phys. Rev. D",
    volume         = "96",
    year           = "2017",
    pages          = "072002",
    doi            = "10.1103/PhysRevD.96.072002",
    reportNumber   = "CERN-EP-2017-038",
    eprint         = "1703.09665",
    archivePrefix  = "arXiv",
    primaryClass   = "hep-ex",
}

@Article{PERF-2016-06,
    author         = "{ATLAS Collaboration}",
    title          = "{Identification and rejection of pile-up jets at high pseudorapidity with the ATLAS detector}",
    journal        = "Eur. Phys. J. C",
    volume         = "77",
    year           = "2017",
    pages          = "580",
    doi            = "10.1140/epjc/s10052-017-5081-5",
    reportNumber   = "CERN-EP-2017-055",
    eprint         = "1705.02211",
    archivePrefix  = "arXiv",
    primaryClass   = "hep-ex",
    related        = "PERF-2016-06-err",
    relatedstring  = "Erratum:",
}

@Article{PERF-2016-07,
    author         = "{ATLAS Collaboration}",
    title          = "{Performance of missing transverse momentum reconstruction with the ATLAS detector using proton--proton collisions at \(\sqrt{s} = 13\,\text{TeV}\)}",
    journal        = "Eur. Phys. J. C",
    volume         = "78",
    year           = "2018",
    pages          = "903",
    doi            = "10.1140/epjc/s10052-018-6288-9",
    reportNumber   = "CERN-EP-2017-274",
    eprint         = "1802.08168",
    archivePrefix  = "arXiv",
    primaryClass   = "hep-ex",
}

@Article{TRIG-2016-01,
    author         = "{ATLAS Collaboration}",
    title          = "{Performance of the ATLAS trigger system in 2015}",
    journal        = "Eur. Phys. J. C",
    volume         = "77",
    year           = "2017",
    pages          = "317",
    doi            = "10.1140/epjc/s10052-017-4852-3",
    reportNumber   = "CERN-EP-2016-241",
    eprint         = "1611.09661",
    archivePrefix  = "arXiv",
    primaryClass   = "hep-ex",
}

@Article{HIGG-2017-08,
    author         = "{ATLAS Collaboration}",
    title          = "{Searches for lepton-flavour-violating decays of the Higgs boson in \(\sqrt{s} = 13\,\text{TeV}\) \(pp\) collisions with the ATLAS detector}",
    journal        = "Phys. Lett. B",
    volume         = "800",
    year           = "2020",
    pages          = "135069",
    doi            = "10.1016/j.physletb.2019.135069",
    reportNumber   = "CERN-EP-2019-126",
    eprint         = "1907.06131",
    archivePrefix  = "arXiv",
    primaryClass   = "hep-ex",
}

@Article{PERF-2017-01,
    author         = "{ATLAS Collaboration}",
    title          = "{Electron reconstruction and identification in the ATLAS experiment using the 2015 and 2016 LHC proton--proton collision data at \(\sqrt{s} = 13\,\text{TeV}\)}",
    journal        = "Eur. Phys. J. C",
    volume         = "79",
    year           = "2019",
    pages          = "639",
    doi            = "10.1140/epjc/s10052-019-7140-6",
    reportNumber   = "CERN-EP-2018-273",
    eprint         = "1902.04655",
    archivePrefix  = "arXiv",
    primaryClass   = "hep-ex",
}

@Article{PERF-2017-03,
    author         = "{ATLAS Collaboration}",
    title          = "{Electron and photon energy calibration with the ATLAS detector using 2015--2016 LHC proton--proton collision data}",
    journal        = "JINST",
    volume         = "14",
    year           = "2019",
    pages          = "P03017",
    doi            = "10.1088/1748-0221/14/03/P03017",
    reportNumber   = "CERN-EP-2018-296",
    eprint         = "1812.03848",
    archivePrefix  = "arXiv",
    primaryClass   = "hep-ex",
}

@Article{DAPR-2018-01,
    author         = "{ATLAS Collaboration}",
    title          = "{ATLAS data quality operations and performance for 2015--2018 data-taking}",
    journal        = "JINST",
    volume         = "15",
    year           = "2020",
    pages          = "P04003",
    doi            = "10.1088/1748-0221/15/04/P04003",
    reportNumber   = "CERN-EP-2019-207",
    eprint         = "1911.04632",
    archivePrefix  = "arXiv",
    primaryClass   = "physics.ins-det",
}

@Article{EGAM-2018-01,
    author         = "{ATLAS Collaboration}",
    title          = "{Electron and photon performance measurements with the ATLAS detector using the 2015--2017 LHC proton--proton collision data}",
    journal        = "JINST",
    volume         = "14",
    year           = "2019",
    pages          = "P12006",
    doi            = "10.1088/1748-0221/14/12/P12006",
    reportNumber   = "CERN-EP-2019-145",
    eprint         = "1908.00005",
    archivePrefix  = "arXiv",
    primaryClass   = "hep-ex",
}

@Article{FTAG-2018-01,
    author         = "{ATLAS Collaboration}",
    title          = "{ATLAS \(b\)-jet identification performance and efficiency measurement with \(t\bar{t}\) events in \(pp\) collisions at \(\sqrt{s} = 13\,\text{TeV}\)}",
    journal        = "Eur. Phys. J. C",
    volume         = "79",
    year           = "2019",
    pages          = "970",
    doi            = "10.1140/epjc/s10052-019-7450-8",
    reportNumber   = "CERN-EP-2019-132",
    eprint         = "1907.05120",
    archivePrefix  = "arXiv",
    primaryClass   = "hep-ex",
}

@Article{HIGG-2018-58,
    author         = "{ATLAS Collaboration}",
    title          = "{Search for the Higgs boson decays \(H \to ee\) and \(H \to e\mu\) in \(pp\) collisions at \(\sqrt{s} = 13\,\text{TeV}\) with the ATLAS detector}",
    journal        = "Phys. Lett. B",
    volume         = "801",
    year           = "2020",
    pages          = "135148",
    doi            = "10.1016/j.physletb.2019.135148",
    reportNumber   = "CERN-EP-2019-184",
    eprint         = "1909.10235",
    archivePrefix  = "arXiv",
    primaryClass   = "hep-ex",
}

@Article{JETM-2018-05,
    author         = "{ATLAS Collaboration}",
    title          = "{Jet energy scale and resolution measured in proton--proton collisions at \(\sqrt{s} = 13\,\text{TeV}\) with the ATLAS detector}",
    journal        = "Eur. Phys. J. C",
    volume         = "81",
    year           = "2020",
    pages          = "689",
    doi            = "10.1140/epjc/s10052-021-09402-3",
    reportNumber   = "CERN-EP-2020-083",
    eprint         = "2007.02645",
    archivePrefix  = "arXiv",
    primaryClass   = "hep-ex",
}

@Article{MUON-2018-03,
    author         = "{ATLAS Collaboration}",
    title          = "{Muon reconstruction and identification efficiency in ATLAS using the full Run~2 \(pp\) collision data set at \(\sqrt{s} = 13\,\text{TeV}\)}",
    journal        = "Eur. Phys. J. C",
    volume         = "81",
    year           = "2021",
    pages          = "578",
    doi            = "10.1140/epjc/s10052-021-09233-2",
    reportNumber   = "CERN-EP-2020-199",
    eprint         = "2012.00578",
    archivePrefix  = "arXiv",
    primaryClass   = "hep-ex",
}

@Article{PIX-2018-001,
    author         = "Abbott, B. and others",
    title          = "{Production and integration of the ATLAS Insertable B-Layer}",
    journal        = "JINST",
    volume         = "13",
    year           = "2018",
    pages          = "T05008",
    doi            = "10.1088/1748-0221/13/05/T05008",
    eprint         = "1803.00844",
    archivePrefix  = "arXiv",
    primaryClass   = "physics.ins-det",
}

@Article{TRIG-2018-01,
    author         = "{ATLAS Collaboration}",
    title          = "{Performance of the ATLAS muon triggers in Run~2}",
    journal        = "JINST",
    volume         = "15",
    year           = "2020",
    pages          = "P09015",
    doi            = "10.1088/1748-0221/15/09/p09015",
    reportNumber   = "CERN-EP-2020-031",
    eprint         = "2004.13447",
    archivePrefix  = "arXiv",
    primaryClass   = "hep-ex",
}

@Article{TRIG-2018-05,
    author         = "{ATLAS Collaboration}",
    title          = "{Performance of electron and photon triggers in ATLAS during LHC Run~2}",
    journal        = "Eur. Phys. J. C",
    volume         = "80",
    year           = "2020",
    pages          = "47",
    doi            = "10.1140/epjc/s10052-019-7500-2",
    reportNumber   = "CERN-EP-2019-169",
    eprint         = "1909.00761",
    archivePrefix  = "arXiv",
    primaryClass   = "hep-ex",
}

@Article{HIGG-2019-09,
    author         = "{ATLAS Collaboration}",
    title          = "{Measurements of Higgs boson production cross-sections in the \(H\to\tau^{+}\tau^{-}\) decay channel in \(pp\) collisions at \(\sqrt{s} = 13\,\text{TeV}\) with the ATLAS detector}",
    journal        = "JHEP",
    volume         = "08",
    year           = "2022",
    pages          = "175",
    doi            = "10.1007/JHEP08(2022)175",
    reportNumber   = "CERN-EP-2021-217",
    eprint         = "2201.08269",
    archivePrefix  = "arXiv",
    primaryClass   = "hep-ex",
}

@Article{HIGG-2019-11,
    author         = "{ATLAS Collaboration}",
    title          = "{Searches for lepton-flavour-violating decays of the Higgs boson into \(e\tau\) and \(\mu\tau\) in \(\sqrt{s} = 13\,\text{TeV}\) \(pp\) collisions with the ATLAS detector}",
    year           = "2023",
    reportNumber   = "CERN-EP-2022-279",
    eprint         = "2302.05225",
    archivePrefix  = "arXiv",
    primaryClass   = "hep-ex",
}

@Article{TRIG-2019-02,
    author         = "{ATLAS Collaboration}",
    title          = "{Performance of the ATLAS Level-1 topological trigger in Run~2}",
    journal        = "Eur. Phys. J. C",
    volume         = "82",
    year           = "2021",
    pages          = "7",
    doi            = "10.1140/epjc/s10052-021-09807-0",
    reportNumber   = "CERN-EP-2021-040",
    eprint         = "2105.01416",
    archivePrefix  = "arXiv",
    primaryClass   = "hep-ex",
}

@Article{TRIG-2019-03,
    author         = "{ATLAS Collaboration}",
    title          = "{The ATLAS Inner Detector Trigger performance in \(pp\) collisions at \(13\,\text{TeV}\) during LHC Run~2}",
    journal        = "Eur. Phys. J. C",
    volume         = "82",
    year           = "2021",
    pages          = "206",
    doi            = "10.1140/epjc/s10052-021-09920-0",
    reportNumber   = "CERN-EP-2021-076",
    eprint         = "2107.02485",
    archivePrefix  = "arXiv",
    primaryClass   = "hep-ex",
}

@Booklet{ATL-SOFT-PUB-2021-001,
    author         = "{ATLAS Collaboration}",
    title          = "{The ATLAS Collaboration Software and Firmware}",
    howpublished   = "{ATL-SOFT-PUB-2021-001}",
    url            = "https://cds.cern.ch/record/2767187",
    year           = "2021",
}

@Booklet{ATL-SOFT-PUB-2021-003,
    author         = "{ATLAS Collaboration}",
    title          = "{ATLAS Computing Acknowledgements}",
    howpublished   = "{ATL-SOFT-PUB-2021-003}",
    url            = "https://cds.cern.ch/record/2776662",
    year           = "2021",
}

@Report{ATLAS-TDR-19,
    author         = "{ATLAS Collaboration}",
    title          = "{ATLAS Insertable B-Layer: Technical Design Report}",
    type           = "ATLAS-TDR-19; CERN-LHCC-2010-013",
    year           = "2010",
    url            = "https://cds.cern.ch/record/1291633",
    related        = "ATLAS-TDR-19-addm",
    relatedstring  = "Addendum:",
}

@Article{CMS-HIG-20-009,
    author         = "{CMS Collaboration}",
    title          = "{Search for lepton-flavor violating decays of the Higgs boson in the \(\mu\tau\) and \(e\tau\) final states in proton--proton collisions at \(\sqrt{s} = 13\,\text{TeV}\)}",
    journal        = "Phys. Rev. D",
    volume         = "104",
    year           = "2021",
    pages          = "032013",
    doi            = "10.1103/PhysRevD.104.032013",
    reportNumber   = "CERN-EP-2021-061",
    eprint         = "2105.03007",
    archivePrefix  = "arXiv",
    primaryClass   = "hep-ex",
}

@Booklet{ATLAS-CONF-2014-018,
    author         = "{ATLAS Collaboration}",
    title          = "{Tagging and suppression of pileup jets with the ATLAS detector}",
    howpublished   = "{ATLAS-CONF-2014-018}",
    url            = "https://cds.cern.ch/record/1700870",
    year           = "2014",
}

@Booklet{ATLAS-CONF-2015-029,
    author         = "{ATLAS Collaboration}",
    title          = "{Selection of jets produced in \(13~\text{TeV}\) proton--proton collisions with the ATLAS detector}",
    howpublished   = "{ATLAS-CONF-2015-029}",
    url            = "https://cds.cern.ch/record/2037702",
    year           = "2015",
}

@Booklet{ATLAS-CONF-2017-029,
    author         = "{ATLAS Collaboration}",
    title          = "{Measurement of the tau lepton reconstruction and identification performance in the ATLAS experiment using \(pp\) collisions at \(\sqrt{s} = 13~\text{TeV}\)}",
    howpublished   = "{ATLAS-CONF-2017-029}",
    url            = "https://cds.cern.ch/record/2261772",
    year           = "2017",
}

@Booklet{ATLAS-CONF-2019-021,
    author         = "{ATLAS Collaboration}",
    title          = "{Luminosity determination in \(pp\) collisions at \(\sqrt{s} = 13\,\text{TeV}\) using the ATLAS detector at the LHC}",
    howpublished   = "{ATLAS-CONF-2019-021}",
    url            = "https://cds.cern.ch/record/2677054",
    year           = "2019",
}

@Booklet{ATL-PHYS-PUB-2014-021,
    author         = "{ATLAS Collaboration}",
    title          = "{ATLAS Pythia~8 tunes to \(7~\text{TeV}\) data}",
    howpublished   = "{ATL-PHYS-PUB-2014-021}",
    url            = "https://cds.cern.ch/record/1966419",
    year           = "2014",
}

@Booklet{ATL-PHYS-PUB-2015-045,
    author         = "{ATLAS Collaboration}",
    title          = "{Reconstruction, Energy Calibration, and Identification of Hadronically Decaying Tau Leptons in the ATLAS Experiment for Run-2 of the LHC}",
    howpublished   = "{ATL-PHYS-PUB-2015-045}",
    url            = "https://cds.cern.ch/record/2064383",
    year           = "2015",
}

@Booklet{ATL-PHYS-PUB-2016-017,
    author         = "{ATLAS Collaboration}",
    title          = "{The Pythia~8 A3 tune description of ATLAS minimum bias and inelastic measurements incorporating the Donnachie--Landshoff diffractive model}",
    howpublished   = "{ATL-PHYS-PUB-2016-017}",
    url            = "https://cds.cern.ch/record/2206965",
    year           = "2016",
}

@Booklet{ATL-PHYS-PUB-2017-007,
    author         = "{ATLAS Collaboration}",
    title          = "{Studies on top-quark Monte Carlo modelling with Sherpa and MG5\_aMC@NLO}",
    howpublished   = "{ATL-PHYS-PUB-2017-007}",
    url            = "https://cds.cern.ch/record/2261938",
    year           = "2017",
}

@Booklet{ATL-PHYS-PUB-2017-013,
    author         = "{ATLAS Collaboration}",
    title          = "{Optimisation and performance studies of the ATLAS \(b\)-tagging algorithms for the 2017-18 LHC run}",
    howpublished   = "{ATL-PHYS-PUB-2017-013}",
    url            = "https://cds.cern.ch/record/2273281",
    year           = "2017",
}

@Booklet{ATL-PHYS-PUB-2019-033,
    author         = "{ATLAS Collaboration}",
    title          = "{Identification of hadronic tau lepton decays using neural networks in the ATLAS experiment}",
    howpublished   = "{ATL-PHYS-PUB-2019-033}",
    url            = "https://cds.cern.ch/record/2688062",
    year           = "2019",
}

@Booklet{ATL-PHYS-PUB-2021-011,
    author         = "{ATLAS Collaboration}",
    title          = "{Evaluating statistical uncertainties and correlations using the bootstrap method}",
    howpublished   = "{ATL-PHYS-PUB-2021-011}",
    url            = "https://cds.cern.ch/record/2759945",
    year           = "2021",
}

\clearpage
 
\begin{flushleft}
\hypersetup{urlcolor=black}
{\Large The ATLAS Collaboration}

\bigskip

\AtlasOrcid[0000-0002-6665-4934]{G.~Aad}$^\textrm{\scriptsize 102}$,
\AtlasOrcid[0000-0002-5888-2734]{B.~Abbott}$^\textrm{\scriptsize 120}$,
\AtlasOrcid[0000-0002-7248-3203]{D.C.~Abbott}$^\textrm{\scriptsize 103}$,
\AtlasOrcid[0000-0002-1002-1652]{K.~Abeling}$^\textrm{\scriptsize 55}$,
\AtlasOrcid[0000-0002-8496-9294]{S.H.~Abidi}$^\textrm{\scriptsize 29}$,
\AtlasOrcid[0000-0002-9987-2292]{A.~Aboulhorma}$^\textrm{\scriptsize 35e}$,
\AtlasOrcid[0000-0001-5329-6640]{H.~Abramowicz}$^\textrm{\scriptsize 151}$,
\AtlasOrcid[0000-0002-1599-2896]{H.~Abreu}$^\textrm{\scriptsize 150}$,
\AtlasOrcid[0000-0003-0403-3697]{Y.~Abulaiti}$^\textrm{\scriptsize 117}$,
\AtlasOrcid[0000-0003-0762-7204]{A.C.~Abusleme~Hoffman}$^\textrm{\scriptsize 137a}$
% ...
\bigskip
\\

% ...
$^{29}$Physics Department, Brookhaven National Laboratory, Upton NY; United States of America.\\
% ...
$^{35}$$^{(a)}$Facult\'e des Sciences Ain Chock, R\'eseau Universitaire de Physique des Hautes Energies - Universit\'e Hassan II, Casablanca;$^{(b)}$Facult\'{e} des Sciences, Universit\'{e} Ibn-Tofail, K\'{e}nitra;$^{(c)}$Facult\'e des Sciences Semlalia, Universit\'e Cadi Ayyad, LPHEA-Marrakech;$^{(d)}$LPMR, Facult\'e des Sciences, Universit\'e Mohamed Premier, Oujda;$^{(e)}$Facult\'e des sciences, Universit\'e Mohammed V, Rabat;$^{(f)}$Institute of Applied Physics, Mohammed VI Polytechnic University, Ben Guerir; Morocco.\\
% ...
$^{55}$II. Physikalisches Institut, Georg-August-Universit\"{a}t G\"ottingen, G\"ottingen; Germany.\\
% ...
$^{102}$CPPM, Aix-Marseille Universit\'e, CNRS/IN2P3, Marseille; France.\\
$^{103}$Department of Physics, University of Massachusetts, Amherst MA; United States of America.\\
% ...
$^{117}$Department of Physics, New York University, New York NY; United States of America.\\
% ...
$^{120}$Homer L. Dodge Department of Physics and Astronomy, University of Oklahoma, Norman OK; United States of America.\\
% ...
$^{137}$$^{(a)}$Departamento de F\'isica, Pontificia Universidad Cat\'olica de Chile, Santiago;$^{(b)}$Millennium Institute for Subatomic physics at high energy frontier (SAPHIR), Santiago;$^{(c)}$Instituto de Investigaci\'on Multidisciplinario en Ciencia y Tecnolog\'ia, y Departamento de F\'isica, Universidad de La Serena;$^{(d)}$Universidad Andres Bello, Department of Physics, Santiago;$^{(e)}$Instituto de Alta Investigaci\'on, Universidad de Tarapac\'a, Arica;$^{(f)}$Departamento de F\'isica, Universidad T\'ecnica Federico Santa Mar\'ia, Valpara\'iso; Chile.\\
% ...
$^{150}$Department of Physics, Technion, Israel Institute of Technology, Haifa; Israel.\\
$^{151}$Raymond and Beverly Sackler School of Physics and Astronomy, Tel Aviv University, Tel Aviv; Israel.\\
% ...
$^{a}$ Also Affiliated with an institute covered by a cooperation agreement with CERN.\\
$^{b}$ Also at An-Najah National University, Nablus; Palestine.\\
$^{c}$ Also at Borough of Manhattan Community College, City University of New York, New York NY; United States of America.\\
$^{d}$ Also at Bruno Kessler Foundation, Trento; Italy.\\
$^{e}$ Also at Center for High Energy Physics, Peking University; China.\\
$^{f}$ Also at Center for Interdisciplinary Research and Innovation (CIRI-AUTH), Thessaloniki ; Greece.\\
% ...

\end{flushleft}


\end{document}